\documentclass{aa}  
\usepackage{graphicx}
\usepackage{amsmath}
\usepackage{amsfonts}
\usepackage{amssymb}
\usepackage{gensymb}
\usepackage{textcomp}
\usepackage[varg]{txfonts}
\usepackage{natbib}
\usepackage{epstopdf}
\bibpunct{(}{)}{;}{a}{}{,}
\begin{document}

\title{Impulsive coronal heating during the interaction of surface magnetic fields in the lower solar atmosphere}

\author{L.~P.~Chitta\inst{1}, H.~Peter\inst{1}, E.~R.~Priest\inst{2}, \and S.~K.~Solanki\inst{1,3}}

\institute{Max-Planck-Institut f\"ur Sonnensystemforschung, Justus-von-Liebig-Weg 3, 37077 G\"ottingen, Germany\\
\email{chitta@mps.mpg.de} 
\and
St Andrews University, Mathematics Institute, St Andrews KY16 9SS, UK
\and
School of Space Research, Kyung Hee University, Yongin 446-701, Gyeonggi, Korea
}

   \date{Received 4 August 2020 / Accepted 23 October 2020}

\abstract
{Coronal plasma in the cores of solar active regions is impulsively heated to more than 5\,MK. The nature and location of the magnetic energy source responsible for such impulsive heating is poorly understood. Using observations of seven active regions from the Solar Dynamics Observatory, we found that a majority of coronal loops hosting hot plasma have at least one footpoint rooted in regions of interacting mixed magnetic polarity at the solar surface. In cases when co-temporal observations from the Interface Region Imaging Spectrograph space mission are available, we found spectroscopic evidence for magnetic reconnection at the base of the hot coronal loops. Our analysis suggests that interactions of magnetic patches of opposite polarity at the solar surface and the associated energy release during reconnection are key to impulsive coronal heating.}

   \keywords{magnetic reconnection --- Sun: chromosphere --- Sun: corona --- Sun: magnetic fields --- Sun: transition region --- Sun: flares}
   \titlerunning{Impulsive coronal heating during the interaction of surface magnetic fields}
   \authorrunning{L. P. Chitta et al.}

   \maketitle

\section{Introduction\label{sec:int}}

The origin of hot plasma in the solar corona is an open question. On average, the quiescent solar corona requires an energy flux of a few $10^5$\,erg\,cm$^{-2}$\,s$^{-1}$ to support the radiative and conductive losses, whereas in active regions that host stronger concentrations of magnetic field at the solar surface, the coronal energy requirements are two orders of magnitude higher \citep[][]{1977ARA&A..15..363W}. Magnetic fields have to supply the required energy flux to sustain hot coronal plasma. A widely accepted hypothesis is that coronal magnetic loops become tangled and braided as their footpoints at the solar surface are slowly moved and stressed by photospheric convective motions. Energy stored in these magnetic braids is then impulsively and intermittently released through reconnection as coronal nanoflares \citep[][]{1988ApJ...330..474P}.
Thus, coronal nanoflares produced by tangled magnetic fields are one possible candidate to explain the intermittent heating of plasma to several million degrees Kelvin, particularly in active regions where energy requirements are higher \citep[][]{1988ApJ...330..474P,2006SoPh..234...41K,2013Natur.493..501C}. 

Observations of transient soft X-ray brightenings \citep[][]{1992PASJ...44L.147S}, the detection of hot plasma over 10\,MK\ in the core of an active region \citep[][]{2017NatAs...1..771I}, and rapid moss variability on timescales below 60\,s\ in the ultraviolet (UV) and extreme-UV (EUV) emission at the footpoints of hot loops with near-simultaneous brightening at both footpoints \citep[][]{2014Sci...346B.315T,2020ApJ...889..124T,2018ApJ...857..137G,2018A&A...615L...9C} are possible signatures of impulsive nanoflares. In the traditional view, it is generally thought that footpoints of impulsively heated nanoflaring loops end in regions of unipolar magnetic fields at the solar surface and that energy is loaded into the magnetic loop by the slow driving of footpoints \citep[][]{2006SoPh..234...41K}. There is some observational evidence for unipolar magnetic fields at the footpoints of hot coronal loops \citep[][]{2005ApJ...621..498K,2008ApJ...689L..77B}, but this conclusion is limited by the spatial resolution and magnetic sensitivity of the instrumentation \citep[][]{2017ApJS..229....4C}. In that case, numerical models suggest that a secondary instability could impulsively heat the loop when the magnetic misalignment angle in the corona exceeds 45\textdegree\ \citep[][]{2005ApJ...622.1191D}. In addition, it has been suggested that the observation that both footpoints of the loop brighten near-simultaneously together with imaging and spectroscopic signatures of rapid moss variability is consistent with coronal braiding nanoflares \citep[][]{2020ApJ...889..124T}.

A closer examination of recently reported nanoflaring loops, however, revealed that their footpoints end in regions of interacting mixed magnetic polarity that exhibit flux cancellation \citep[][]{2018A&A...615L...9C}. A similar process of interacting mixed polarities and flux cancellation is argued to trigger intermittent coronal brightenings in magnetic braids \citep[][]{2014ApJ...795L..24T} and flare-like transients in an emerging flux region \citep[][]{2011ApJ...726...12E}. Furthermore, observations suggest that bright coronal loops in soft X-rays in active regions are apparently rooted near polarity inversion lines \citep[][see also \citealt{2017ApJ...843L..20T}]{1997ApJ...482..519F}. Based on a visual inspection, it has been recently suggested that the Fe\,{\sc xviii} EUV emission intensity of the brightest loops is correlated with the presence of sunspots, along with the emergence or cancellation of magnetic flux \citep[][]{2019ApJ...881..107A}.

\begin{table*}
\caption{Overview of the observed active regions.\label{tab:over}}
\centering
\begin{tabular}{l c c}
\hline\hline
NOAA No. & Period & No. of analysed events \\
\hline  
12665 & 2017 July 11 UT 00:00 to 2017 July 12 UT 10:00 & 21 \\
12692 & 2017 December 23 UT 00:00 to 2017 December 26 UT 00:00 & 22 \\
12699 & 2018 February 11 UT 00:00 to 2018 February 12 UT 00:00 & 11 \\
12712 & 2018 May 27 UT 18:00 to 2018 June 02 UT 18:00 & 32 \\
12713 & 2018 June 17 UT 00:00 to 2018 June 18 UT 10:00 & 14 \\
12733 & 2019 January 24 UT 00:00 to 2019 January 26 UT 00:00 & 18 \\
12738 & 2019 April 12 UT 00:00 to 2019 April 14 UT 00:00 & 19\\
\hline  
\end{tabular}
\end{table*}

\begin{table*}
\caption{Overview of the IRIS observations.\label{tab:irisover}}
\centering
\begin{tabular}{l c c c c c}
\hline\hline
NOAA No. & Start time & Position & Scan step & Exposure time & SJI cadence \\
\hline
12692 & 2017 December 23 UT 08:39 & $(-200\arcsec,328\arcsec)$ & 0.35\arcsec & 8\,s & 37\,s \\ 
12692 & 2017 December 23 UT 10:16 & $(-186\arcsec,329\arcsec)$ & 0.35\arcsec & 8\,s & 37\,s \\
12699 & 2018 February 11 UT 23:26 & $(213\arcsec,-11\arcsec)$ & 0.35\arcsec & 8\,s & 37\,s \\
12712\tablefootmark{a} & 2018 May 29 UT 00:00 & $(-293\arcsec,274\arcsec)$ & $-$ & $-$ & 125\,s \\
12712 & 2018 May 29 UT 14:57 & $(-176\arcsec,280\arcsec)$ & 0.35\arcsec & 15\,s & 67\,s \\
12712 & 2018 May 31 UT 06:07 & $(152\arcsec,280\arcsec)$ & 0.35\arcsec & 8\,s & 37\,s \\
12713\tablefootmark{b} & 2018 June 17 UT 14:23 & $(-128\arcsec,56\arcsec)$ & 1\arcsec & 60\,s & 245\,s \\
12738 & 2019 April 11 UT 23:52 & $(-380\arcsec,171\arcsec)$ & 0.35\arcsec & 8\,s & 37\,s \\
\hline
\end{tabular}
\tablefoot{
\tablefoottext{a}{Spectroscopic data not considered as the IRIS slit crossed the loop almost 60\,minutes after the event}
\tablefoottext{b}{SJI data not considered due to very low cadence}
}
\end{table*}

Inspired by recent discoveries that flux cancellation is much more common than had previously been recognised and that bright loops often have footpoints in regions where opposite-polarity magnetic flux at the solar surface is cancelling, a theoretical model for the creation of nanoflares by flux cancellation rather than magnetic braiding has been developed \citep[][]{2018ApJ...862L..24P,syntelis19a,syntelis20}. It considers what happens when opposite-polarity fragments approach one another in an ambient magnetic field. On the Sun, flux emergence could naturally lead to such conditions when magnetic fragments of one polarity from an emerging flux feature approach and interact with the ambient magnetic field of opposite polarity \citep[e.g.][]{2019A&A...623A.176C}. Initially, the fragments are so far apart that they are not linked magnetically. As soon as they become linked, their approach drives magnetic reconnection in the overlying atmosphere in an initial pre-cancellation phase at a separator or quasi-separator, which rises to a maximum height and then moves down to the photosphere. The resulting heating and jet acceleration can occur in the photosphere, chromosphere, transition region, or corona in a way that is evaluated in the model and depends on the sizes of the magnetic fluxes, their separation, and the strength of the overlying magnetic field. The height of the separator and the rate of energy release are calculated analytically and verified in 2D and 3D computational experiments \citep[][]{syntelis19a,syntelis20}. Then a second cancellation phase occurs as the opposite-polarity fragments undergo actual cancellation and thus drive reconnection at the photospheric footpoints and in the overlying atmosphere. This `cancellation nanoflare model', conditions for which could be set by previous flux emergence events, is able to qualitatively account for chromospheric and coronal heating and for a wide variety of dynamic jet-like phenomena throughout the atmosphere \citep[][]{2018ApJ...862L..24P,syntelis19a,syntelis20}. 

Here we explore the magnetic landscape at the solar surface around the footpoints of hot loops to shed light on the nature and location of the magnetic energy source that is likely to be responsible for impulsive coronal heating to several million Kelvin. In particular, we  address the role of emergence and cancellation of magnetic flux and associated reconnection at the foot points of impulsive hot loops from an observational perspective. We go beyond isolated case studies \citep[e.g.][]{2017ApJS..229....4C,2018A&A...615L...9C} and instead consider a large number of events in an attempt to estimate how commonly signatures of magnetic reconnection occur at loop footpoints in conjunction with impulsive coronal heating. 

\section{Observations and methods\label{sec:obs}}

We investigate the magnetic roots of impulsively heated loops in the cores of active regions (ARs), observed with the Solar Dynamics Observatory \citep[SDO;][]{2012SoPh..275....3P}. We consider seven isolated ARs during the declining phase of solar cycle 24 between 2017-19. Each AR is chosen to be the only prominent one visible on the solar disk during the chosen period of time (Table\,\ref{tab:over}). To identify impulsive hot loops, we use a combination of EUV images obtained with the 94\,\AA, 171\,\AA, and 193\,\AA\ filters of the Atmospheric Imaging Assembly \citep[AIA;][]{2012SoPh..275...17L} on board SDO. The AIA 171\,\AA\ filter is sensitive to plasma emission just below 1\,MK. The AIA 193\,\AA\ channel images coronal features with temperatures around 1.6\,MK. The AIA 94\,\AA\ filter response has contribution from warmer plasma around 1\,MK and hot plasma at 7.1\,MK \citep[][]{2012SoPh..275...41B}. To investigate the lower-atmosphere connection of the loops, we use AIA 1600\,\AA\ and 1700\,\AA\ UV filters. To identify the magnetic roots of hot loops we use line-of-sight magnetic field maps from the Helioseismic and Magnetic Imager \citep[HMI;][]{2012SoPh..275..207S} on board SDO. The SDO data are retrieved from the Joint Science Operations Center (JSOC) web interface\footnote{\url{http://jsoc.stanford.edu}}. Patches covering an area of 37.5\textdegree$\times$37.5\textdegree\ in longitude and latitude are extracted and remapped using Plate Carr\'{e}e map projection with map scale set to 0.05\textdegree\,pixel$^{-1}$ (corresponding to roughly 0.833\arcsec\,pixel$^{-1}$ at the disk centre; $1\arcsec\approx725$\,km at the disk centre). These patches are tracked at the Carrington rate to remove the rotation of the Sun. The Carrington longitude and latitude and the reference time for tracking correspond to those when the AR is closest to the central meridian. These processing steps are all completed using the JSOC data export module. These AIA EUV, UV, and HMI magnetic field data have cadences of 120\,s, 96\,s, and 90\,s, respectively. The AIA EUV, and UV data are normalised by their respective exposure times. We note that AIA produces data at higher native cadence (EUV images at 12\,s and UV images at 24\,s cadence), and HMI produces line-of-sight magnetic field observations at 45\,s cadence. Here we work with lower-cadence data mainly to keep the data volume manageable.

For coronal diagnostics of impulsive heating, we extract the hot plasma component (due to Fe\,{\sc xviii} emission) present in the 94\,\AA\ channel using an empirical method \citep[][]{2012ApJ...759..141W}. The method removes the warm plasma component (around 1\,MK) from the emission detected in the 94\,\AA\ channel. The Fe\,{\sc xviii} line is generally thought to form at a temperature of about 7.1\,MK, which would indicate the presence of hot plasma in the active region core \citep[e.g.][]{2012ApJ...759..141W}, although in some cases, the Fe\,{\sc xviii} line is formed at lower temperatures of 3\,MK \citep[][]{2013A&A...558A..73D}. Because of the quiescent nature of the selected ARs, the solar-disk integrated X-ray flux recorded by the 1-8\,\AA\ band of the Geostationary Operational Environmental Satellite (GOES) was weak, at B-class level ($10^{-7}$\,W\,m$^{-2}$) or lower. This relative quiescence helps us to associate reasonably well the impulsive heating events in the active region core with any X-ray events detected by GOES, the faintest of which (A-class events) reach temperatures in excess of 5\,MK \citep[][]{1996ApJ...460.1034F}. 

Our aim is to connect the footpoints of hot loops in AR cores to the underlying surface magnetic field in an automated way, in order to investigate the energy source of impulsive coronal heating to several million Kelvin. This automated method consists of three main steps. (1) Identification of impulsive heating events in AR cores. (2) Identifying the footpoints of impulsively heated loops. (3) Analysing surface magnetic field distribution and its evolution during the course of impulsive heating.

\begin{figure*}
\begin{center}
\includegraphics[width=0.5\textwidth]{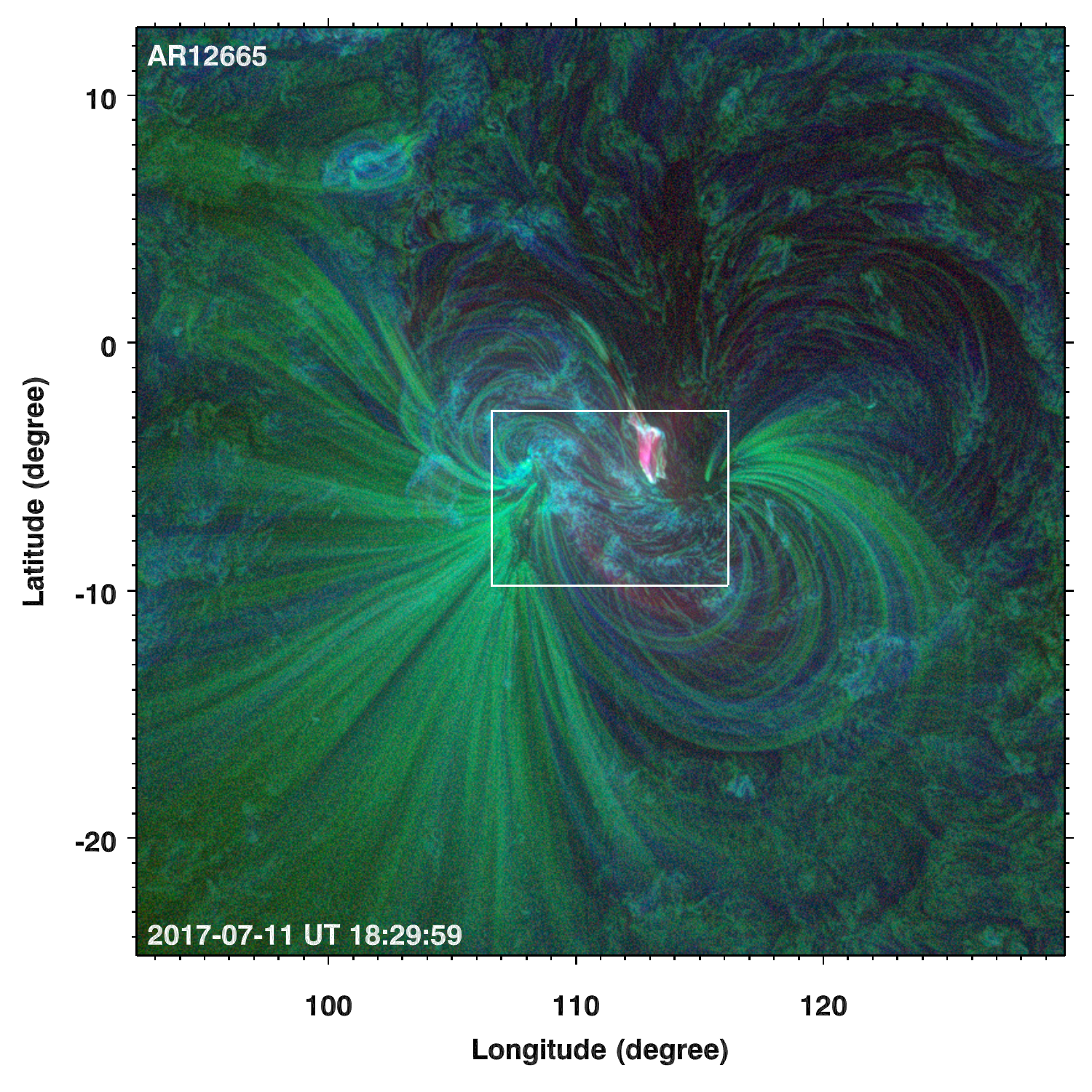}
\includegraphics[width=\textwidth]{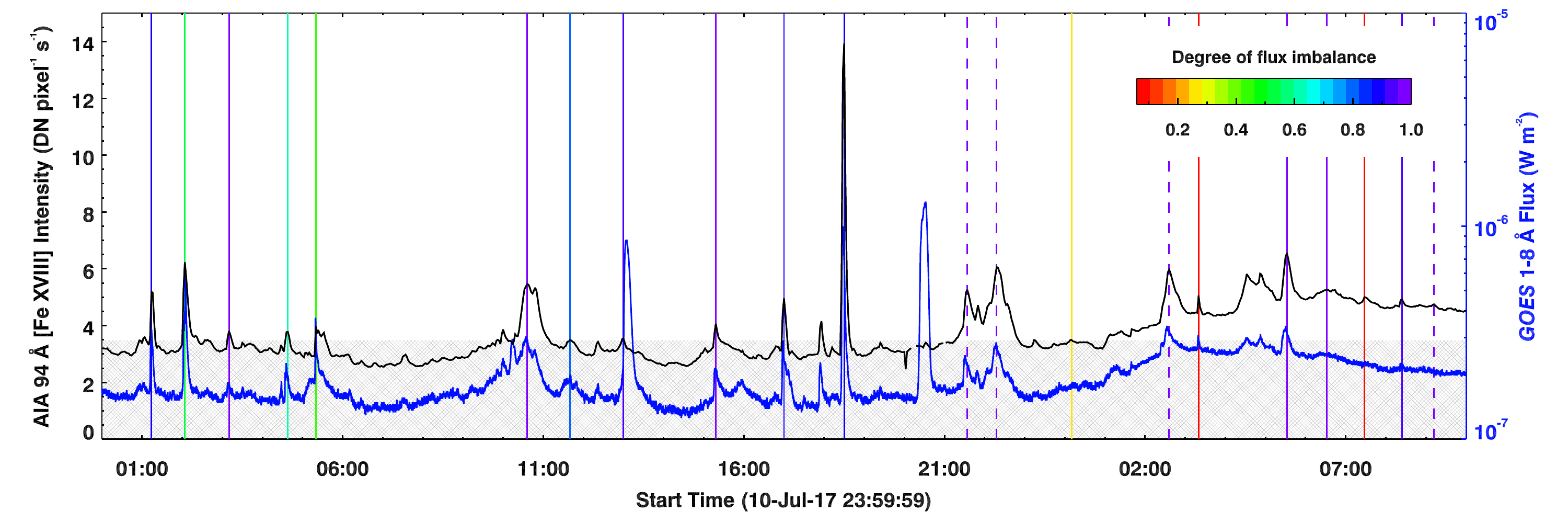}
\caption{Overview of AR 12665 and impulsive heating events observed therein. The upper panel displays a three-channel composite map of the AR observed with SDO/AIA on 2017 July 11. The EUV emission recorded by three filters, AIA 171\,\AA\ (green), AIA 193\,\AA\ (blue), and AIA 94\,\AA\ Fe\,{\sc xviii} proxy (red), highlight various regions in the AR. The core of the AR is marked by a white rectangle. The lower panel shows the time series of core-integrated Fe\,{\sc xviii} EUV emission (black curve) for a period of 34\,hours. The blue curve represents the time series of solar disk integrated GOES 1-8\,\AA\ X-ray flux. Various vertical lines mark local peaks identified in the Fe\,{\sc xviii} light curve. Based on corresponding SDO/HMI observations, the solid vertical lines mark mixed polarity events (i.e. when at least one footpoint of the brightening loops had a mixed polarity), dashed lines identify unipolar events. These vertical lines are colour-coded with the degree of photospheric magnetic flux imbalance. Any Fe\,{\sc xviii} peak entirely within the grey shaded band is not considered in our analysis. See Sect.\,\ref{sec:obs} and Appendix \,\ref{sec:met} for details.\label{fig:over1}}
\end{center}
\end{figure*}

\begin{figure*}
\begin{center}
\includegraphics[width=\textwidth]{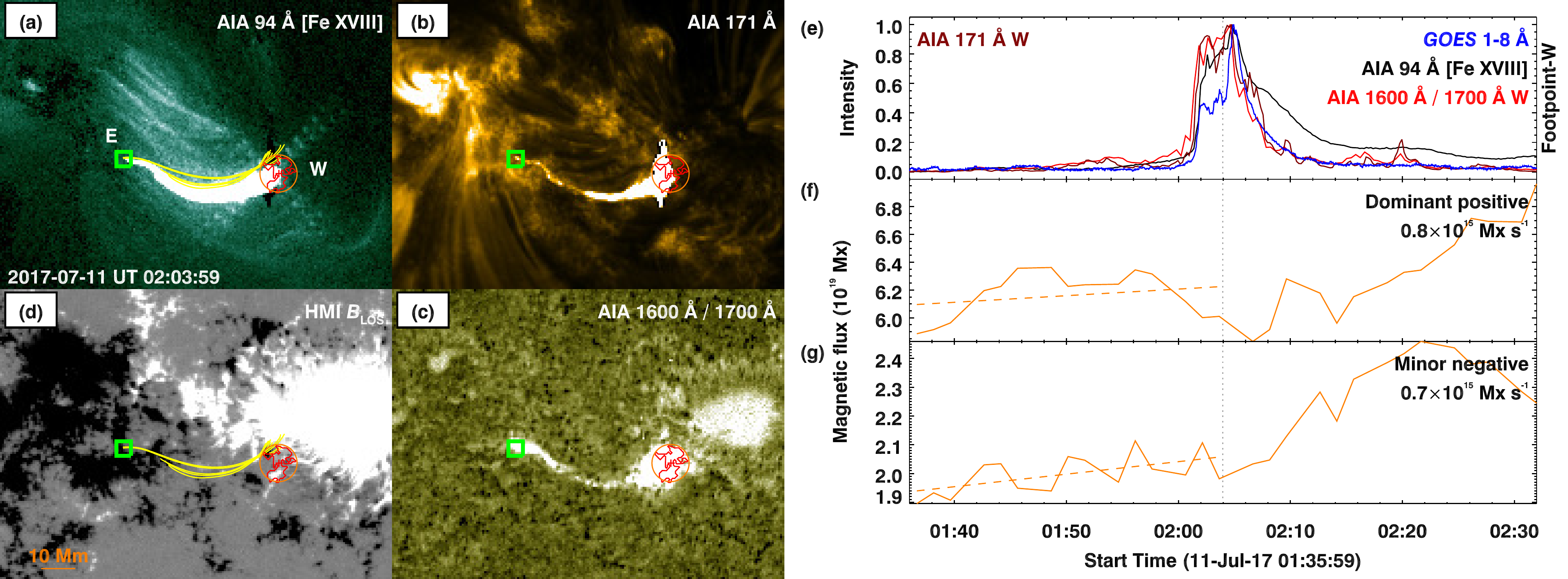}
\includegraphics[width=\textwidth]{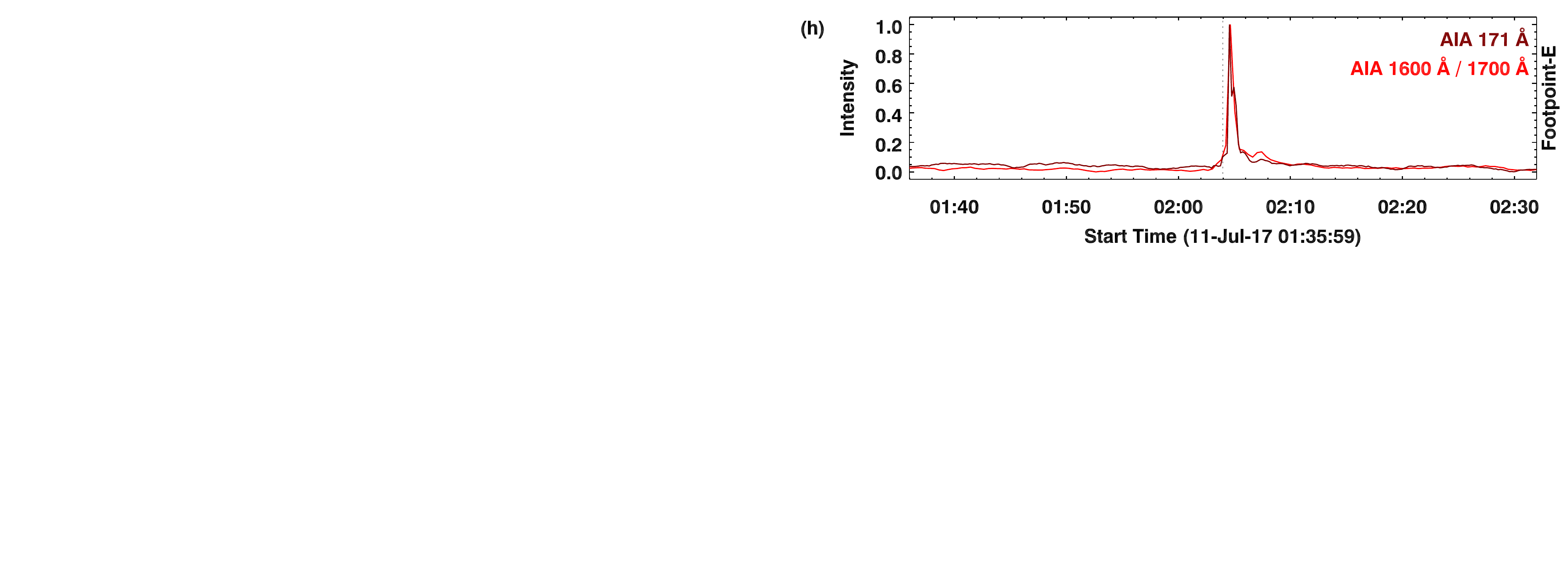}
{\vskip-4cm
\caption{Impulsive heating in the core of AR 12665. Panels (a)-(d) are snapshots of various observables in the core close to the peak in Fe\,{\sc xviii} intensity (see Fig.\,\ref{fig:over1}). Maps of AIA 94\,\AA\ Fe\,{\sc xviii} (panel a), AIA 171\,\AA\ (panel b), ratio of AIA 1600\,\AA\ to 1700\,\AA\ (UV ratio; panel c), and with a grey-scale an HMI line-of-sight magnetic field map (panel d) are displayed. The magnetic flux density is saturated at $\pm300$\,G. White (dark) shaded regions represent positive (negative) magnetic polarity. In panel (a), E and W label eastern and western footpoints of the loop. The red contour outlines a contiguous patch in the UV ratio map detected by our method. The plus symbol marks the centroid of the contoured region. The circle with radius $\approx$5.4\,Mm, about the centroid, outlines the region used for magnetic analysis. A 10\,Mm scale is overlaid. In panels (a) and (d), the yellow coloured curves represent magnetic field lines traced from a linear force free field extrapolation. Panels (e)-(g) show the time series of various observables. Panel (e) shows disk-integrated GOES 1-8\,\AA\ X-ray flux (blue), Fe\,{\sc xviii} emission integrated over the AR's core (black), UV ratio (red) and AIA 171\,\AA\ (maroon) emission integrated over the contoured region at the western footpoint. Panels (f) and (g) show integrated magnetic flux as a function of time of the dominant (positive) and minor (negative) magnetic polarities within the circled region overlaid on panel (d). The dashed line is a linear fit to the magnetic flux curve, with its slope quoted in the top right corner. Panel (h) shows the UV ratio and AIA 171\,\AA\ emission from the eastern (E) footpoint regions marked in panel (a). The vertical dotted line in panels (e)-(h) marks the peak of Fe\,{\sc xviii} as identified in the 120\,s cadence data. For display purposes, in panels (e) and (h) we plot AIA data at their native cadence (12\,s EUV and 24\,s UV). Animation of panels (a) to (g) is available online. See Sects.\,\ref{sec:obs}, \ref{sec:simp} and Appendix\,\ref{sec:met} for details. \label{fig:AR12665c1}}}
\end{center}
\end{figure*}

In the first step, we begin with the time series of Fe\,{\sc xviii} emission integrated over the core of an AR. The size and location of the core are chosen by visual inspection. An overview of one AR along with the selected core region is presented in the upper panel of Fig.\,\ref{fig:over1} \citep[the contrast of AIA images is enhanced using a multi-scale Gaussian normalisation technique developed by][]{2014SoPh..289.2945M}. Our selection of the core is validated by comparing the Fe\,{\sc xviii} time series with the GOES X-ray flux (lower panel of Fig.\,\ref{fig:over1}). Both diagnostics show correlated intensity variations, suggesting that the selected core to be the source region of events seen in the disk integrated GOES X-ray flux produced by hot plasma at temperatures over 5\,MK. We identify local maxima in the core-integrated Fe\,{\sc xviii} series to detect episodes of impulsive heating. We impose a peak separation of 40\,minutes or longer and also set a lower intensity threshold while selecting the local maxima (grey shaded band in the lower panel of Fig.\,\ref{fig:over1}). A detected peak is discarded if it is smaller than an adjacent local peak within $\pm30$\,minutes. The remaining peaks selected for analysis are marked by vertical lines in the bottom panel of Fig.\,\ref{fig:over1}. In total, we identified 137 impulsive hot loops from seven active regions (the Fe\,{\sc xviii} peak times in UT of all the identified events are listed in Tables \ref{tab:AR12665}, \ref{tab:AR12692}, \ref{tab:AR12699}, \ref{tab:AR12712}, \ref{tab:AR12713}, \ref{tab:AR12733}, and \ref{tab:AR12738}). For each of these peaks, we extract 1\,hour segments (i.e. $\pm30$\,minutes) of AIA and HMI data of the core region. A closer look at the spatial maps of the core of AR 12665 in various SDO diagnostics is presented in Fig.\,\ref{fig:AR12665c1}. 

One way to locate the footpoints of hot loops is by visual means, that is, examining and manually locating footpoints in all cases. Given the large number of impulsive events we identified, this visually locating footpoints is rather impractical and time consuming. We therefore resort to an automated method for footpoint identification using images from the AIA 1600\,\AA\ and 1700\,\AA\ filters. In quiescent-Sun regions including quiescent plages, AIA UV emission in the 1600\,\AA\ and 1700\,\AA\ filters is dominated by photospheric continuum \citep[][]{2012SoPh..275...41B,2019ApJ...870..114S}. In flaring conditions, however, both filters record chromospheric and transition region emission. In particular, the AIA 1600\,\AA\ filter is dominated by emission from the C\,{\sc iv} 1550\,\AA\ doublet forming at $10^5$\,K and the Si\,{\sc i} continuum \citep[][]{2019ApJ...870..114S}. Since the ratio of their filter responses peaks sharply at around 1550\,\AA, the intensity ratio of the 1600\,\AA\ to 1700\,\AA\ filters (henceforth, UV ratio) traces the signal mostly from the C\,{\sc iv} doublet. Signatures of UV bursts and rapid moss variability in the lower atmosphere can be traced in these UV ratio maps \citep{2018A&A...615L...9C,2018A&A...617A.128S}.
We use these UV ratio maps in the second step to identify the footpoints of hot loops in the lower atmosphere. From the extracted 1-hour segments of UV ratio maps covering the core, we cross-correlate the time series of UV ratio signal at each spatial pixel with the corresponding time series of core-integrated Fe\,{\sc xviii} emission. The resulting 2D cross-correlation map is further processed to identify a single contiguous patch containing at least nine spatial pixels with cross-correlation values of at least 0.5 or higher (see Appendix\,\ref{sec:lowat} for details). Based on visual inspection, we found that such patches are indeed associated with the footpoint regions of overlying hot coronal loops. In Figs.\,\ref{fig:AR12665c1}(a)-(d) we overlay the (red) contours of the detected patch to demonstrate its association with the overlying hot loop in the core of AR 12665. The centroid position of the patch is marked with a plus symbol. Similarly, the centroid position (Solar $X,Y$, with disk centre as the origin) of all the detected patches associated with hot loops in the analysed seven ARs are listed in Tables \ref{tab:AR12665}, \ref{tab:AR12692}, \ref{tab:AR12699}, \ref{tab:AR12712}, \ref{tab:AR12713}, \ref{tab:AR12733}, and \ref{tab:AR12738}). 

Finally, to investigate the magnetic field distribution at the footpoints of hot loops, in this third step, we analyse the polarity of the line-of-sight magnetic field (i.e. whether unipolar or mixed-polarity) in a circular region of radius 5.4\,Mm, using the centroid position as its centre. The size of the circle is chosen to encompass the footpoints of hot loops, which typically extend over 10\,Mm \citep[e.g.][]{2000SoPh..193...53K}. In particular, we compute the degree of magnetic flux imbalance at the detected footpoint region to analyse whether the line-of-sight component of the local magnetic field is unipolar or mixed. For the computation of flux imbalance, we consider only those HMI pixels with a significant magnetic flux density of more than 30\,G (corresponding to $10^{17}$\,Mx). Furthermore, to ensure that the flux imbalance measure is robust, it is computed using a 30-minute time-averaged line-of-sight magnetic field map corresponding to each event. The details of the magnetic field analysis are presented in Appendix\,\ref{sec:fluxim}. We label the overlying impulsive heating event type as unipolar or mixed, depending on whether the surface magnetic field distribution at the detected footpoint is unipolar or mixed. Those cases in which the footpoint could not be determined in the UV ratio map are labelled as unclear. The event type of all the analysed hot loops is listed in Tables \ref{tab:AR12665}, \ref{tab:AR12692}, \ref{tab:AR12699}, \ref{tab:AR12712}, \ref{tab:AR12713}, \ref{tab:AR12733}, and \ref{tab:AR12738}). If the event is identified as mixed (i.e. the detected footpoint overlies mixed magnetic polarities), we determine the dominant and minor magnetic polarities and compute rate of change of magnetic flux of the respective polarities over a period of 30\,minutes leading to the peak of core-integrated Fe\,{\sc xviii} emission for a given event (for details see Appendix\,\ref{sec:rate}). The rate of change of magnetic fluxes thus computed for the dominant (\textbf{D}) and minor (\textbf{M}) are listed in Tables \ref{tab:AR12665}, \ref{tab:AR12692}, \ref{tab:AR12699}, \ref{tab:AR12712}, \ref{tab:AR12713}, \ref{tab:AR12733}, and \ref{tab:AR12738}). Our method is applied to the lower cadence SDO data. In some cases, post-analysis, we use higher cadence AIA data for display purposes; these cases are noted in the respective figure captions. 

We also consider level-2 imaging and spectroscopic data from the Interface Region Imaging Spectrograph\footnote{\url{https://iris.lmsal.com/}} \cite[IRIS;][]{2014SoPh..289.2733D} for eight of the heating events (see Table\,\ref{tab:irisover}). The timing of these IRIS observations overlaps with the analysed impulsive heating episodes. The footpoint regions are examined using Si\,{\sc iv}\,1394\,\AA\ spectral line that traces the transition region around 0.08\,MK, and Mg\,{\sc ii}\,k and Mg\,{\sc ii}\,triplet lines that sample chromospheric plasma around $10^4$\,K. We also make use of IRIS slit-jaw 1400\,\AA\ and 2796\,\AA\ time series to analyse footpoint regions of hot loops.

\section{Examples for impulsive heating in active region cores\label{sec:cases}}

We analysed a large variety of impulsive hot loop brightenings in the cores of seven ARs along with their magnetic field evolution. Though the detailed morphology of hot loops and the distribution of underlying magnetic field differs from case to case, the analysed impulsive heating events and their association with footpoint magnetic fields can be broadly grouped into simple and complex categories. Here we discuss two representative examples from each category and present more (similar) cases in Appendix\,\ref{sec:exam}.

\subsection{An example with an apparently simple photospheric magnetic structure\label{sec:simp}}

We followed AR 12665 for a period of 34\,hours and analysed 21 impulsively heated loops and their footpoint magnetic field evolution (see Fig.\,\ref{fig:over1} and Table\,\ref{tab:AR12665} for an overview of events in this AR). We first present an example from AR 12665 that at first sight seems to fit well into the traditional view of nanoflares initiated by stressing the magnetic field at the foot points, with only a single magnetic polarity per foot point.
Fig.\,\ref{fig:AR12665c1}  shows one case from the set of 21 analysed heating events from AR 12665. A rapid brightening of a loop is observed in the Fe\,{\sc xviii} emission map that peaked on 2017 July 11 around 02:04\,UT\ and connects footpoints E and W (panel a). The Fe\,{\sc xviii} emission from the loop suggests that it would have reached temperatures of at least 3\,MK and possibly even more than 7\,MK. The AIA 171\,\AA\ and UV ratio maps display footpoint brightenings associated with the loop (panels b and c). The line-of-sight magnetic field map shows that footpoint-E is rooted in a negative magnetic polarity region and footpoint-W in a dominant positive magnetic polarity region (panel d). Thus at first sight, the loop appears to fit the traditional picture of a magnetic loop connecting two unipolar regions at the solar surface. 

Both the disk-integrated GOES X-ray flux and the core-integrated Fe\,{\sc xviii} emission show concurrent and rapid intensity enhancements associated with the heating of the loop (panel e)\footnote{For display purpose we show AIA light curve data at their highest native cadence, for the analysis, however, we used lower cadence AIA data, as described in Sect.\,\ref{sec:obs}.}. The whole event lasted for about 10\,minutes. The GOES X-ray flux displayed short-term ($\approx$60\,s) fluctuations around the peak time, which are also seen in the AIA EUV and UV diagnostics. In this case, our automated method detected a contiguous patch in the UV ratio map directly at footpoint-W (red contour in panel c). The UV ratio signal from this patch correlates well with the core integrated Fe\,{\sc xviii} emission (red curve in panel e). The evolution of AIA 171\,\AA\ emission from the same patch is similar to its counterparts (maroon curve in panel e). The near-simultaneous brightening in the UV, EUV and X-ray diagnostics suggests that the impulsive loop heating is multi-thermal in this case.

\begin{figure*}
\begin{center}
\includegraphics[width=0.5\textwidth]{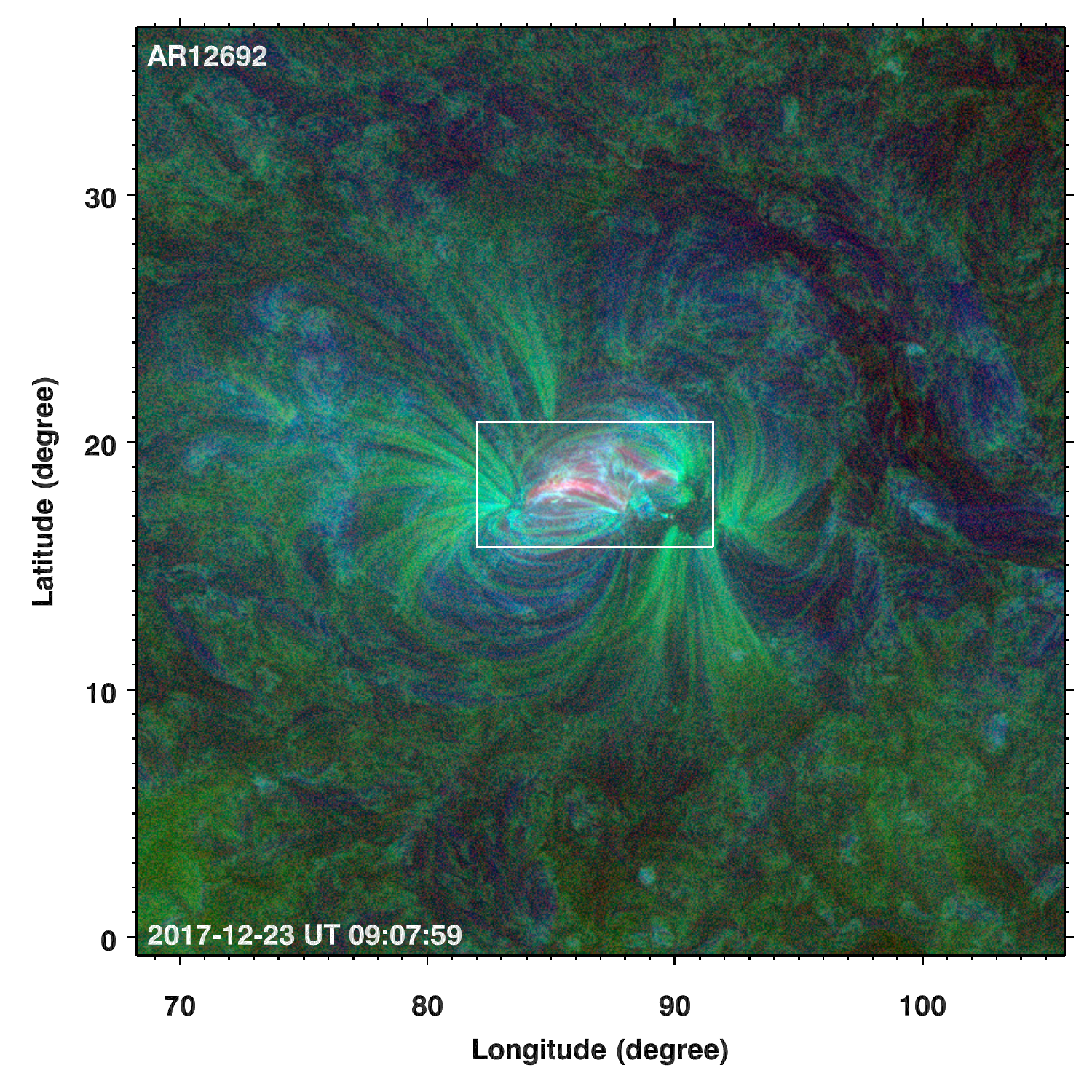}
\includegraphics[width=\textwidth]{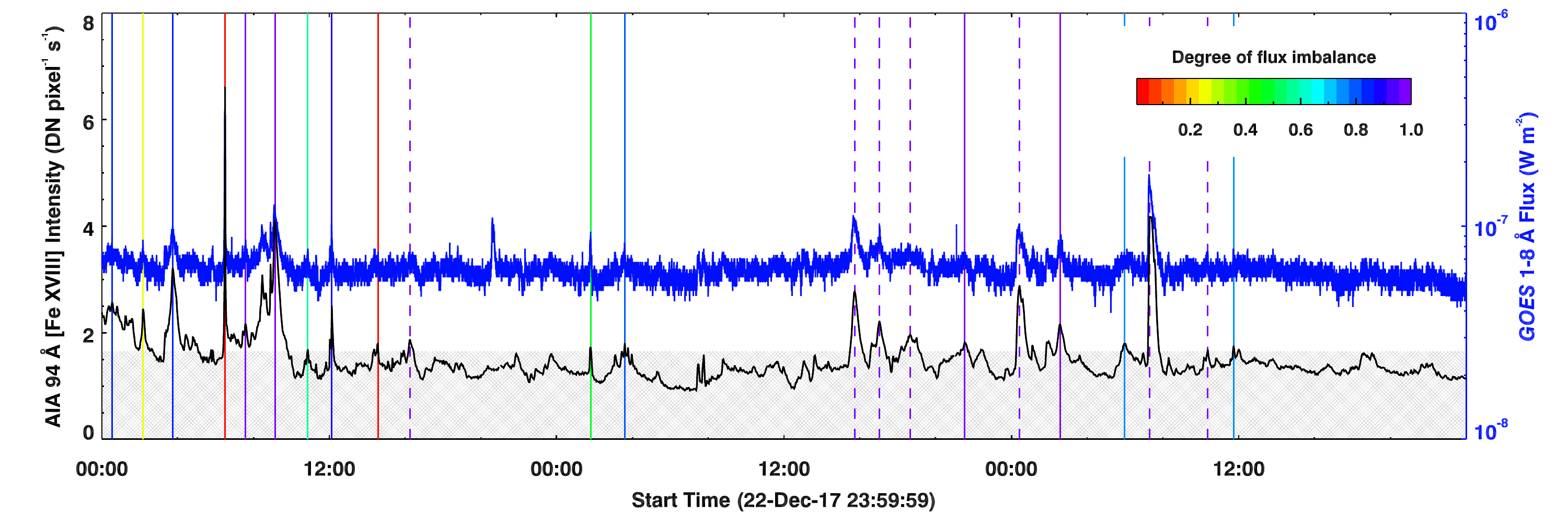}
\caption{Overview of impulsive heating events observed in AR 12692. The format is the same as in Fig.\,\ref{fig:over1}. \label{fig:over2}}
\end{center}
\end{figure*}

After the automated-identification of footpoint-W, we visually located the conjugate footpoint of this hot loop to compare the two regions. This conjugate footpoint-E exhibits short-term rapid intensity variations in the AIA 171\,\AA\ and the UV ratio diagnostics (panel h). Such fluctuations are typical signatures of rapid moss variability associated with impulsive coronal heating \citep[][]{2014Sci...346B.315T}. However, the short-term intensity variation at footpoint-E is markedly different from its conjugate. Furthermore, though both footpoints peak near-simultaneously, the intensity disturbances first originate at footpoint-W and propagate towards footpoint E (see online animation). This is consistent with energy injection and transfer from footpoint-W to footpoint-E.

Our magnetic field analysis (see Appendix\,\ref{sec:met}) at footpoint-W revealed that underlying the detected UV ratio patch, in addition to the dominant positive polarity magnetic field, a minor patch of embedded negative polarity magnetic field is also present. To qualitatively illustrate the magnetic connectivity between footpoints W and E, we employ magnetic field extrapolations. Here we select the line-of-sight magnetic field map (spanning the entire field of view shown in Fig.\,\ref{fig:over1}) that is nearest in time to the Fe\,{\sc xviii} emission peak as the lower boundary condition and extrapolate coronal magnetic field under the linear force free field (lfff) approximation. A sample of traced magnetic field lines connecting footpoints W and E is plotted in panels (a) and (d), in which we choose the force-free $\alpha$ parameter so that the field lines outline the hot loop reasonably well. The field lines from the dominant positive polarity magnetic field near footpoint-W form a dome-like topology with the embedded minor negative polarity patch. Along with the long field lines that outline the main loop, there are shorter closed loops below the dome, creating a complex magnetic structure near footpoint-W. Moreover, integrated fluxes of both polarities within the marked circular region are observed to be changing. In addition to the flux variations on granular timescales of 5-10\,minutes, there is a longer-term trend over the course of 30\,minutes until Fe\,{\sc xviii} peaks. A simple linear fit to the time series of magnetic fluxes during this 30\,minute period shows that in this case both the minor and the major polarities increase at a rate of $0.7-0.8 \times 10^{15}$\,Mx\,s$^{-1}$. These mixed polarity magnetic field changes at footpoint-W could be the source of energy to heat the loop.

\subsection{A more complex example\label{sec:comp}}

\begin{figure*}[!ht]
\begin{center}
\includegraphics[width=\textwidth]{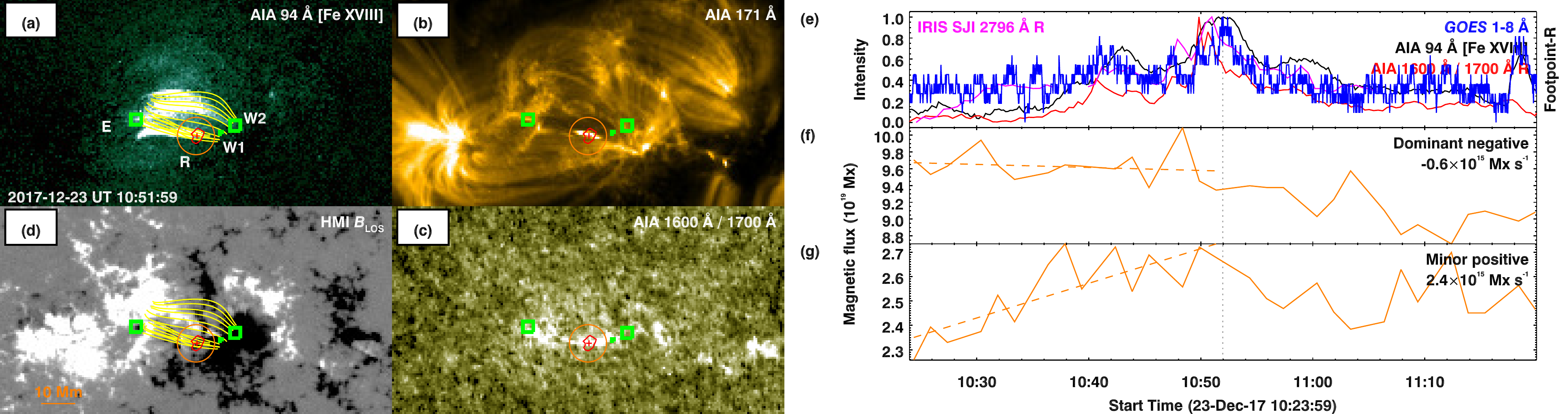}
\includegraphics[width=\textwidth]{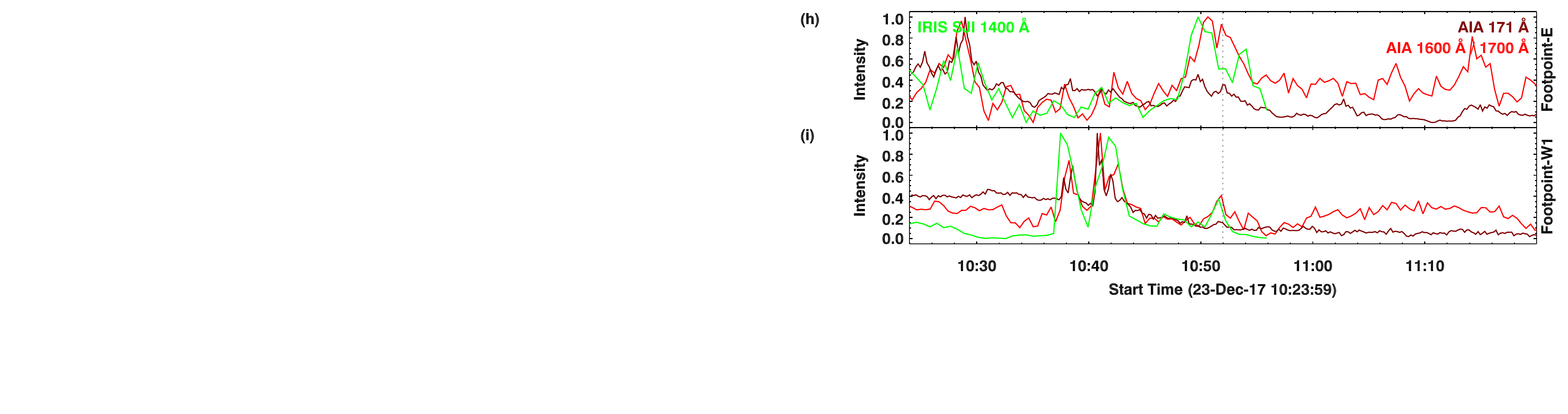}
{\vskip-4.85cm
\includegraphics[width=0.45\textwidth]{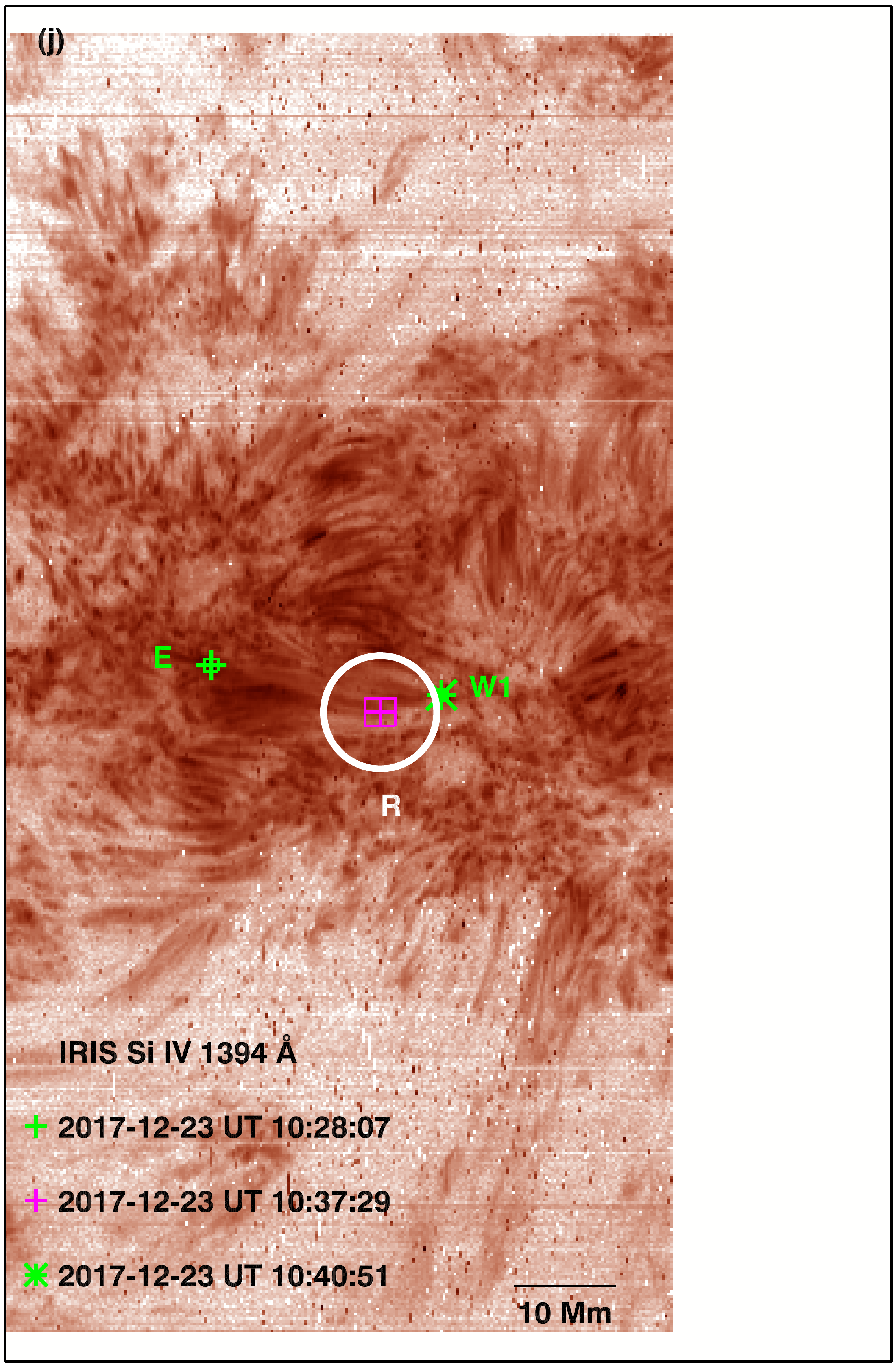}}
\includegraphics[width=0.5\textwidth]{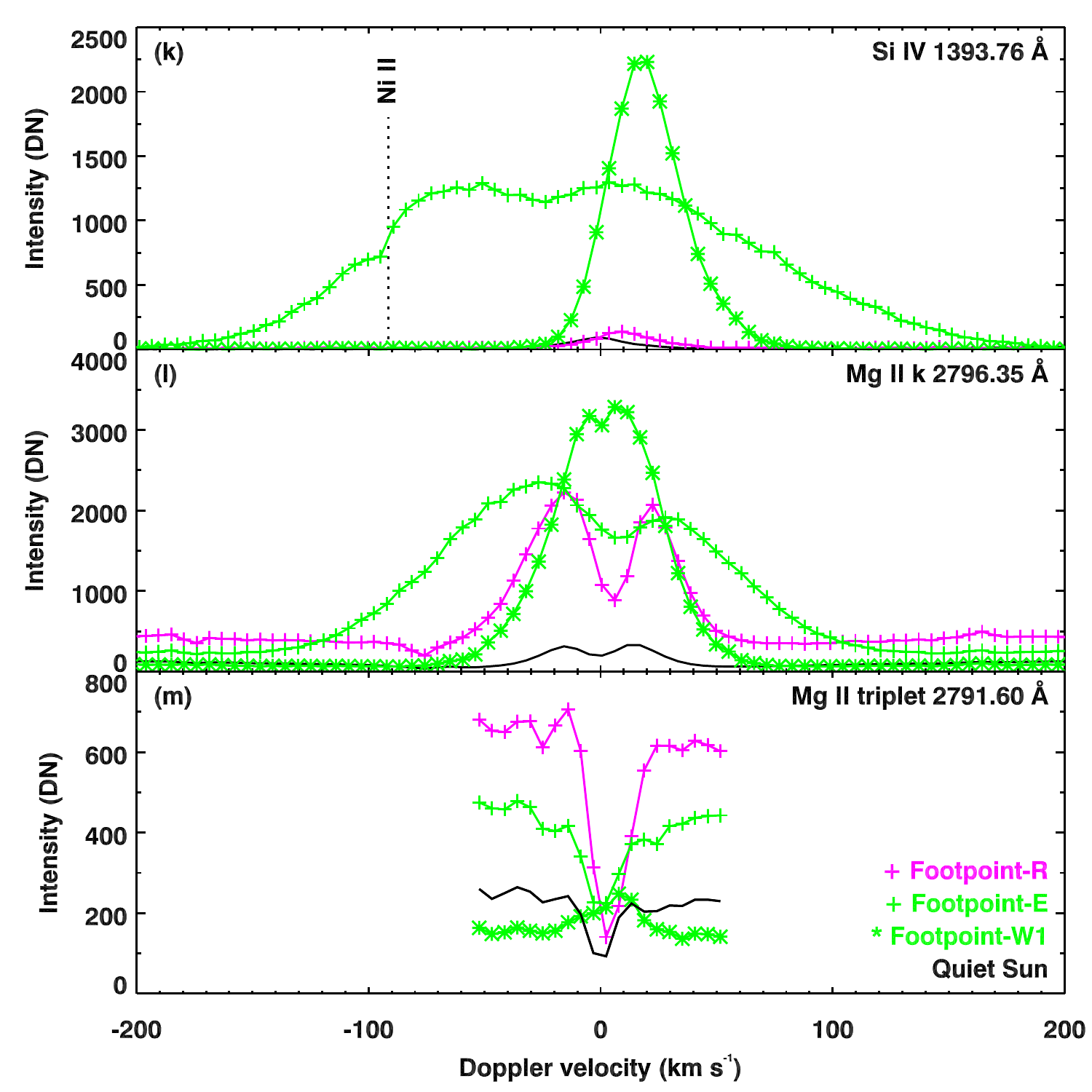}
\caption{Impulsive heating in the core of AR 12692. Panels (a)-(g) are similar to their counterparts in Fig.\,\ref{fig:AR12665c1}. Panels (h) and (i) show time series of light curves from footpoint regions E and W1 (marked in panel a). Panel (j) shows an IRIS Si\,{\sc iv}\,1394\,\AA\ raster map in an inverted colour scheme. The overlaid circle is roughly at the same location as in panels (a)-(d). Footpoint regions E-R-W1 are marked. A 10\,Mm scale is overlaid. The IRIS SJI 2796\,\AA\ light curve from footpoint-R (magenta box) is displayed in panel (e). SJI 1400\,\AA\ light curves from footpoints E and W1 (green boxes in panel j) are shown in panels (h) and (i). Panels (k)-(m) show Si\,{\sc iv}, Mg\,{\sc ii}\,k, and Mg\,{\sc ii}\,triplet spectral profiles (footpoint-R in magenta-plus; footpoint-E in green-plus; footpoint-W in green-asterisk; average quiet Sun profiles in black; in panel (k) the quiet Sun profile is multiplied by ten and the dotted line identifies a Ni\,{\sc ii} absorption profile). The timestamps of the spectral profiles in UT are given in panel (j). AIA data are plotted at their native cadence (EUV at 12\,s and UV at 24\,s). Animation of panels (a) to (g) is available online. See Sects.\,\ref{sec:obs}, \ref{sec:comp} and Appendix\,\ref{sec:met} for details. \label{fig:AR12692c1}}
\end{center}
\end{figure*}

The example discussed in Fig.\,\ref{fig:AR12665c1} represents a relatively simple case of impulsive heating in which the mixed polarities are found directly at one of the footpoints. Visual inspection of the analysed loops in the Fe\,{\sc xviii} emission and the associated footpoint signatures in the AIA 171\,\AA\ and UV ratio diagnostics, however, reveals that this type is not the most common. We found that many of these hot loops have complex morphology with more than two footpoints. These complex systems could be composed of multiple loops that exhibit sympathetic brightening.

We discuss the case of a complex, multi-footpoint hot loop system observed in the core of AR 12692. In Fig.\,\ref{fig:over2} the core region of the AR 12692 is marked (upper panel), along with the time series of core-integrated Fe\,{\sc xviii} emission and GOES X-ray flux that show a similar evolution over a period of 72\,hours (lower panel). The AR exhibited impulsive heating of a loop system on 2017 December 23 at 10:52\,UT (Fig.\,\ref{fig:AR12692c1}). In the Fe\,{\sc xviii} emission, the loop system connects a pair of visually located footpoints E and W and is separated into two main sections by a dark feature (panel a). This dark feature is composed of filamentary material seen in EUV absorption (panel b). Footpoint W is split into W1 and W2 on either side of this dark feature. The UV ratio signal displayed scattered bright patches under the loop system (panel c). Broadly, footpoints E and W1-2 are rooted in dominant positive and negative polarity regions (panel d). 

The GOES X-ray flux and the core integrated Fe\,{\sc xviii} emission displayed a gradual increase with multiple pulses starting around 10:38\,UT (panel e). Our method detected a UV ratio patch (labelled R; contoured regions in panels a-d) close to W1 that exhibited intensity variations concurrent and correlated with coronal diagnostics. This strongly indicates that the identified patch is also another footpoint of the loop system along with E and W1-2. Our magnetic field analysis revealed that footpoint R is rooted in a mixed polarity region (panel d). At footpoint-R, the flux of dominant negative polarity gradually decreases while the flux of minor positive polarity field increases for a period of 30\,minutes until the Fe\,{\sc xviii} peak around 10:52\,UT (panels f-g). In this case, footpoint-E also overlies a mixed polarity region (with a minor negative polarity patch next to the dominant positive polarity field region (panel d)). It is also conceivable that this complex system is actually composed of three different loops, namely, E-R, E-W1, and E-W2 that exhibit sympathetic brightening. Magnetic field lines traced from the lfff extrapolations showcase these different connections in this loop system (panels a and d).

This region is covered by IRIS, whose slit scanned the footpoint system before the GOES peak (panel j). Both AIA and IRIS diagnostics show consistent evolution at footpoints E-R-W1 (panels e, h-i; the signal is either too weak or indistinguishable at W2). IRIS SJI movies show footpoints E-R to be compact bright sources persistent for a longer duration. Footpoint-E (over the mixed polarity region) shows two episodes of brightenings in the SJI 1400\,\AA\ time series, the first at the start of the time series (10:28\,UT) and the second concurrent with the loop brightening (10:52\,UT). 

The slit crossed this footpoint-E around 10:28\,UT, where we detected a broad, multi-component Si\,{\sc iv} spectral profile, superimposed with a Ni\,{\sc ii} absorption feature (panel k). The line emission is observed beyond Doppler shifts of $\pm150$\,km\,s$^{-1}$, strongly suggesting that these flows are driven by episodes of chromospheric/transition-region reconnection \citep[][]{1997Natur.386..811I,2014Sci...346C.315P,2018SSRv..214..120Y}. The Mg\,{\sc ii}\,k and Mg\,{\sc ii}\,triplet chromospheric emissions are also enhanced compared to the average quiet-Sun profile (panels l-m), suggesting reconnection and plasma heating in the chromosphere \citep[][]{2015ApJ...806...14P}. Around 10:38\,UT the slit crossed footpoint-R, where the wings of the Mg\,{\sc ii}\,k and Mg\,{\sc ii}\,triplet and local continuum are enhanced compared to the quiet-Sun, suggesting a temperature increase in the deep chromosphere. At this time, the Si\,{\sc iv} profile is also strong and broader than the quiet-Sun profiles, but weaker than the profile from footpoint-E. Both the AIA UV ratio and  IRIS SJI 2796\,\AA\ diagnostics begin to increase around 10:38\,UT.  All this corroborating evidence suggests the onset of deep chromospheric reconnection at footpoint-R around this time (some 14\,minutes prior to the coronal peak time). 

In contrast, footpoint W1 exhibited brightenings that are  consistent with rapid moss variability in response to the impulsive loop heating \citep[e.g.][]{2014Sci...346B.315T,2020ApJ...889..124T}. When the slit crossed the region around 10:41\,UT (around the time when the coronal signal has pulse-like enhancements), the IRIS profiles are markedly different. Si\,{\sc iv} is narrower than the profile at footpoint-E. Both Mg\,{\sc ii} profiles show emission with weak absorption. This suggests a rapid temperature increase with height in the chromosphere \citep[][]{2015ApJ...806...14P}. The rapid moss variability suggests energy deposition at that loop footpoint-W1 \citep[][]{2020ApJ...889..124T}.

\begin{figure*}
\begin{center}
\includegraphics[width=\textwidth]{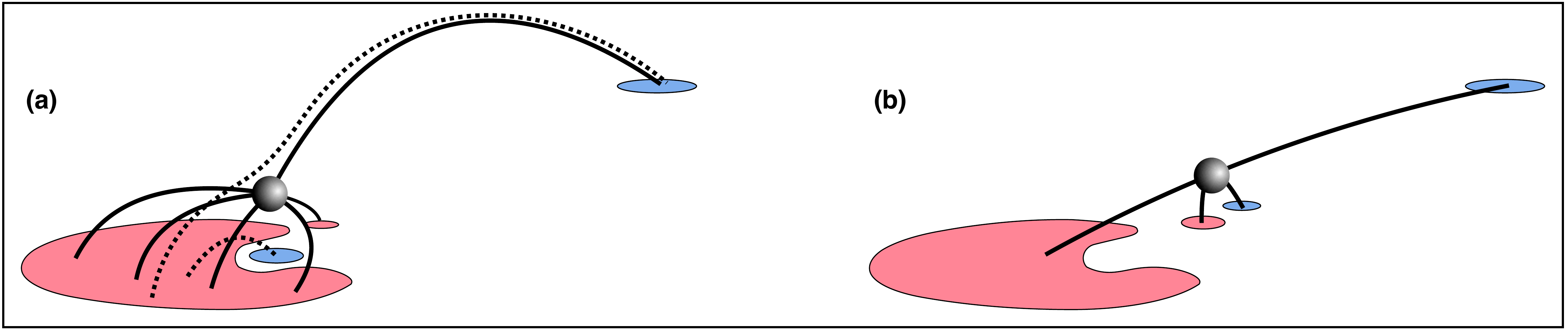}
\caption{Illustration of the initial configuration of two magnetic topologies potentially linked to hot loops. Panel (a) shows a magnetic feature (blue patch) of one polarity embedded in a predominantly unipolar magnetic environment of the opposite polarity (red patches) at one footpoint of the loop. The expected magnetic connectivity is indicated by black curves. The sphere indicates null point, the potential site of reconnection in this type of magnetic configuration. The long and short dotted curves are magnetic field lines that lie above and below the null point. The segment overlying the null point then represents a seemingly simple hot loop with two footpoints. In panel (b) magnetic topology of a complex hot loop system is displayed. An emerging or cancelling magnetic bipole (smaller red and blue patches) reconnects with the overlying magnetic field and leads to the formation of a hot loop connecting two larger magnetic patches or footpoints. See Sect.\,\ref{sec:magp} for discussion.
\label{fig:cart}}
\end{center}
\end{figure*}

\section{Statistics of impulsive heating events\label{sec:summ}}

Other examples of mixed impulsive coronal heating episodes exhibiting different degrees of complexity in their evolution are illustrated in Appendix\,\ref{sec:exam}. These loops are often of the multi-footpoint variety,  rooted in more than two isolated (or extended) magnetic patches in the lower atmosphere. In all these examples, we show the light curves of the detected footpoint along with the evolution of the underlying mixed polarity magnetic flux. In a few select cases, like Fig.\,\ref{fig:AR12692c1}, we present detailed light curves of atmospheric emission from other prominent (visually located) footpoints associated with the hot loops (Figs.\,\ref{fig:AR12692c2}, \ref{fig:AR12699c1}, \ref{fig:AR12699c2}, \ref{fig:AR12712c1}, and \ref{fig:AR12712c2}). Common to all these loops is the feature that the core-integrated emission shows a rise and fall in intensity  concurrent with the corresponding GOES X-ray time series. At least one of the footpoints of these hot loops is rooted in mixed polarity  magnetic fields. In addition, we observed clear magnetic flux changes in either dominant, minor or both polarities. Time series of the UV ratio signal from the respective (detected) patches overlying the evolving mixed polarity regions displayed brightenings that are correlated with coronal diagnostics. IRIS spectroscopic observations, whenever available, show significantly enhanced chromospheric and broad transition region line profiles in the vicinity of the detected UV patches over the mixed polarity regions.\footnote{Given that the shape, size, and exact location of a detected UV ratio patch result from a cross-correlation analysis of a 1-hour time series (see Appendix\,\ref{sec:lowat}), it will not capture the instantaneous morphology (or evolution) of  rapidly evolving reconnection source regions as seen by IRIS. Moreover, we do not have repeated IRIS spectroscopic observations of the detected patches to investigate the spatial morphology and temporal evolution of the location exhibiting broad spectral profiles. These factors could explain apparent spatial offsets between the detected UV ratio patches and the associated IRIS reconnection signals (e.g. Fig.\,\ref{fig:AR12692c2}).} For instance, chromospheric Mg\,{\sc ii}\,k and triplet profiles are strongly enhanced relative to the typical quiet Sun emission. At the same time, Si\,{\sc iv} line profiles emanating from the nearly $10^5$\,K plasma exhibit enhanced wing emission exceeding 100\,km\,s$^{-1}$ in either blue, red, or in both wings. These IRIS diagnostics are consistent with local (chromospheric) heating and outflows driven by magnetic reconnection. These additional examples provide further evidence for magnetic reconnection near at least one of the (multiple) footpoints of hot loops.

In all, we analysed 137 impulsively heated loops and their footpoints in seven different ARs (see Sect.\,\ref{sec:obs} and Appendix\,\ref{sec:met} for details). Summaries of the events are tabulated in Tables \ref{tab:AR12665}, \ref{tab:AR12692}, \ref{tab:AR12699}, \ref{tab:AR12712}, \ref{tab:AR12713}, \ref{tab:AR12733}, and \ref{tab:AR12738} (one for each AR), while overviews of these ARs are displayed in Figs.\,\ref{fig:over1}, \ref{fig:over2}, \ref{fig:over3}, \ref{fig:over4}, \ref{fig:over5}, \ref{fig:over6}, and \ref{fig:over7}. 

\subsection{Magnetic properties of hot loops\label{sec:magp}}

\begin{figure}
\begin{center}
\includegraphics[width=0.48\textwidth]{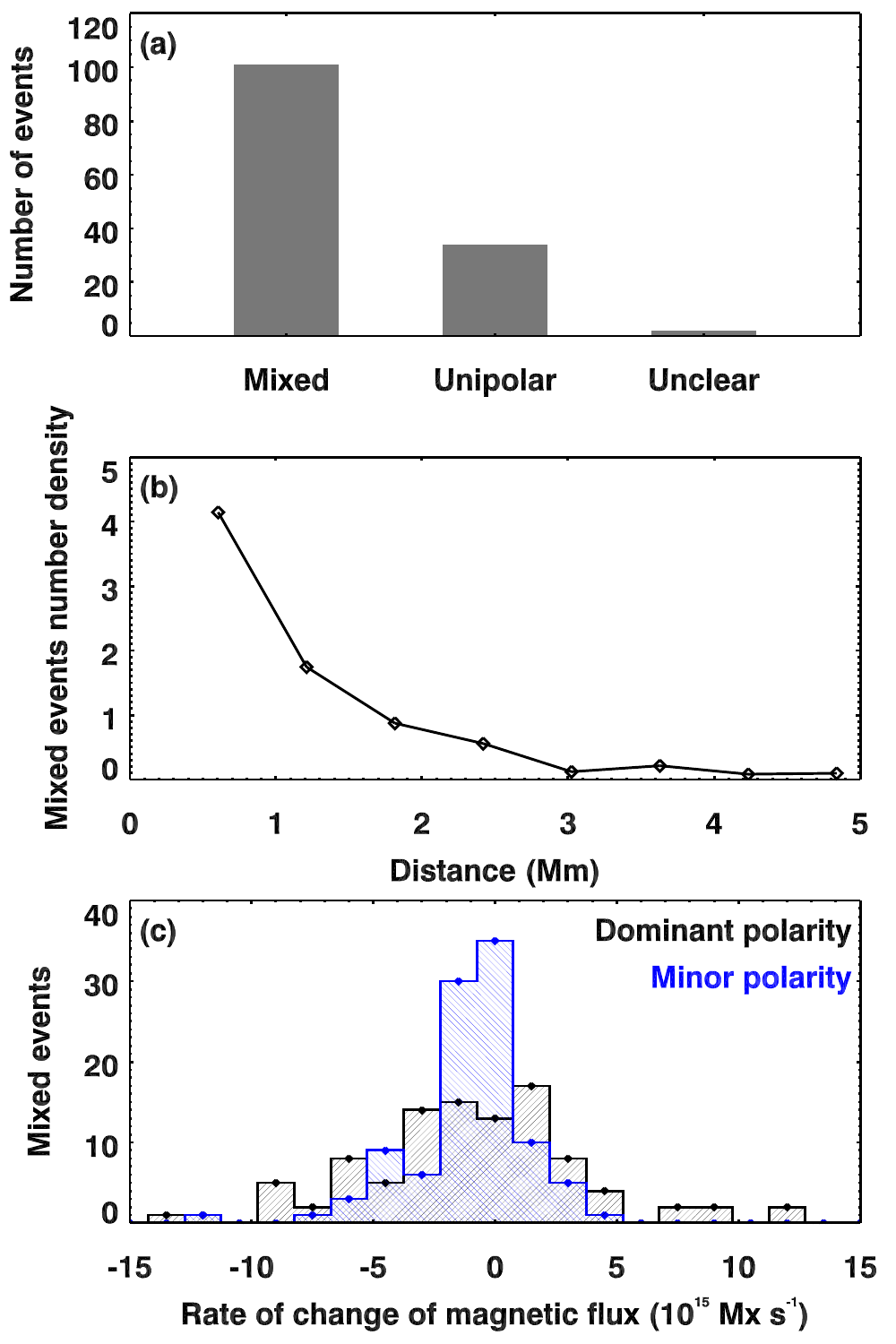}
\caption{Overview of magnetic properties at the base of impulsively heated loops. Panel (a) shows the number of heating events for each type of magnetic structure at the footpoints of hot loops. Only those events with significant minor polarity field within a 5.4\,Mm circle about the centroid of the detected UV ratio patches, associated with hot loops, are counted as mixed-polarity cases. For mixed event, area-normalised distribution of the minimum distance of the minor polarity field from the centroid of the patch detected in the UV ratio maps is plotted in panel (b). Panel (c) shows the histogram of the rate of change of magnetic flux, separately for dominant (black) and minor (blue) magnetic polarities, for events labelled as mixed. See Sect.\,\ref{sec:magp} for discussion.
\label{fig:mag_hist}}
\end{center}    
\end{figure}

Though the hot loops display varying degrees of complexity in their spatial morphology and varying trends of their temporal evolution, it is possible to obtain the essential qualitative characteristics of their relation with the underlying magnetic field. In Fig.\,\ref{fig:cart}, we illustrate the initial magnetic skeleton of hot loops for both apparently simple (Fig.\,\ref{fig:cart}a) and complex (Fig.\,\ref{fig:cart}b) cases. In the simple case, at initial stages, one of the footpoints of a magnetic loop is rooted in a predominantly unipolar magnetic environment containing an embedded patch of minor opposite-polarity magnetic field. The minor magnetic polarity could be brought into the system through flux emergence in the vicinity, although other sources are also conceivable. This configuration results in a separatrix dome topology with a null point in the solar atmosphere \citep[][]{pontin13a}. Observations and simulations suggest that when embedded magnetic patches are perturbed either through emergence, cancellation, advection or shear, it could lead to reconnection at the null point that impulsively heats the plasma \citep[e.g.][]{2010ApJ...724.1083G,pontin13a,2015ApJ...801...83C,2017A&A...605A..49C,2019A&A...628A...8P}. This is similar to the case discussed in Fig.\,\ref{fig:AR12665c1}. Enhanced activity observed in the AIA (E)UV diagnostics at footpoint-W over the mixed polarity region is consistent with magnetic reconnection at that footpoint. In the complex case (Fig.\,\ref{fig:cart}b), flux emergence or cancellation occurs beneath an overlying magnetic structure. Such a configuration could be related to the so-called serpentine fields during early phases of flux emergence or AR formation \citep[e.g.][]{2002ApJ...575..506G}. Alternatively, the overlying magnetic field could be related to the filamentary structures that overlie polarity inversion lines \citep[e.g.][]{1989ApJ...343..971V,2001ApJ...552..833M}. In either of these scenarios, subsequent magnetic reconnection injects energy into the system and leads to impulsive coronal heating. In this case, the dominant footpoints at both ends might show rapid moss variability, whereas the site of flux emergence or cancellation displays signatures consistent with magnetic reconnection, as in Fig.\,\ref{fig:AR12692c1}.

Quantitatively, out of the 137 analysed impulsive brightenings, we found 101 loops (i.e. $>73$\% of analysed events) to have at least one of their footpoints rooted in regions of mixed magnetic polarity. The detected footpoints in 34 loops ($\approx$25\%) are associated with unipolar magnetic field regions, in the sense that any opposite-polarity flux associated with these footpoints is below our threshold of 30\,G per pixel (i.e. the flux imbalance is larger than our threshold, see Appendix\,\ref{sec:fluxim} for a discussion). The magnetic setting underlying two loops could not be determined. A histogram of the analysed impulsive heating events labelled according to the underlying magnetic setting is displayed in Fig.\,\ref{fig:mag_hist}a.

For the mixed cases, we measure the minimum distance from the centroid position where the opposite-polarity patch is first encountered. The results are displayed in Fig.\,\ref{fig:mag_hist}b. 
The number density of mixed events  per unit area decreases rapidly with distance. Therefore, in the mixed cases, there is a clear tendency for the opposite-polarity magnetic patch to concentrate in the inner part of the circle, at a distance of $3$\,Mm or less from the centroid. Thus, in spite of the rather conservative assignment of mixed magnetic polarity to loop footpoints, we find that a majority of hot loops have opposite-polarity field within 3\,Mm of at least one of its footpoints. This provides additional support for our thesis that the presence of an opposite polarity plays a role in producing the heating events. It also suggests that the reconnection preferably happens low in the atmosphere (i.e. when the two polarities are located very close to each other).

For the mixed polarity cases, we measure the rate of change of magnetic flux near the loop footpoints. Its histogram is displayed in Fig.\,\ref{fig:mag_hist}c. The minority polarity features exhibit a narrower distribution for the rate of change of magnetic flux than the dominant polarity magnetic field. The mean of both distributions is negative, suggesting on average flux cancellation at the solar surface. However, many loop heating events are associated with an increase in flux (e.g. by emergence), or only a minor change in magnetic flux. We note, however, that magnetic reconnection does not imply a big change in flux as seen at the solar surface. The crucial point is that in all these scenarios, interaction is expected between the dominant and minor polarities that is likely to drive magnetic reconnection near the base of hot loops. Moreover, the linear fit used to retrieve the flux emergence and cancellation rates here indicates only the trend of the magnetic flux change over a 30-minute period. Consequently, the rate will be low if, for instance, a magnetic polarity exhibits flux increase and decrease of comparable magnitudes during the course of these 30\,minutes. In Fig.\,\ref{fig:AR12665c1}, for example, magnetic flux of the dominant polarity increases for the first 10\,minutes, then levels off, and decreases over another 10\,minutes, during which the atmosphere is impulsively heated. It is clear that the rate of cancellation of the dominant polarity during these 10\,minutes at the time of impulsive heating is higher than the rate determined from the trend of magnetic flux over the 30-minute period. Though ignored in our study, such short-term variations will likely regulate magnetic reconnection and the resulting energy injection at the coronal base of hot loops.

Based on visual inspection, we found that some of the 34 unipolar cases could actually be associated with mixed magnetic polarity at one of the footpoints of the loop. This is because, while the footpoint detected by our method itself is not rooted in mixed magnetic polarities, at least one of the remaining conjugate footpoints of the corresponding hot loop, meaning, those footpoints not detected by our method, are found to be rooted in mixed polarity regions. In Appendix\,\ref{sec:uni} we discuss an example of a so-called unipolar case with the undetected footpoints rooted in mixed-polarity regions that display signatures consistent with reconnection in the lower solar  atmosphere. Thus, the percentage of mixed cases identified through our analysis is likely an underestimate. We expect that more sophisticated methods that could detect all the footpoints associated with hot loops would lead to a higher percentage of mixed cases. This enhances even further the strong case for the integral role of surface magnetic interactions in impulsive coronal heating. 

\begin{figure}
\begin{center}
\includegraphics[width=0.48\textwidth]{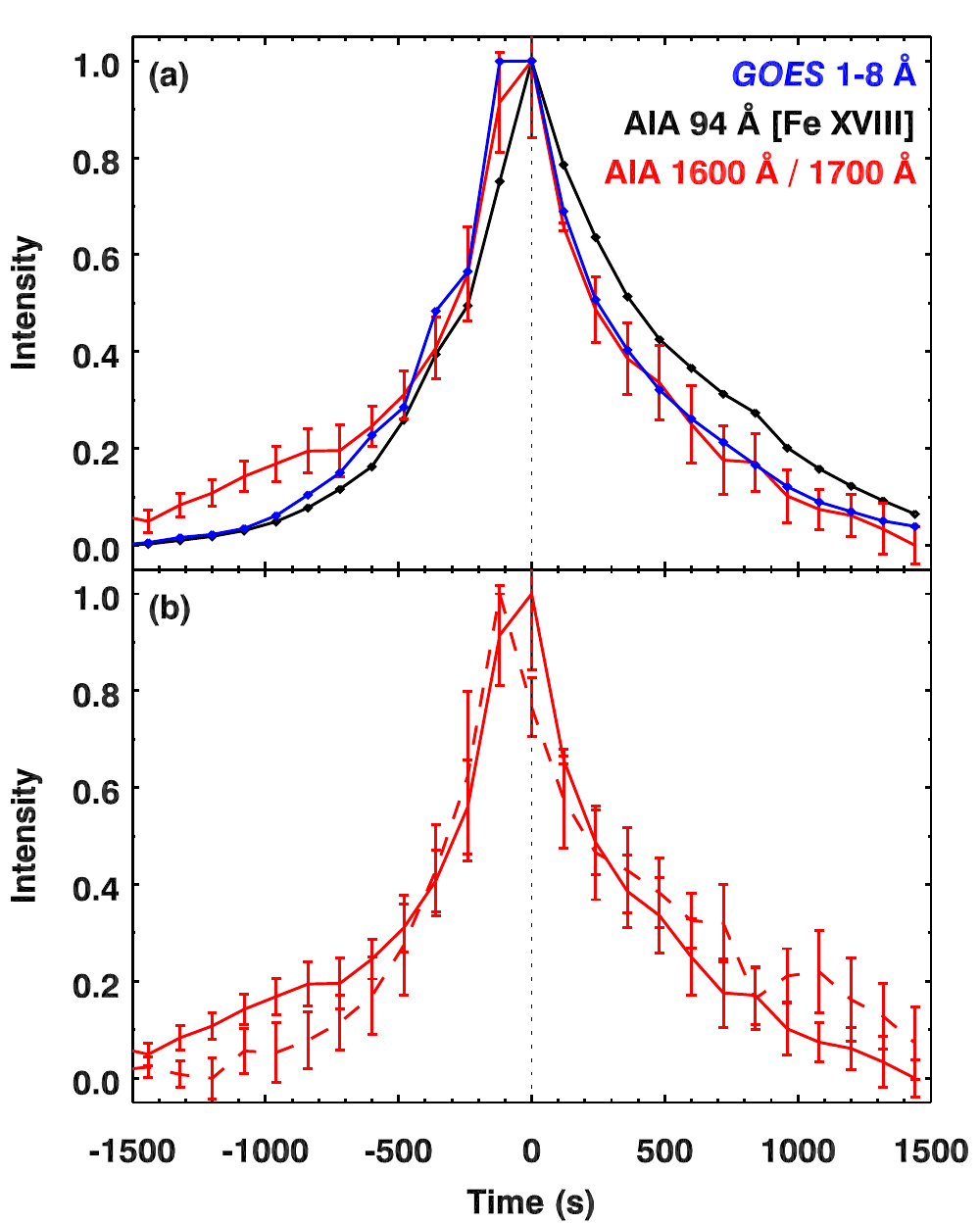}
\caption{Statistical overview of impulsive heating events observed in the cores of seven different ARs. An integrated light curve derived from all the impulsive heating events associated with mixed magnetic polarities at the loop footpoints is displayed in panel (a). The disk integrated GOES 1-8\,\AA\ X-ray flux, core integrated Fe\,{\sc xviii} emission and the corresponding UV ratio signal from contoured patches, in bins of 120\,s (i.e. the considered cadence of AIA EUV observations) as a function of time are shown. Here the elapsed time is in seconds with respect to the time of Fe\,{\sc xviii} peak. The respective error bars associated with the UV ratio signal represent $1-\sigma$\ standard deviation at a given bin, derived from 1000 realisations of light curves, each produced by a blind integration of random number of individual light curves. In panel (b) the UV ratio signal from the mixed cases (solid curve; the same as in the top panel ) is compared with its counterpart from the unipolar cases (dashed curve). See Sect.\,\ref{sec:atmp} for discussion.}
\label{fig:atm_hist}
\end{center}
\end{figure}

\subsection{Atmospheric properties of hot loops\label{sec:atmp}}

In all the events, the atmospheric diagnostics exhibited a rise and fall of intensity associated with impulsive heating. To visualise the general progression of such an impulsive heating episode in UV, EUV, and X-ray diagnostics, we created an integrated light curve for the mixed polarity cases and another for the unipolar cases.

For the 101 mixed polarity cases, the integrated light curve is created by integrating the GOES 1-8\,\AA\ X-ray flux, core integrated Fe\,{\sc xviii} emission, and the corresponding UV ratio signal from detected patches, in bins of 120\,s (i.e. the cadence we used for AIA EUV observations). For each event, the reference time is chosen to be the time when the Fe\,{\sc xviii} emission reaches its peak. The elapsed time is then in seconds with respect to the reference time. The results are displayed in Fig.\,\ref{fig:atm_hist}a. This represents an impulsive light curve seen in various diagnostics from all 101 mixed polarity events, with the timing of each brightening shifted such that they all reach the Fe\,{\sc xviii} emission peak at the same time. After an initial slow rise phase of 1000\,s, the coronal signatures, namely, the GOES X-ray flux and Fe\,{\sc xviii} emission, rapidly rise for 500\,s before the light curve reaches its peak. When individual cases are examined, the GOES light curve precedes the Fe\,{\sc xviii} in some events (e.g. Fig.\,\ref{fig:AR12699c2}) and coincides in other cases (e.g. Fig.\,\ref{fig:AR12665c1}). In many events however, both these diagnostics exhibit peaks at similar times (as shown in the lower panels of overview figures, such as in Fig.\,\ref{fig:over1}). The resulting integrated light curves of these diagnostics from all the 101 events thus peak near-simultaneously. During the initial phase, the UV ratio signal (chromospheric and transition region signature) from the mixed polarity cases (solid red curve) shows a clear bump, consistent with the enhanced activity seen over the mixed polarity fields in the examples discussed. This excess UV emission suggests activation of the chromosphere and transition region at the footpoint rooted in the mixed polarity region prior to the impulsive heating episode. 

For the 34 unipolar cases, the integrated UV signal also exhibits a rapid rise over 500\,s before the light curve peak (dashed red curve in Fig.\,\ref{fig:atm_hist}b), but, unlike the mixed polarity light curve, there is no UV enhancement or bump prior to the rise phase. This is further checked by a blind integration of a random number of events. We created 1000 random light curves from which we computed $1-\sigma$ standard deviation at each temporal bin after normalisation (plotted as red coloured error bars in Fig.\,\ref{fig:atm_hist}). These results show that the integrated UV ratio signal from the unipolar events is clearly distinct from its mixed events counterpart. This suggests that the UV increase at a unipolar footpoint is probably due to energy transfer to that region, perhaps from the other footpoint. 

\section{Discussion\label{sec:disc}}

One of the key issues to consider is the response to impulsive heating of the footpoints that show rapid variability. Models that invoke a coronal energy source (such as produced by braiding) as the main driver of impulsive heating suggest that non-thermal particles accelerated from a coronal reconnection site deposit energy at the footpoints. This would then give rise to the observed rapid variability and plasma flows observed in the low atmosphere. Such models produce Si\,{\sc iv} Doppler shifts of about $\pm40$\,km\,s$^{-1}$ and line widths of up to 20\,km\,s$^{-1}$ \citep[][]{2020ApJ...889..124T}. Provided one assumes an appropriate distribution for the energy flux carried by non-thermal particles accelerated at a coronal site, such a model can explain both the intermittency and the spectral properties of the UV emission at one footpoint in one of our examples (e.g. Fig.\,\ref{fig:AR12692c1}k; footpoint\,W1; for additional examples see Appendix\,\ref{sec:exam}).  However, it is difficult to see how it could explain the presence of broad spectral profiles with wing enhancements exceeding $\pm100$\,km\,s$^{-1}$ and with superimposed chromospheric absorption lines at the other footpoint  (Fig.\,\ref{fig:AR12692c1}k; footpoint\,E; additional examples in Appendices\,\ref{sec:exam} and \ref{sec:uni}). Such profiles are similar to those of a UV burst \cite[][]{2014Sci...346C.315P} with a signature of energy deposition deep in the chromosphere. A separate local or main source of reconnection and heating would be necessary to explain the observed chromospheric behaviour in conjunction with impulsive coronal heating. Therefore, our observations suggest that coronal reconnection due to braiding without localised heating at the coronal base is likely not the only driver of impulsive heating. Reconnection in the low atmosphere (caused by interactions between opposite magnetic polarities at the mixed-polarity footpoint of the heated loop) could play an additional role in atmospheric heating. 

Our observations show that a majority of hot loops are rooted in mixed polarity regions that could drive reconnection in the overlying atmosphere and at the footpoints. But do these interacting mixed magnetic polarities play a dynamically important role in coronal heating? Imaging and spectroscopic evidence presented here suggests that they are indeed associated with enhanced activity and high-speed plasma flows, both suggestive of reconnection at the coronal base. Current sheets could form in the overlying atmosphere \citep[e.g.][]{2003Natur.425..692S,priest14a} and at the interface of interacting mixed polarity magnetic fields, where reconnection is initiated and liberates magnetic energy that could impulsively heat coronal plasma. Depending on the magnetic configuration (Fig.\,\ref{fig:cart}), such interactions will be initiated  either by flux emergence or by cancellation of a minor magnetic polarity as it approaches magnetic patches of opposite polarity in its vicinity. 

The total magnetic energy released during flux emergence or cancellation is $E_{\rm mag} = B\Phi L/(8\pi)$, where $B$ is the magnetic field strength, $\Phi$ is the total cancelled magnetic flux, and $L$ is the length of the current sheet. Assuming a typical rate of change of magnetic flux for the minor polarity of $10^{15}$\,Mx\,s$^{-1}$ (Fig.\,\ref{fig:mag_hist}c), the total emerged or cancelled minor magnetic flux over a period of 30\,minutes is $\Phi=1.8\times 10^{18}$\,Mx. In our study, this time period covers the initial 30-minute period until the Fe\,{\sc xviii} peak in each case. Based on the broad spectral line profiles, indicative of magnetic reconnection at the coronal base, we assume here for simplicity that the current sheet is confined completely to chromospheric and transition region heights, meaning, $L$=2\,Mm. Then an evolving minor magnetic feature with field strengths of $B$=100\,G to 1000\,G releases an energy of the order of $10^{27}$\,erg to $10^{28}$\,erg as its magnetic flux content changes over a period of 30\,minutes. This timescale covers only the rise phase of the impulsive heating. Reconnecting flux patches at the solar surface may magnetically decouple from the overlying atmosphere at the end of this phase as the separator moves down to the photosphere. However, if they remain coupled to the atmosphere, further evolution of mixed polarity magnetic patches will continue to supply energy also during the fall in intensity of the event. In addition, if more than one loop footpoint is rooted in dynamically important mixed polarity regions, the energy injection could exceed this value. Overall, the released magnetic energy would be sufficient to power a small B-class flare in the solar corona \citep[][]{2011SSRv..159..263H}. Thanks to the overall lower disk-integrated soft X-ray background, all the impulsive heating events that we analysed here could be associated with the GOES A- or B-class flares. Therefore, on timescales of tens of minutes, the magnetic energy liberated through interaction of magnetic flux patches of opposite polarity could impulsively heat the coronal core of an active region  \citep[e.g.][]{2018ApJ...853L..26H,2018A&A...615L...9C,2018ApJ...862L..24P,2019ApJ...881..107A}. Furthermore, it is possible that when mixed-polarity fields persist for several hours at a given location, their continued interaction and cancellation, combined with local photospheric flows could lead to intermittent (and persistent) coronal heating at the same location \citep[][]{2014ApJ...795L..24T}. 

Our study highlights the complex magnetic setting at the footpoints of impulsively heated coronal loops. That these impulsive loops are not outlined by a simple bipolar magnetic configuration with unipolar footpoints at both ends may hold the key to understanding the origins of hot plasma in active region cores. In our statistical analysis, we mainly focused on the magnetic and atmospheric properties, namely, the magnetic flux changes at the base of hot loops and the integrated UV, EUV, and soft X-ray intensities, which are the quantities that can be compared among various events. However, we have not quantified some other potentially important aspects related to these hot loops. For example, in complex cases where interactions of mixed-polarity magnetic fields are observed at more than one footpoint (e.g. as is the case at footpoints E and R in Fig.\,\ref{fig:AR12692c1}), we have not statistically quantified which of those, if any, would be the primary contributor to  impulsive heating. Related to this, we have not analysed any spatial morphological evolution of hot loops in the corona, an aspect that cannot be grouped into a statistical study because of its case-dependent nature. Similarly, we have also not explored the physics of reconnection onset itself which is likely governed by highly dynamic current sheets and plasma evolution \citep[e.g.][]{1974SoPh...36..433P,1977ApJ...216..123H,2020ApJ...890L...2C}. Nevertheless, as a first step, our observations emphasise the additional role of footpoint reconnection driven by interacting mixed magnetic polarities in impulsive atmospheric heating, including coronal heating to several million degrees Kelvin. The main future observational challenge is then to verify whether it is the dominant process. 

\begin{acknowledgements}
We thank the anonymous referee for providing constructive comments that helped improve the manuscript. L.P.C thanks Ignacio Ugarte-Urra, NRL, for kindly providing IDL procedures to calculate Fe\,{\sc xviii} emission from AIA images and for useful discussions. \textit{SDO} data are  courtesy of \textit{NASA}/\textit{SDO} and the AIA, EVE, and HMI science teams. \textit{IRIS} is a NASA small explorer mission developed and operated by LMSAL with mission operations executed at NASA Ames Research Center and major contributions to downlink communications funded by ESA and the Norwegian Space Centre. We are grateful to the GOES team for making the data publicly available. This project has received funding from the European Research Council (ERC) under the European Union’s Horizon 2020 research and innovation programme (grant agreement No. 695075) and has been supported by the BK21 plus program through the National Research Foundation (NRF) funded by the Ministry of Education of Korea. This research has made use of NASA’s Astrophysics Data System. 
\end{acknowledgements}

\clearpage
\newpage
\begin{appendix}

\section{Methods\label{sec:met}}

\subsection{Connecting heating events to the solar lower atmosphere\label{sec:lowat}}

We cross-correlate the 1-hour UV ratio signal at each pixel with the respective 1-hour time series of the integrated Fe\,{\sc xviii} core emission. We first spatially smooth the UV ratio maps with $3\times3$ pixels to suppress any spikes. Then we consider only those pixels in the UV ratio maps that satisfy the following conditions. The 1-hour time-averaged magnetic flux density in that pixel should be greater than 30\,G (corresponding to magnetic flux of $\sim10^{17}$\,Mx, which is typical of quiet Sun regions. Furthermore, it is at least three times larger than the nominal HMI noise level\footnote{\url{http://jsoc.stanford.edu/HMI/Magnetograms.html}} of between 5\,G and 10\,G). The average Fe\,{\sc xviii} intensity in that pixel within $\pm5$\,minutes covering the identified peak time should be greater than the average Fe\,{\sc xviii} intensity within the core patch over the selected time window. We found that these two conditions generally exclude the quiet-Sun pixels that are of no interest to us. This step results in a 2D map of cross-correlation values at each pixel that satisfy the aforementioned criteria. A cross-correlation value of zero is assigned to the rest of the pixels.

\begin{figure*}
\begin{center}
\includegraphics[width=0.5\textwidth]{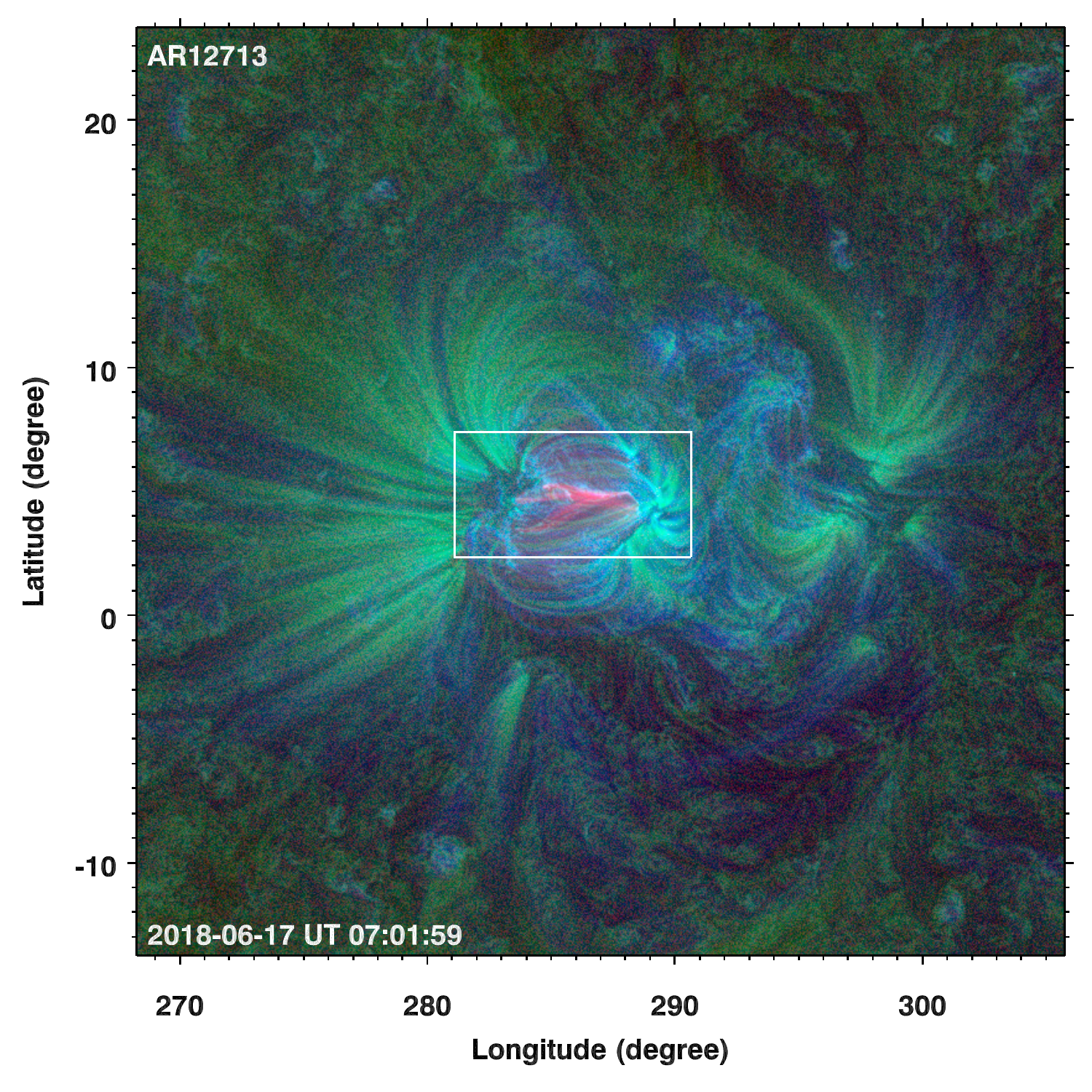}
\includegraphics[width=\textwidth]{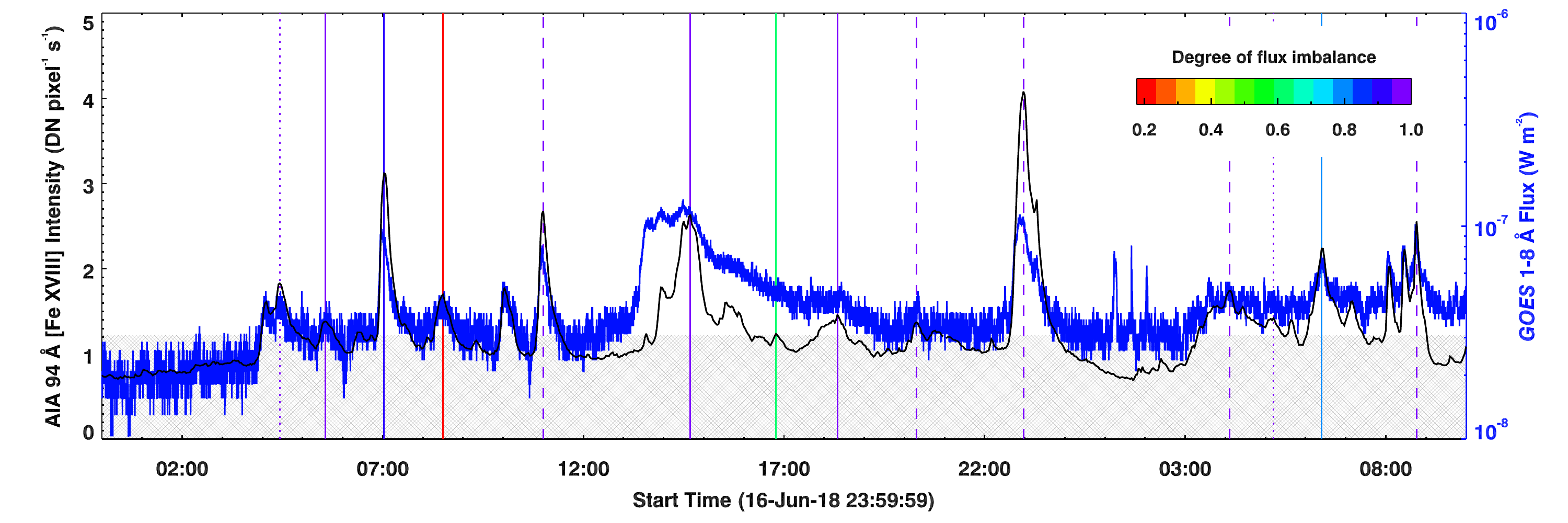}
\caption{Overview of impulsive heating events observed in AR 12713. The format is the same as in Fig.\,\ref{fig:over1}. In the bottom panel, the dotted lines mark events whose magnetic footpoints could not be determined (see Sect.\,\ref{sec:met} for details).\label{fig:over5}}
\end{center}
\end{figure*}

From the resulting 2D maps we identify all the contiguous regions where the value of cross-correlation per pixel is at least 0.5 or higher and which have a minimum of nine pixels. The lower limit on the size of the region discards any spurious results of high cross-correlations from single isolated pixels. We have not set any criteria on the shape of the feature itself. Thus in principle, as an extreme example, a linear patch of nine pixels would be allowed. When the 2D maps do not have regions that meet minimum size and cross-correlation value thresholds, we label the event as unclear. This means that we are unable to locate any footpoint regions of that particular loop in the lower atmosphere. No further analysis is carried out on that heating event. In total, we found two such unclear cases out of the 137 events from seven active regions (events identified with dotted lines in the bottom panel of Fig.\,\ref{fig:over5}). 

The next step in the process is to isolate a single patch from the 2D cross-correlation map that relates to the coronal heating event. If there is only one region in the map that has a minimum of nine pixels with minimum cross-correlation of 0.5 at each of those pixels, we flag that region for further analysis. In many cases however, there could be several discrete regions in the 2D map that satisfy both conditions. In that case, for each patch, we calculate the average intensity as a function of time. We once again cross-correlate these distinct UV ratio time series with the 1-hour (core-integrated) Fe\,{\sc xviii} coronal time series. At the same time we compute time-integrated intensity for each of the distinct UV ratio time series. Finally, for each patch we multiply the cross-correlation with the time-integrated intensity to obtain the respective intensity-weighted cross-correlation. We select the patch that has the highest intensity-weighted cross-correlation. This further ensures that we only analyse those patches that best correspond to a coronal signal. We compute the centroid of the selected patch for each event (the $X$ and $Y$ centroid positions in arcsec of all the identified events are listed in Tables. \ref{tab:AR12665}, \ref{tab:AR12692}, \ref{tab:AR12699}, \ref{tab:AR12712}, \ref{tab:AR12713}, \ref{tab:AR12733}, and \ref{tab:AR12738}; for the cases labelled unclear, the $X$ and $Y$ positions are left blank).  

The main limitation of our method is that we designed it to find only a single patch for analysis. This means that when multiple hot loops are present during a given time window, as is the case in some of the events we identified, we analyse the magnetic roots of only one of the loops during that 1-hour period. Through visual inspection we found that the resulting patches do correspond to the footpoint regions of overlying coronal brightenings. We consider that the sample size of 137 heating events from seven different active regions forms a representative group for the variety of impulsive heating events observed in the solar corona. 

\subsection{Analysis of magnetic roots of impulsive heating events\label{sec:mag}}

We begin with the centroid position of the patch identified through our procedure (Appendix\,\ref{sec:lowat}) to investigate the underlying magnetic structure of hot loops. We consider a circular region with radius of about 5.4\,Mm (nine HMI pixels), with the centroid as its centre. We focus on the evolution of the magnetic field for the 30\,minute period preceding the peak Fe\,{\sc xviii} intensity of the heating event.

To be sure that any detection of mixed magnetic polarities in the footpoint regions of the loops is robust, we first time-average the line-of-sight magnetic field in that circular region over this 30\,minute period. From the 30-minute time-averaged map, we determine the magnetic polarity (either positive or negative) that has the highest magnetic flux (above 30\,G per pixel) in that circular region. We label that polarity as the dominant one. Then we check if the circular region contains a persistent opposite magnetic polarity with time-averaged value of above 30\,G per pixel (corresponding to $10^{17}$\,Mx). If no opposite-polarity magnetic patch is found, we label that particular case as unipolar (such events are represented by dashed vertical lines in the bottom panel of Fig.\,\ref{fig:over1}). If there is an opposite-polarity patch over the 30\,G threshold, we label that event as mixed and label the opposite magnetic polarity as the minor polarity (solid vertical lines in the bottom panel of Fig.\,\ref{fig:over1}). Our choice of time-averaging the magnetic field only over the initial 30\,minutes ensures that we do not misclassify unipolar cases as mixed cases in events when new flux emergence brings up persistent opposite polarity in the region after the heating event peak. At the same time we might underestimate the number of mixed cases when the opposite polarity is weak and transient in the first 30-minute window \citep[see examples of transient opposite-polarity patches in unipolar regions presented in][]{2019A&A...623A.176C}. 

We found 101 mixed and 34 unipolar events in our sample of 137 events (the remaining two events are unclear as discussed in Appendix\,\ref{sec:lowat}). Our method is designed to detect only one footpoint of the loop (Appendix.\,\ref{sec:lowat}). In general, however, coronal loops have at least two footpoints. Through visual inspection, we found that some of the mixed cases have both footpoints rooted in mixed polarity regions. In some of the unipolar events, we found that the conjugate footpoint (i.e. the footpoint not detected by our method) is rooted in a mixed polarity region (also from visual inspection). A possible reason for this non-detection of those footpoints is that the underlying patches in the UV ratio maps did not meet our detection criteria (Appendix.\,\ref{sec:uni}). However, we restrict our magnetic field analysis only to automatically detected patches, and we do not analyse further the magnetic field properties of the events marked unipolar.

\subsubsection{Degree of magnetic flux imbalance in mixed polarity events\label{sec:fluxim}}

For the events identified as mixed, we investigate the degree of minimum magnetic flux imbalance from the 30-minute time averaged magnetic field maps. To quantify how much of persistent minor polarity is present with respect to the dominant polarity and to find the minimum distance from the centroid position to an opposite-polarity feature, we adopt the measure of flux imbalance $\psi$ as a function of radius, $r$, defined as
\begin{equation}
    \psi(r) = \left|\frac{\Phi_{\rm{p}}(r) - |\Phi_{\rm{n}}(r)|}{\Phi_{\rm{p}}(r) + |\Phi_{\rm{n}}(r)|}\right|,
\end{equation}
where $\Phi_{\rm{p}}(r)$ is the spatially integrated magnetic flux density of the positive polarity feature in the circular region of radius, $r$, and $\Phi_{\rm{n}}(r)$ is the absolute value of the spatially integrated magnetic flux density of the negative polarity feature in the same circular region. Only those pixels with magnetic flux density above 30\,G are considered for spatial integration. In our analysis, $0\,{\rm Mm} \leqslant r \leqslant 5.4\,{\rm Mm}$. By definition, at $r=0$, $\psi(0)=1$. For $0\,{\rm Mm} \leqslant r \leqslant 5.4\,{\rm Mm}$, we compute the minimum value of $\psi(r)$, and define it as the degree of flux imbalance for each case. We colour-code each event (i.e. mixed, unipolar and unclear) with the degree of flux imbalance (e.g. Fig.\,\ref{fig:over1}). The unipolar cases always have $\psi=1$.

A similar measure was employed to investigate the degree of mixed polarity at the footpoints of hot loops in one active region from observations spanning roughly 80\,minutes \citep[][]{2005ApJ...621..498K}. The study yielded the result that the magnetic field is essentially unipolar at the footpoints of hot loops in that active region. However, it is based on ground-based magnetic field observations affected by variable seeing. That, together with their relatively large scan step suggests that the spatial resolution is unlikely to have been better than ours and was possibly worse. Here we conducted an extensive statistical analysis of 137 impulsive heating events from seven different active regions covering 404\,hours of solar evolution and found that about 73\% of cases (101 out of 137) have mixed polarity magnetic fields with flux content over $10^{17}$\,Mx at the footpoints of hot loops. Due to the limited spatial resolution and sensitivity of the HMI instrument, the number of mixed events is likely to be an underestimate as shown in examples where the mixed polarity nature of the fields remains hidden to  HMI observation \citep[][]{2017ApJS..229....4C,2019A&A...623A.176C}.

\subsubsection{Rate of change of magnetic flux in mixed polarity events\label{sec:rate}}

Our final step in the process is to determine the rate of change of magnetic flux at the footpoint regions of hot loops. If the rate of change of magnetic flux is associated with reconnection, the magnetic energy liberated in the process could produce impulsive heating events. To quantify if the magnetic energy liberated is sufficient to heat the loops, we first quantify the rate of flux change ($\dot\Phi_{i}^{\rm cir}$) in the circular region,  where index $i$ denotes either dominant (\textbf{D}) or minor (\textbf{M}) polarities. To quantify the flux change rate until the peak time of the heating event, we use a simple linear fit to the first 30\,minute of the time series of the magnetic flux (its absolute value for negative polarities). The slope of the line resulting from the linear fit gives the rate of change of magnetic flux. Here the spatial integration of magnetic flux densities for dominant and minor polarities is performed over a circular region of radius 5.4\,Mm; and only those pixels with magnetic flux densities above 30\,G are considered. Depending on whether the slope of the line is positive or negative and through its magnitude, we can quantify flux enhancements (due to emergence) or reductions (due to cancellation or submergence) within the circular regions near the footpoints of hot loops.

However, it could be argued that any flux change in the region is simply due to either flux entering (mimicking emergence) or leaving (mimicking cancellation) the domain. In principle, this effect could be corrected if we know how different flux elements within the domain move. A possible way to do so is to follow the magnetic flows using a local correlation tracing technique and to correct for the flux loss or gain in the domain appropriately. However, such techniques have some shortcomings. Typically, flows are measured by smoothing the data using 2D Gaussian kernels with widths of several pixels. This results in velocities at much coarser resolution. Such velocities have to be temporally averaged to reduce noise. Moreover, at every instance, the flow velocity component normal to the tangent at every point along the circle has to be taken into account. This requires interpolation of data. Though feasible conceptually, in practice this method may not be accurate enough to determine flux loss or gain through the boundary.   

Here we implement an approximate method to account for the flux loss or gain through the boundary. Assume that $\dot\Phi_{i}^{\rm cir}$ is solely due to flux leaving or entering the domain. Magnetic features at the solar surface typically move at speeds of 1\,km\,s$^{-1}$ \citep[e.g.][]{2012ApJ...752...48C}. At these speeds, in 90\,s (the HMI cadence) the element moves about 90\,km. Even if we assume that the features are moving at local sound speeds of about 7\,km\,s$^{-1}$ for 90\,s, they would traverse a distance of 630\,km (about one HMI pixel). At such high speeds, magnetic features near the boundary can cross it (inwards or outwards). To account for this effect, we consider an annulus of width three pixels that  overlaps the circumference of the circular domain. Then, depending on whether the previously determined 30\,minute slope is positive or negative, at a given time step $t$ we subtract from or add to $\Phi_{i}^{\rm cir}(t)$ the change of magnetic flux within the annulus $\Phi_{i}^{\rm a}$ from $t$ to $t+1$. The updated flux $\Phi_{i}^{\rm up}(t)$, for $t>0$ is then given by
\[
    \Phi_{i}^{\rm up}(t+1)= 
\begin{cases}
    \Phi_{i}^{\rm cir}(t+1) - \left[\Phi_{i}^{\rm a}(t+1)-\Phi_{i}^{\rm a}(t)\right],& \text{if } \dot\Phi_{i}^{\rm cir} > 0\\
\Phi_{i}^{\rm cir}(t+1) + \left[\Phi_{i}^{\rm a}(t+1)-\Phi_{i}^{\rm a}(t)\right],& \text{if } \dot\Phi_{i}^{\rm cir} < 0.
\end{cases}
\]

We use $\Phi_{i}^{\rm up}$ to determine the rate of change of magnetic flux ($\dot\Phi_{i}^{\rm up}$) in the circular domain near the loop footpoints, by approximately accounting for the magnetic flux entering or leaving the domain (the values of \textbf{D} and \textbf{M} in units of $10^{15}$\,Mx\,s$^{-1}$ of all the mixed events are listed in Tables. \ref{tab:AR12665}, \ref{tab:AR12699}, \ref{tab:AR12712}, \ref{tab:AR12713}, \ref{tab:AR12733}, and \ref{tab:AR12738}; for unipolar and unclear events we assign a value of zero).

\section{Further details and additional examples of impulsive heating events\label{sec:exam}}

\begin{figure*}
\begin{center}
\includegraphics[width=\textwidth]{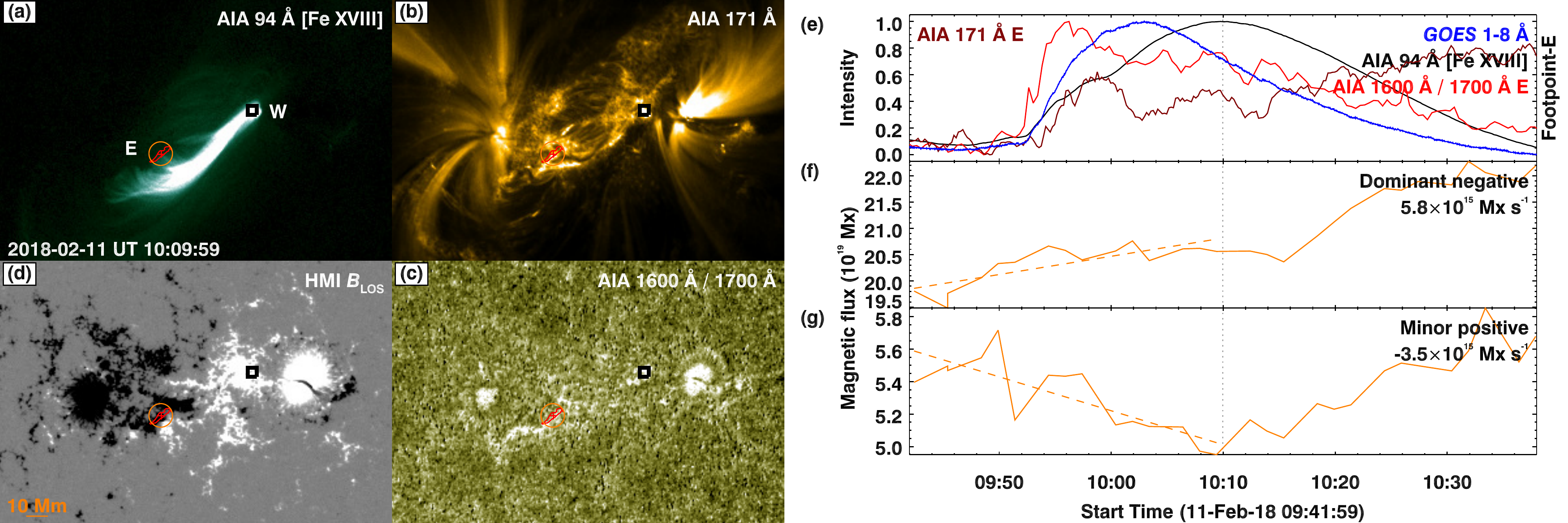}
\includegraphics[width=\textwidth]{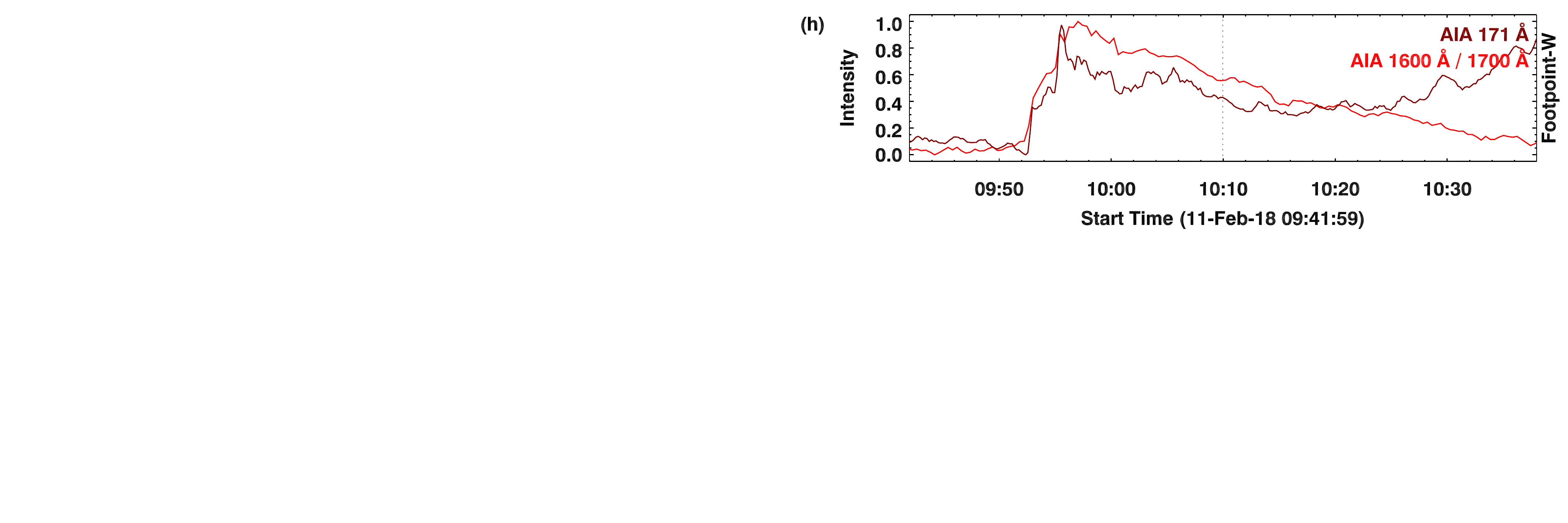}
{\vskip-2.75cm
\caption{Impulsive heating in the core of AR 12699. The format is the same as in Fig.\,\ref{fig:AR12665c1}. Animation of panels (a) to (g) is available online. See Appendix\,\ref{sec:exam} for details.\label{fig:AR12699c1}}}
\end{center}
\end{figure*}

\begin{figure*}
\begin{center}
\includegraphics[width=\textwidth]{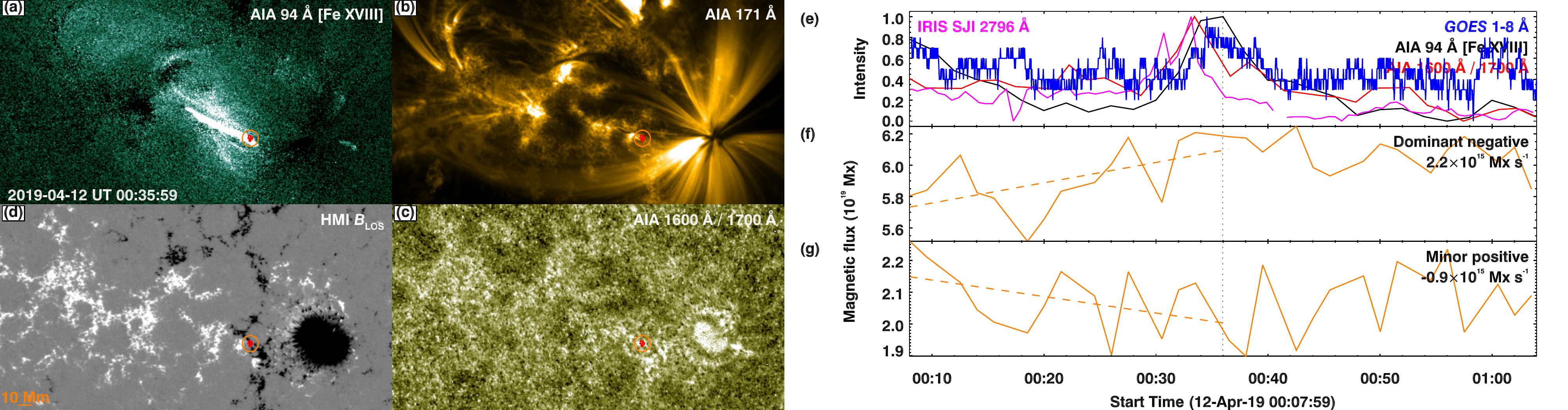}
\includegraphics[width=0.45\textwidth]{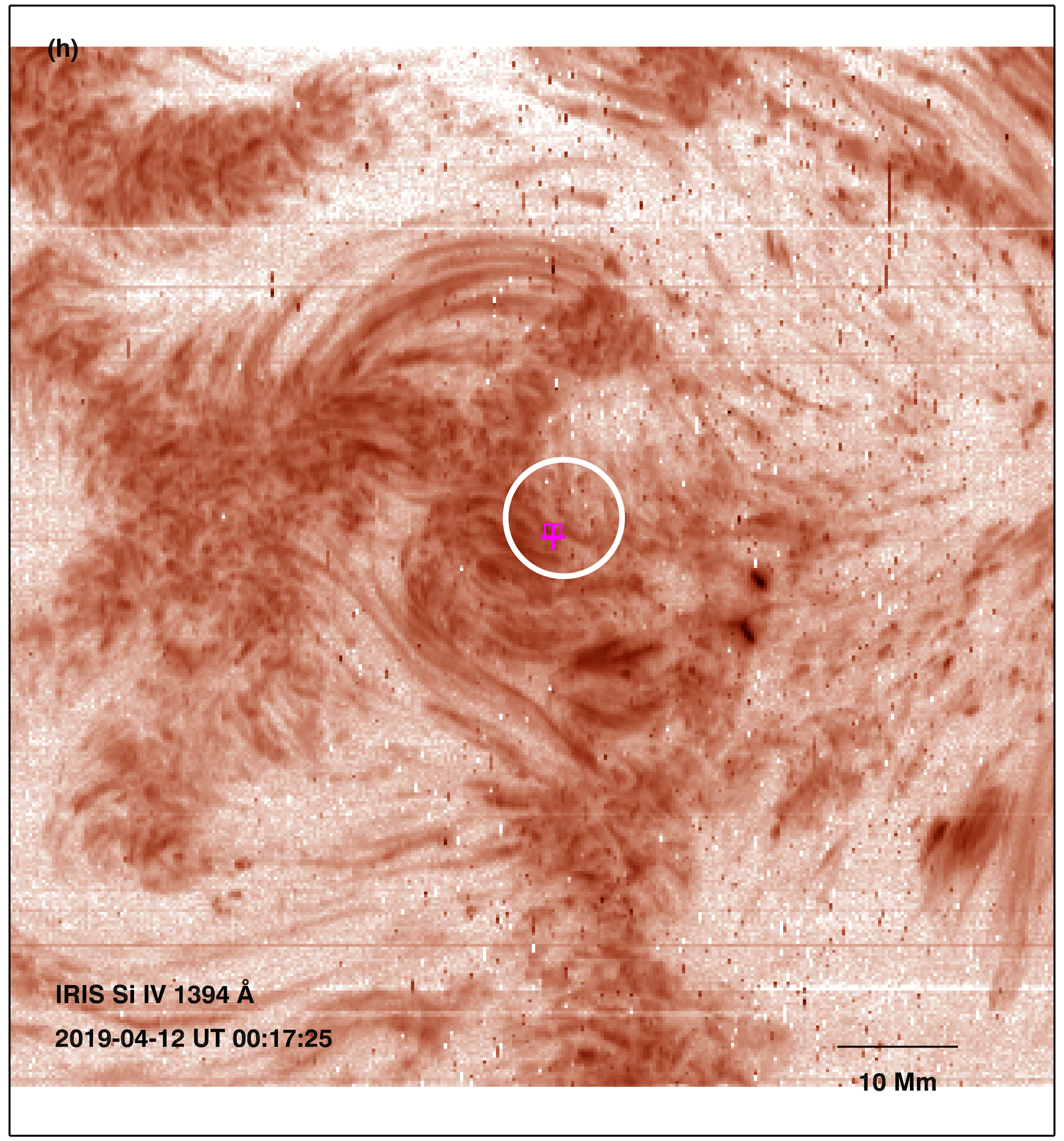}
\includegraphics[width=0.45\textwidth]{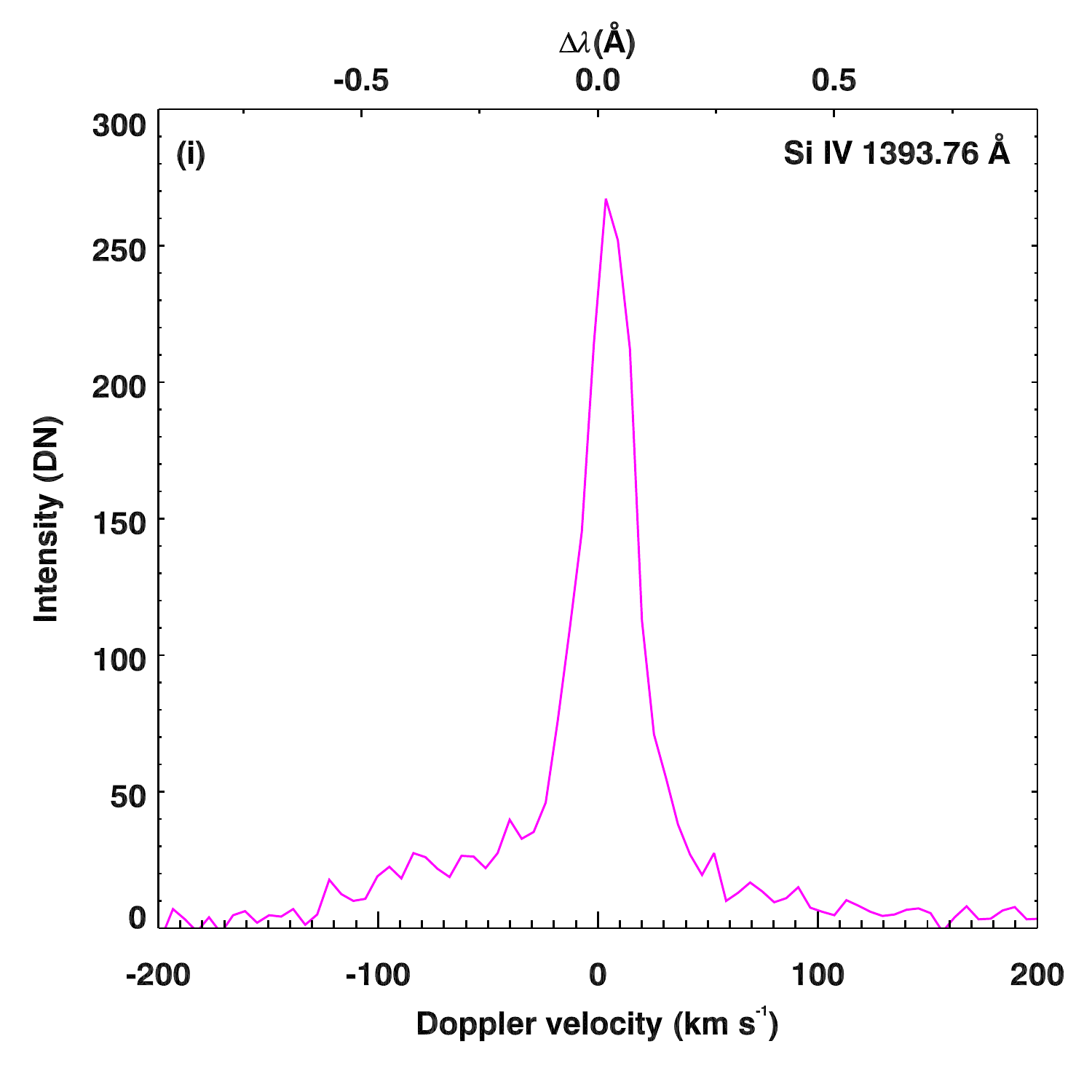}
\caption{Impulsive heating in the core of AR 12738. The format is the same as in Fig.\,\ref{fig:AR12692c1}. AIA light curves are shown at lower cadence. Animation of panels (a) to (g) is available online. See Appendix\,\ref{sec:exam} for details.\label{fig:AR12738c1}}
\end{center}
\end{figure*}

In the main text we presented overview of two ARs along with two representative examples. Here we present overview and representative examples from the remaining ARs including tabulated details of all the impulsive heating events analysed in the seven ARs. Just as for the representative examples in the main text, we overlay magnetic field lines (in yellow) traced from a linear force free field  (lfff) extrapolation in all but the two heating events displayed in Figs.\,\ref{fig:AR12699c1} and \ref{fig:AR12738c1}. In these two cases, though the spatial extent of the hot loop seen in the Fe\,{\sc xviii} emission and its relation to the footpoints, including the one detected by our method is clear (see also supporting online animations), the lfff extrapolations did not yield field lines that  trace the hot loop. This is probably because of shortcoming in the simple lfff extrapolation methodology employed here. A detailed parametric study and magnetic modelling is required in the future to properly evaluate why lfff extrapolations fail during these impulsive heating episodes. Perhaps more complex non-linear force free field extrapolations would do a better job, even though they also possess limitations when modelling complex active region magnetic fields \citep[e.g.][]{2008SoPh..247..269M,2015A&A...584A..68P}. Or perhaps magnetohydrostatic or magnetohydrodynamic modelling is necessary, but these are outside the scope of the present paper.

\subsection{AR 12665}

\begin{table*}
\begin{center}
\caption{Overview of impulsive heating events in the core of active region AR12665.\label{tab:AR12665}}
\begin{tabular}{l c c  c c c}
\hline\hline
Event No. & Fe\,{\sc xviii} peak time (UT) & Solar (\textit{X}, \textit{Y}) & Event type & \textbf{D} (10$^{15}$ Mx s$^{-1}$) & \textbf{M} (10$^{15}$ Mx s$^{-1}$) \\
\hline
1   & 2017-07-11 UT 01:13:59   & (-140.1\arcsec, -147.0\arcsec)   & Mixed   &    2.2   &   -0.4   \\
2   & 2017-07-11 UT 02:03:59   & (-130.4\arcsec, -178.9\arcsec)   & Mixed   &    0.8   &    0.7   \\
3   & 2017-07-11 UT 03:09:59   & (-159.7\arcsec, -179.5\arcsec)   & Mixed   &    0.7   &   -0.0   \\
4   & 2017-07-11 UT 04:37:59   & (-108.9\arcsec, -184.1\arcsec)   & Mixed   &   -1.4   &    0.7   \\
5   & 2017-07-11 UT 05:19:59   & (-103.5\arcsec, -184.2\arcsec)   & Mixed   &    1.2   &    0.9   \\
6   & 2017-07-11 UT 10:35:59   & (-117.6\arcsec, -151.0\arcsec)   & Mixed   &    0.5   &   -0.1   \\
7   & 2017-07-11 UT 11:39:59   & (-117.4\arcsec, -191.6\arcsec)   & Mixed   &   -2.5   &    0.6   \\
8   & 2017-07-11 UT 12:59:59   & (-28.3\arcsec, -152.3\arcsec)   & Mixed   &   -5.4   &    0.5   \\
9   & 2017-07-11 UT 15:17:59   & (-11.5\arcsec, -154.1\arcsec)   & Mixed   &    2.1   &   -0.7   \\
10   & 2017-07-11 UT 16:59:59   & (3.2\arcsec, -152.6\arcsec)   & Mixed   &    3.7   &    0.8   \\
11   & 2017-07-11 UT 18:29:59   & (15.2\arcsec, -148.6\arcsec)   & Mixed   &   -3.1   &   -1.6   \\
12   & 2017-07-11 UT 21:33:59   & (-15.5\arcsec, -132.6\arcsec)   & Unipolar   & \textendash   & \textendash   \\
13   & 2017-07-11 UT 22:17:59   & (49.9\arcsec, -168.5\arcsec)   & Unipolar   & \textendash   & \textendash   \\
14   & 2017-07-12 UT 00:09:59   & (30.5\arcsec, -170.5\arcsec)   & Mixed   &    1.8   &    3.1   \\
15   & 2017-07-12 UT 02:35:59   & (102.6\arcsec, -181.5\arcsec)   & Unipolar   & \textendash   & \textendash   \\
16   & 2017-07-12 UT 03:19:59   & (96.8\arcsec, -190.6\arcsec)   & Mixed   &   -0.1   &   -4.7   \\
17   & 2017-07-12 UT 05:31:59   & (119.0\arcsec, -179.2\arcsec)   & Mixed   &    3.0   &    0.2   \\
18   & 2017-07-12 UT 06:31:59   & (97.8\arcsec, -156.7\arcsec)   & Mixed   &    3.4   &    0.7   \\
19   & 2017-07-12 UT 07:27:59   & (136.9\arcsec, -191.3\arcsec)   & Mixed   &    1.9   &   -1.7   \\
20   & 2017-07-12 UT 08:23:59   & (92.3\arcsec, -158.5\arcsec)   & Mixed   &   -4.9   &   -0.7   \\
21   & 2017-07-12 UT 09:11:59   & (100.8\arcsec, -186.9\arcsec)   & Unipolar   & \textendash   & \textendash   \\
\hline
\end{tabular}
\end{center}
\end{table*}
\clearpage

\newpage
\subsection{AR 12692}

\begin{figure*}
\begin{center}
\includegraphics[width=\textwidth]{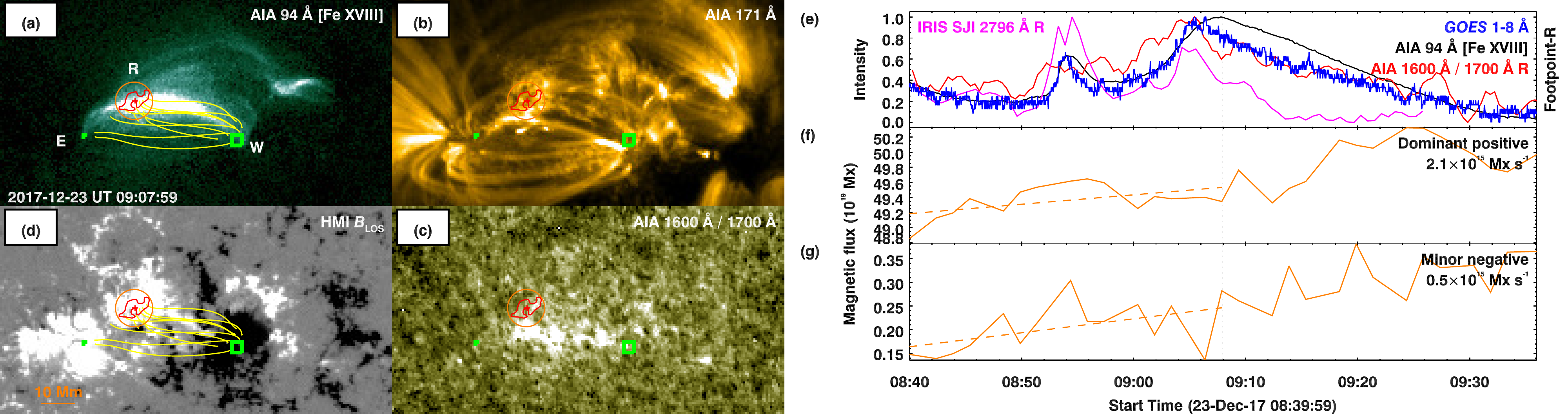}
\includegraphics[width=\textwidth]{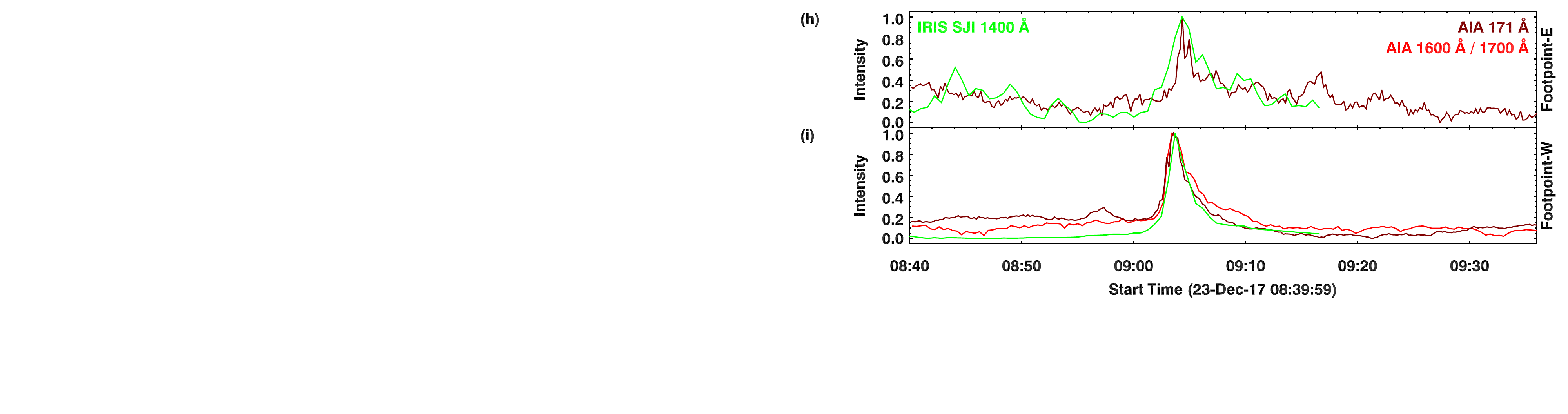}
{\vskip-4.75cm
\includegraphics[width=0.45\textwidth]{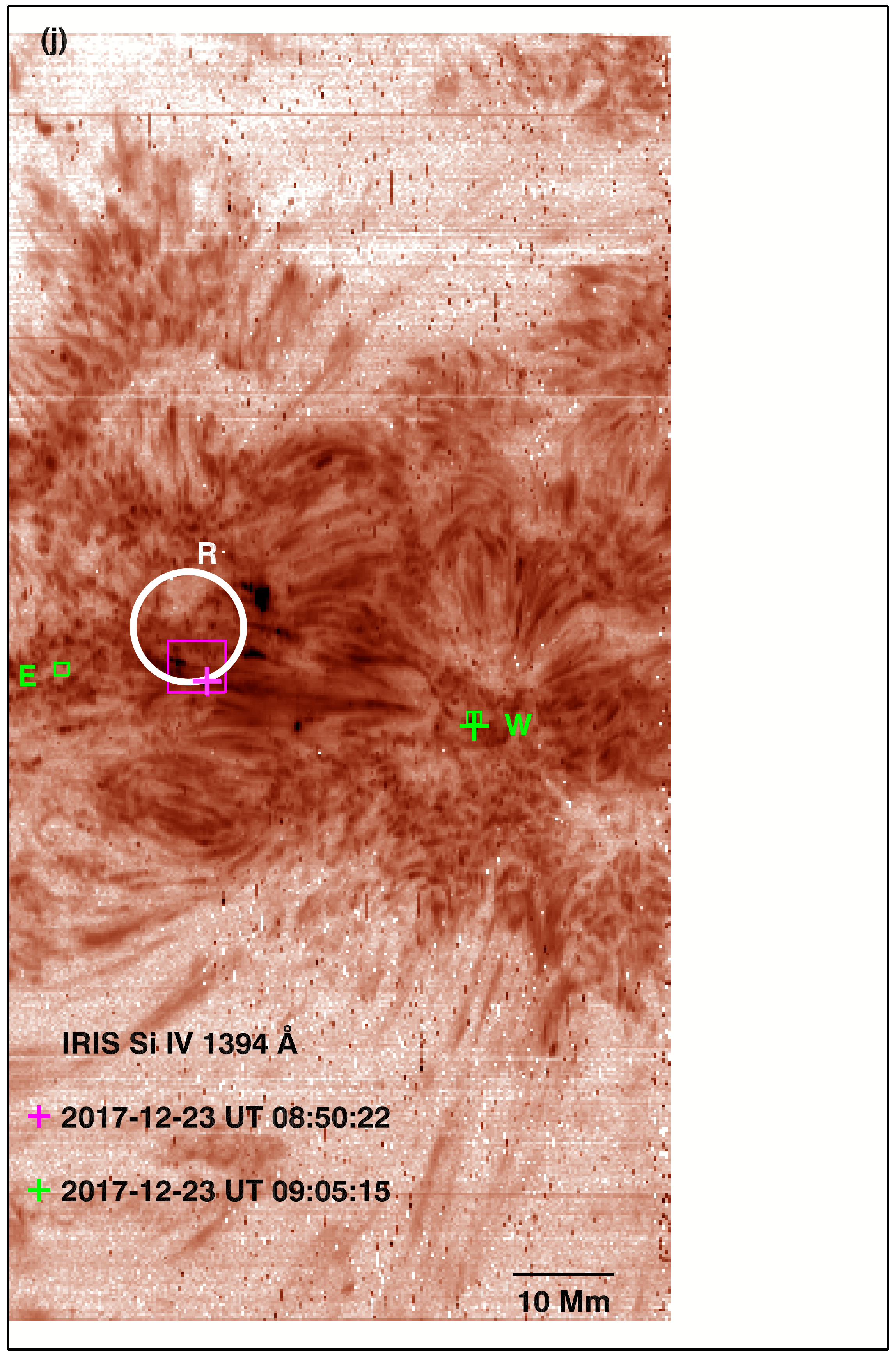}}
\includegraphics[width=0.45\textwidth]{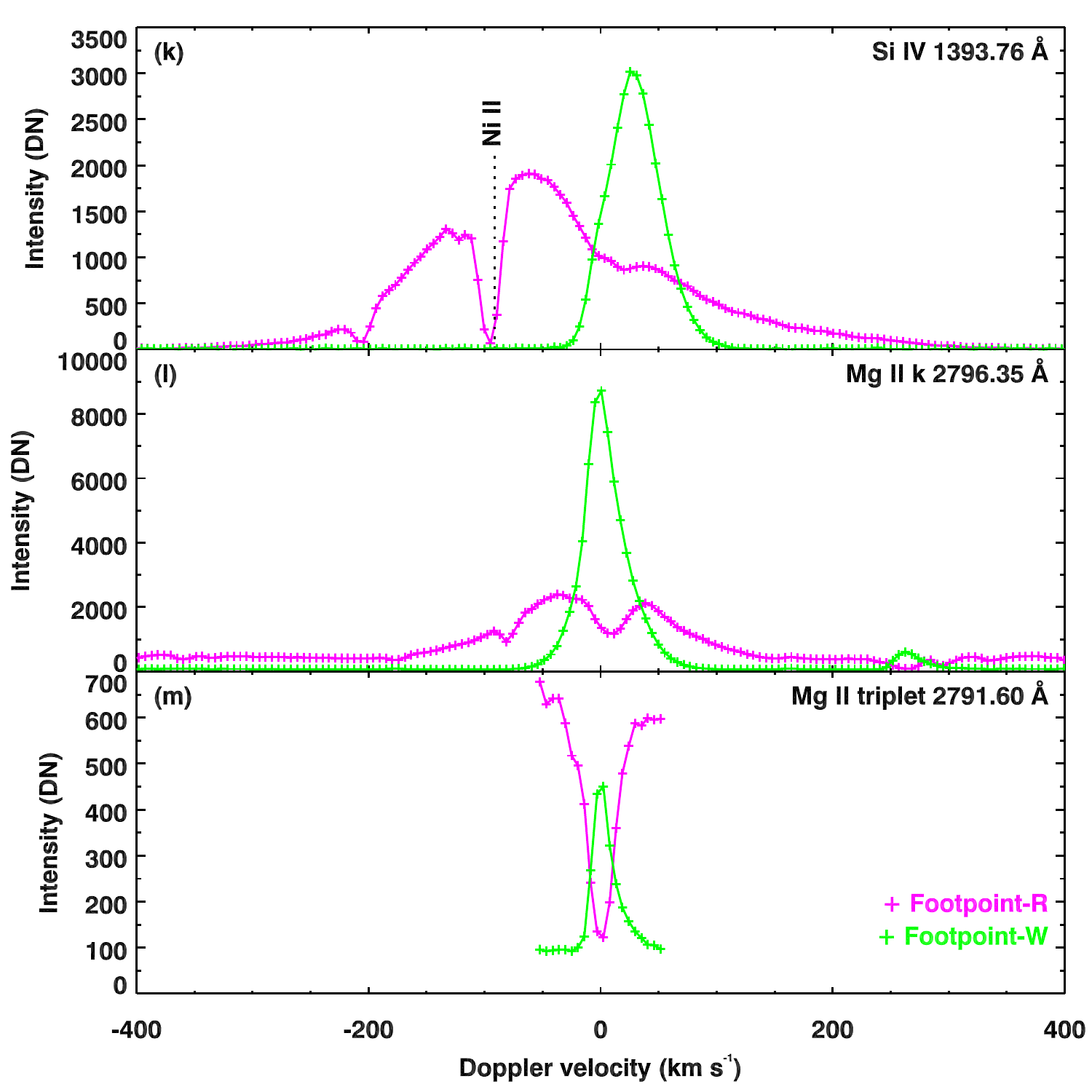}
\caption{Impulsive heating in the core of AR 12692. The format is the same as in Fig.\,\ref{fig:AR12692c1}. Animation of panels (a) to (g) is available online. \label{fig:AR12692c2}}
\end{center}
\end{figure*}

\begin{table*}
\begin{center}
\caption{Overview of impulsive heating events in the core of active region AR12692.\label{tab:AR12692}}
\begin{tabular}{l c c  c c c}
\hline\hline
Event No. & Fe\,{\sc xviii} peak time (UT) & Solar (\textit{X}, \textit{Y}) & Event type & \textbf{D} (10$^{15}$ Mx s$^{-1}$) & \textbf{M} (10$^{15}$ Mx s$^{-1}$) \\
\hline
1   & 2017-12-23 UT 00:31:59   & (-279.2\arcsec, 321.3\arcsec)   & Mixed   &    1.9   &    0.3   \\
2   & 2017-12-23 UT 02:09:59   & (-264.0\arcsec, 325.7\arcsec)   & Mixed   &   -7.7   &   -0.2   \\
3   & 2017-12-23 UT 03:43:59   & (-244.9\arcsec, 343.6\arcsec)   & Mixed   &    0.5   &    0.9   \\
4   & 2017-12-23 UT 06:29:59   & (-236.6\arcsec, 328.7\arcsec)   & Mixed   &    3.7   &    2.1   \\
5   & 2017-12-23 UT 07:33:59   & (-205.8\arcsec, 329.8\arcsec)   & Mixed   &    3.0   &   -0.5   \\
6   & 2017-12-23 UT 09:07:59   & (-231.5\arcsec, 338.4\arcsec)   & Mixed   &    2.1   &    0.5   \\
7   & 2017-12-23 UT 10:51:59   & (-192.5\arcsec, 325.3\arcsec)   & Mixed   &   -0.6   &    2.4   \\
8   & 2017-12-23 UT 12:07:59   & (-173.0\arcsec, 327.1\arcsec)   & Mixed   &   -0.8   &   -0.6   \\
9   & 2017-12-23 UT 14:33:59   & (-192.5\arcsec, 314.8\arcsec)   & Mixed   &    9.4   &    0.9   \\
10   & 2017-12-23 UT 16:15:59   & (-164.6\arcsec, 337.0\arcsec)   & Unipolar   & \textendash   & \textendash   \\
11   & 2017-12-24 UT 01:47:59   & (-2.1\arcsec, 305.9\arcsec)   & Mixed   &   -0.5   &   -0.9   \\
12   & 2017-12-24 UT 03:35:59   & (14.8\arcsec, 304.4\arcsec)   & Mixed   &   -2.9   &   -1.2   \\
13   & 2017-12-24 UT 15:43:59   & (73.7\arcsec, 329.8\arcsec)   & Unipolar   & \textendash   & \textendash   \\
14   & 2017-12-24 UT 17:01:59   & (81.9\arcsec, 332.3\arcsec)   & Unipolar   & \textendash   & \textendash   \\
15   & 2017-12-24 UT 18:39:59   & (64.7\arcsec, 317.0\arcsec)   & Unipolar   & \textendash   & \textendash   \\
16   & 2017-12-24 UT 21:31:59   & (83.9\arcsec, 333.4\arcsec)   & Mixed   &   -4.9   &   -1.4   \\
17   & 2017-12-25 UT 00:25:59   & (160.4\arcsec, 334.3\arcsec)   & Unipolar   & \textendash   & \textendash   \\
18   & 2017-12-25 UT 02:33:59   & (135.1\arcsec, 337.8\arcsec)   & Mixed   &    1.4   &   -0.1   \\
19   & 2017-12-25 UT 05:57:59   & (221.9\arcsec, 346.2\arcsec)   & Mixed   &   -0.8   &    0.2   \\
20   & 2017-12-25 UT 07:17:59   & (178.3\arcsec, 347.4\arcsec)   & Unipolar   & \textendash   & \textendash   \\
21   & 2017-12-25 UT 10:21:59   & (240.2\arcsec, 330.3\arcsec)   & Unipolar   & \textendash   & \textendash   \\
22   & 2017-12-25 UT 11:43:59   & (220.0\arcsec, 346.5\arcsec)   & Mixed   &   -2.9   &    0.5   \\
\hline
\end{tabular}
\end{center}
\end{table*}
\clearpage

\newpage
\subsection{AR 12699}

\begin{figure*}
\begin{center}
\includegraphics[width=0.5\textwidth]{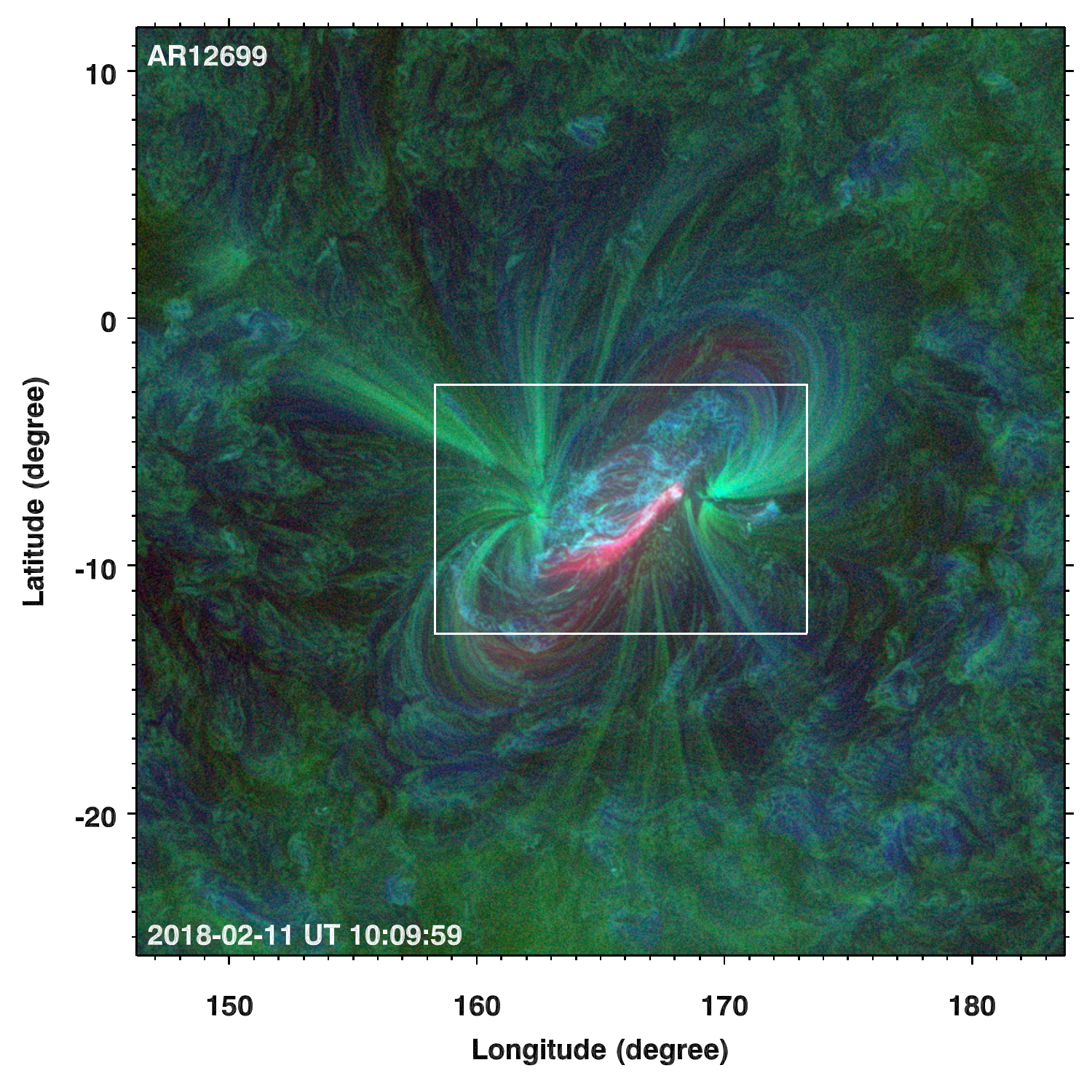}
\includegraphics[width=\textwidth]{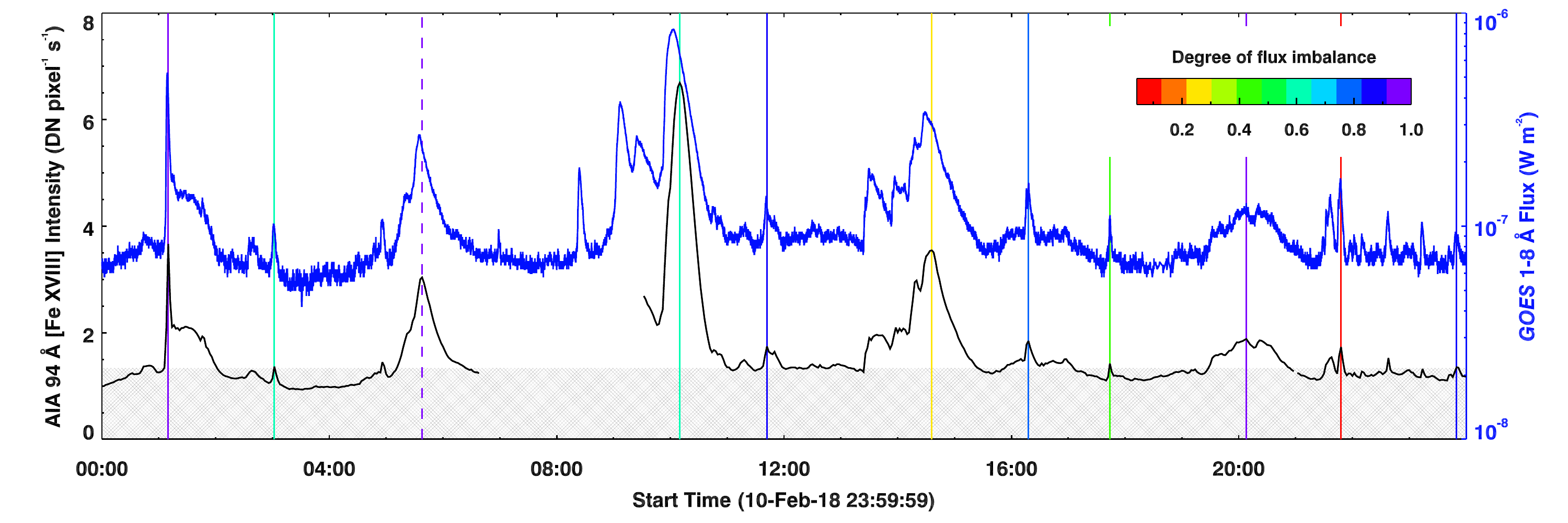}
\caption{Overview of impulsive heating events observed in AR 12699. The format is the same as in Fig.\,\ref{fig:over1}. \label{fig:over3}}
\end{center}
\end{figure*}

\begin{figure*}
\begin{center}
\includegraphics[width=\textwidth]{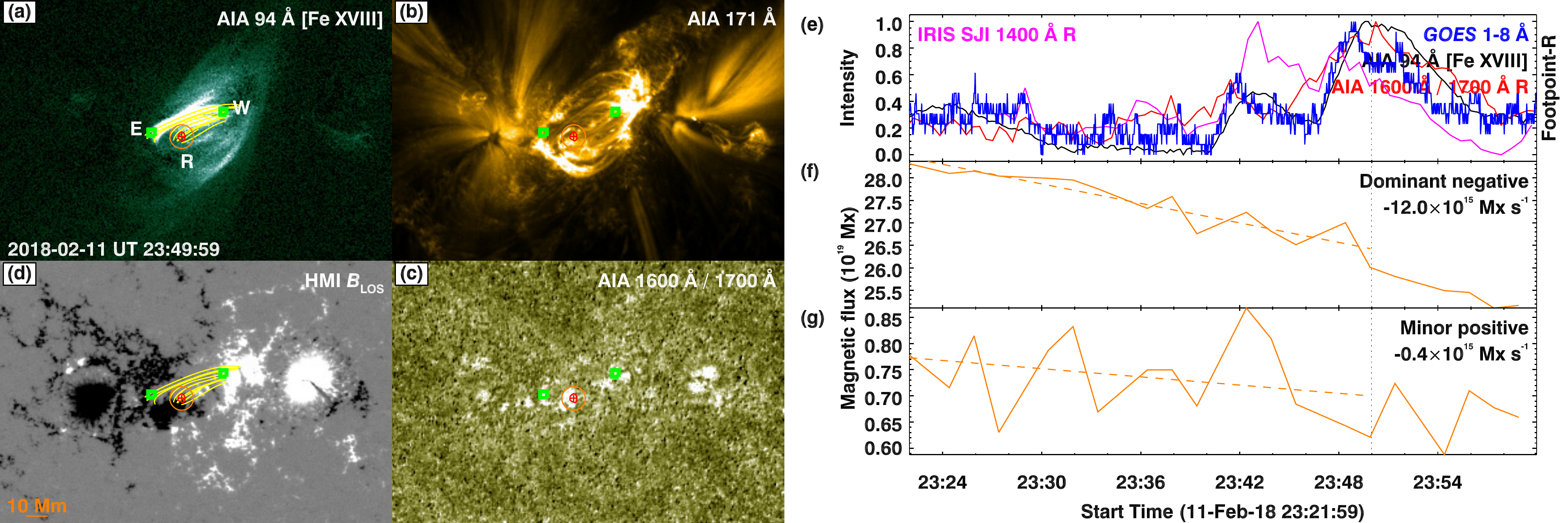}
\includegraphics[width=\textwidth]{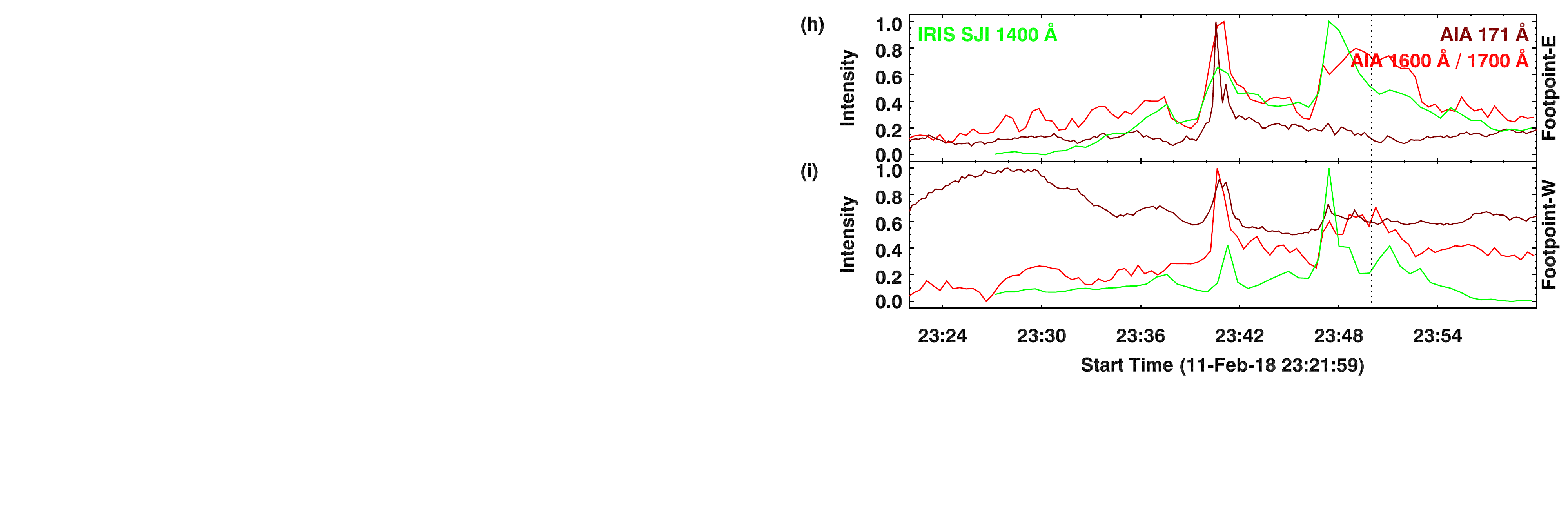}
{\vskip-5.75cm
\includegraphics[width=0.45\textwidth]{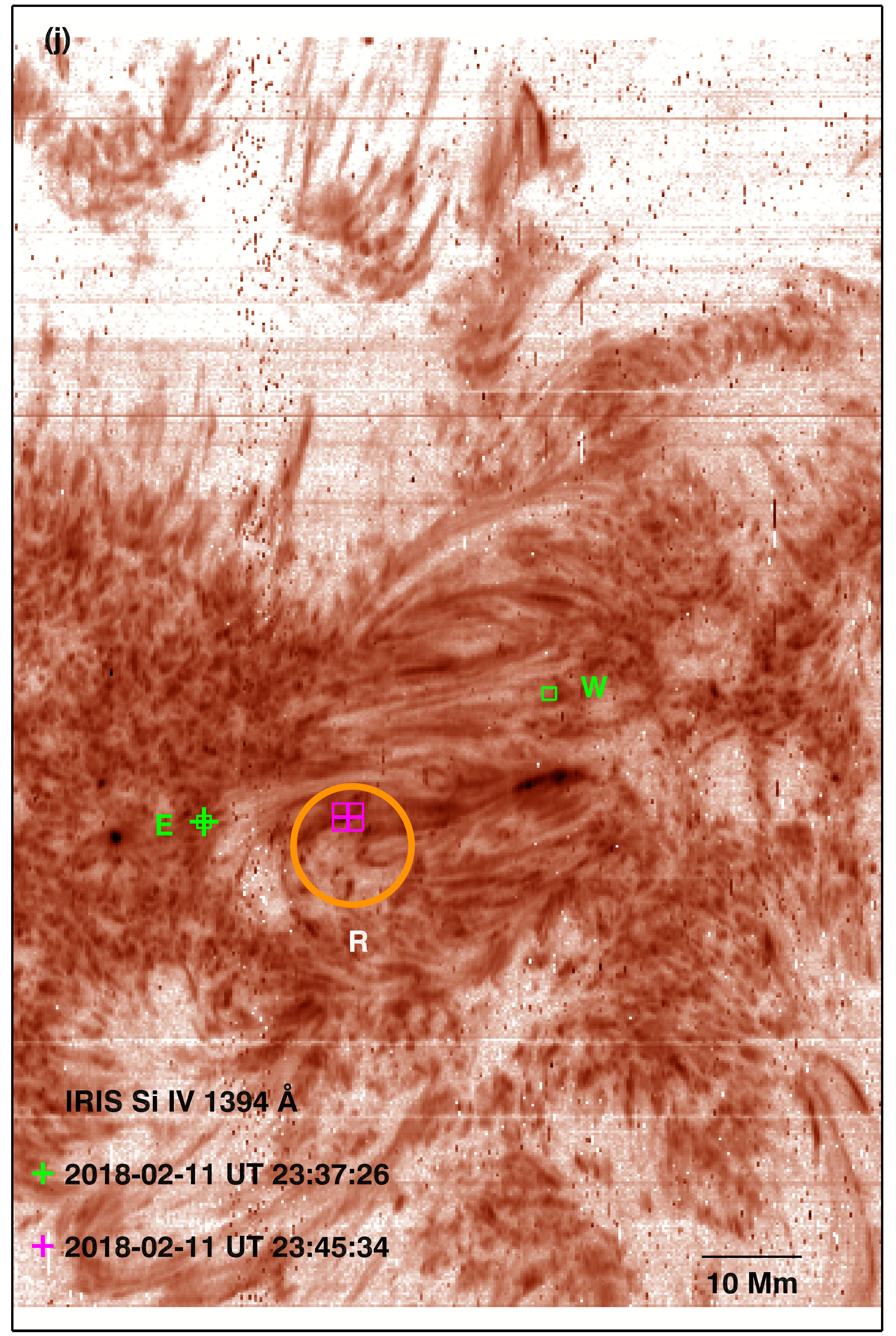}}
\includegraphics[width=0.45\textwidth]{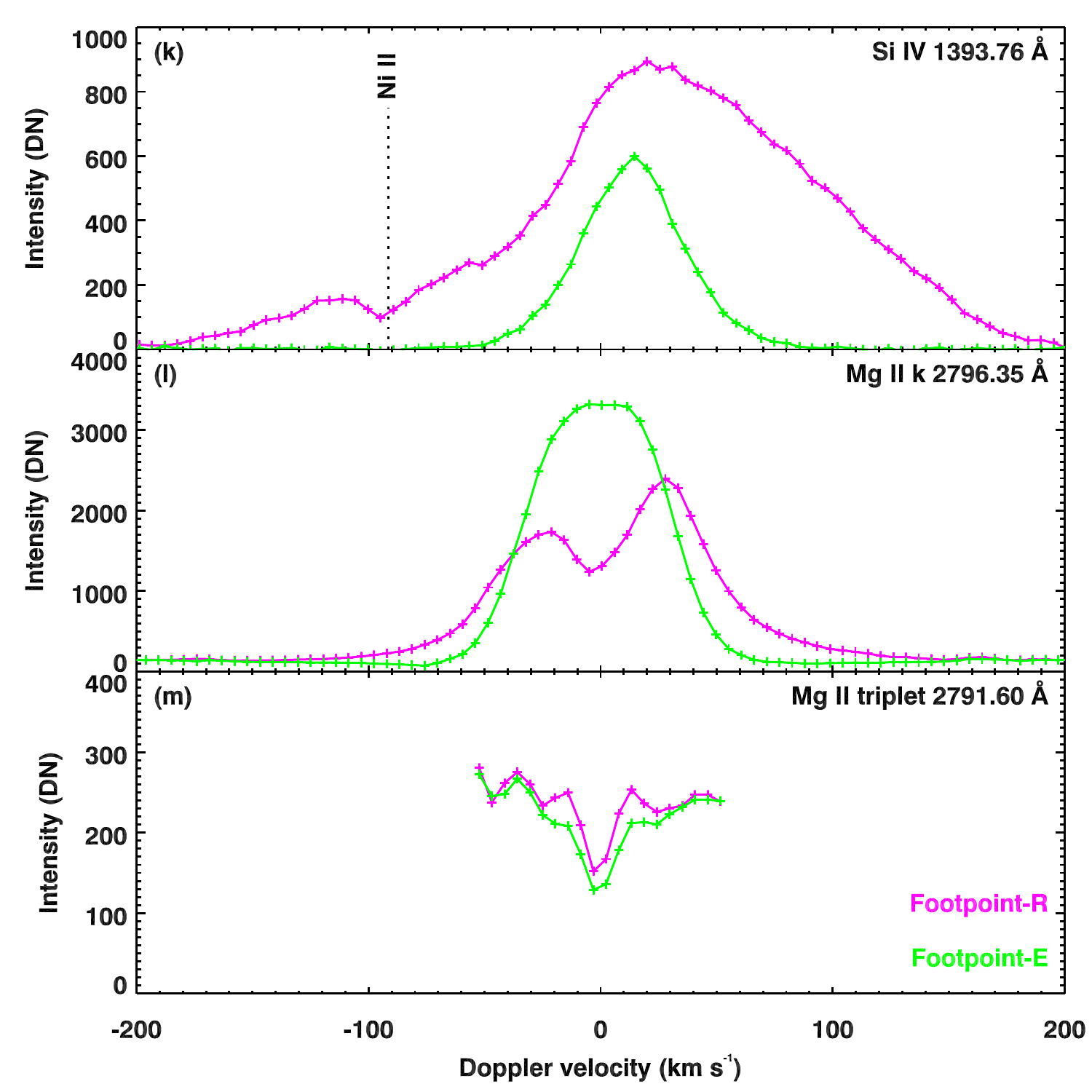}
\caption{Impulsive heating in the core of AR 12699. The format is the same as in Fig.\,\ref{fig:AR12692c1}. Animation of panels (a) to (g) is available online. \label{fig:AR12699c2}}
\end{center}
\end{figure*}

\begin{table*}
\begin{center}
\caption{Overview of impulsive heating events in the core of active region AR12699.\label{tab:AR12699}}
\begin{tabular}{l c c  c c c}
\hline\hline
Event No. & Fe\,{\sc xviii} peak time (UT) & Solar (\textit{X}, \textit{Y}) & Event type & \textbf{D} (10$^{15}$ Mx s$^{-1}$) & \textbf{M} (10$^{15}$ Mx s$^{-1}$) \\
\hline
1   & 2018-02-11 UT 01:09:59   & (75.5\arcsec, -17.4\arcsec)   & Mixed   &   -7.2   &    0.2   \\
2   & 2018-02-11 UT 03:01:59   & (108.0\arcsec, -15.0\arcsec)   & Mixed   &    0.4   &    1.2   \\
3   & 2018-02-11 UT 05:37:59   & (77.6\arcsec, -11.7\arcsec)   & Unipolar   & \textendash   & \textendash   \\
4   & 2018-02-11 UT 10:09:59   & (64.7\arcsec, -33.9\arcsec)   & Mixed   &    5.8   &   -3.5   \\
5   & 2018-02-11 UT 11:41:59   & (83.7\arcsec, -42.6\arcsec)   & Mixed   &   -1.3   &   -1.3   \\
6   & 2018-02-11 UT 14:35:59   & (134.4\arcsec, -22.8\arcsec)   & Mixed   &   12.7   &   -3.1   \\
7   & 2018-02-11 UT 16:17:59   & (99.8\arcsec, -41.7\arcsec)   & Mixed   &   -0.4   &   -3.6   \\
8   & 2018-02-11 UT 17:43:59   & (119.8\arcsec, -43.6\arcsec)   & Mixed   &   -3.7   &   -6.6   \\
9   & 2018-02-11 UT 20:07:59   & (178.3\arcsec, -29.2\arcsec)   & Mixed   &   -5.1   &   -0.4   \\
10   & 2018-02-11 UT 21:47:59   & (120.3\arcsec, 2.1\arcsec)   & Mixed   &   -7.9   &    0.6   \\
11   & 2018-02-11 UT 23:49:59   & (203.8\arcsec, -24.4\arcsec)   & Mixed   &  -12.0   &   -0.4   \\
\hline
\end{tabular}
\end{center}
\end{table*}

\clearpage

\newpage
\subsection{AR 12712}

\begin{figure*}
\begin{center}
\includegraphics[width=0.5\textwidth]{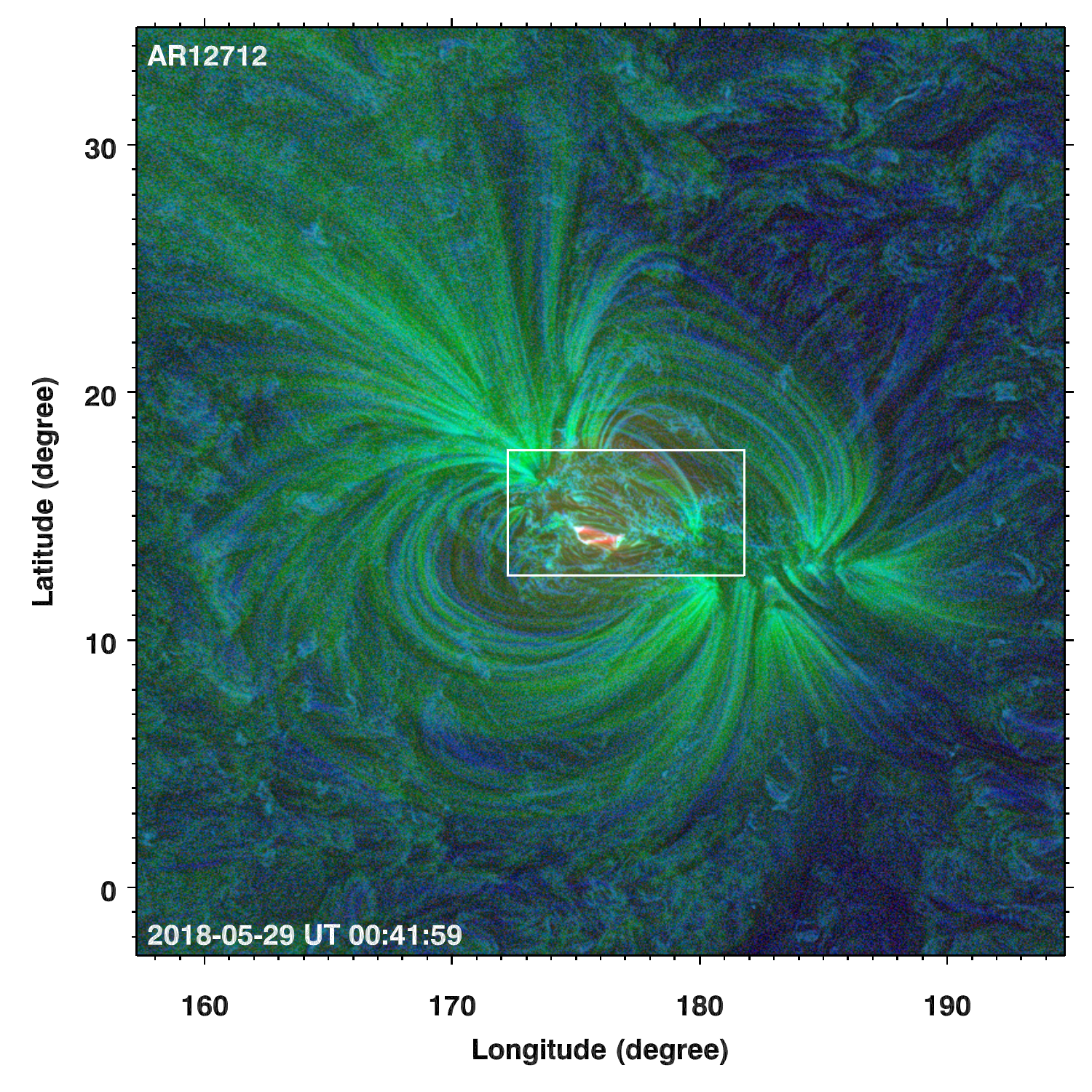}
\includegraphics[width=\textwidth]{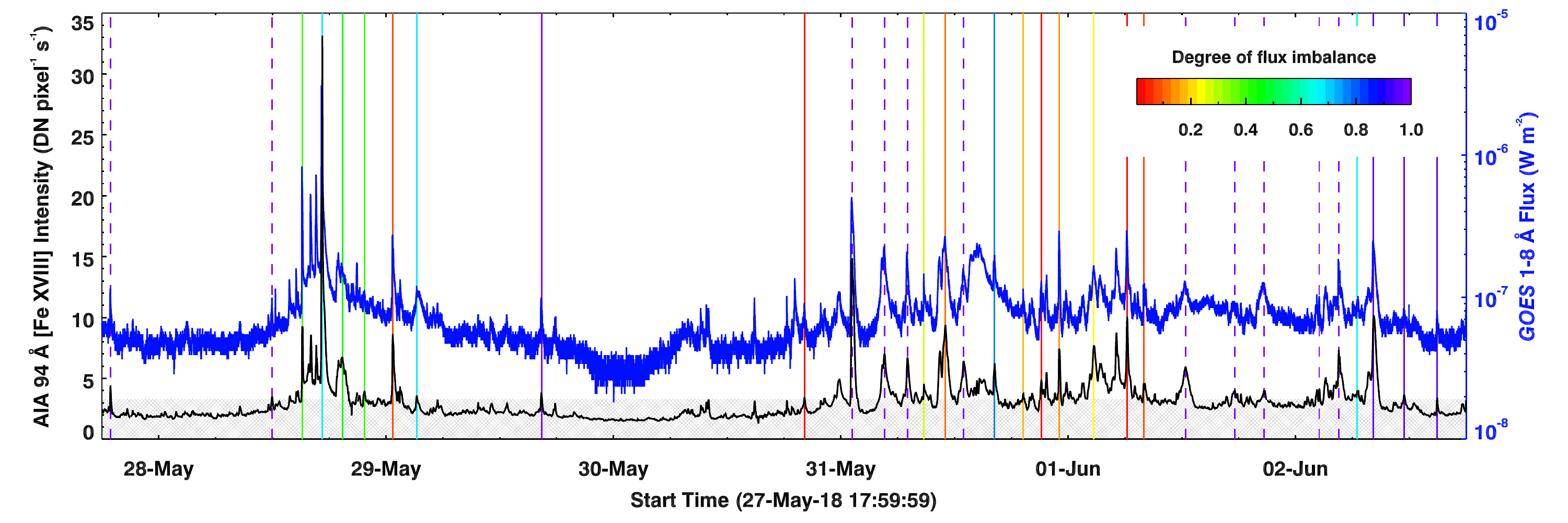}
\caption{Overview of impulsive heating events observed in AR 12712. The format is the same as in Fig.\,\ref{fig:over1}. \label{fig:over4}}
\end{center}
\end{figure*}

\begin{figure*}
\includegraphics[width=\textwidth]{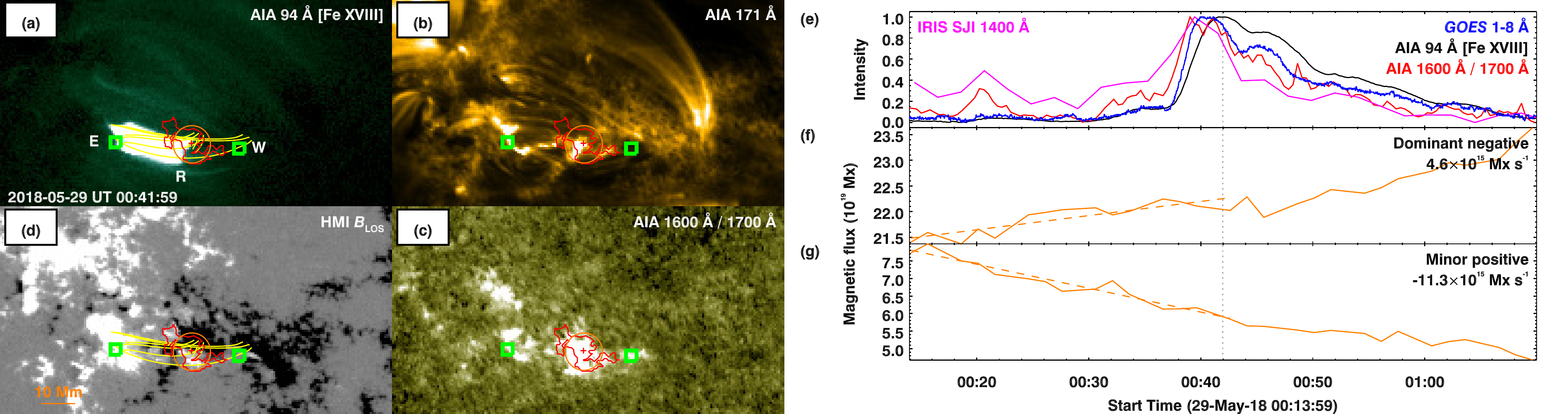}
\includegraphics[width=\textwidth]{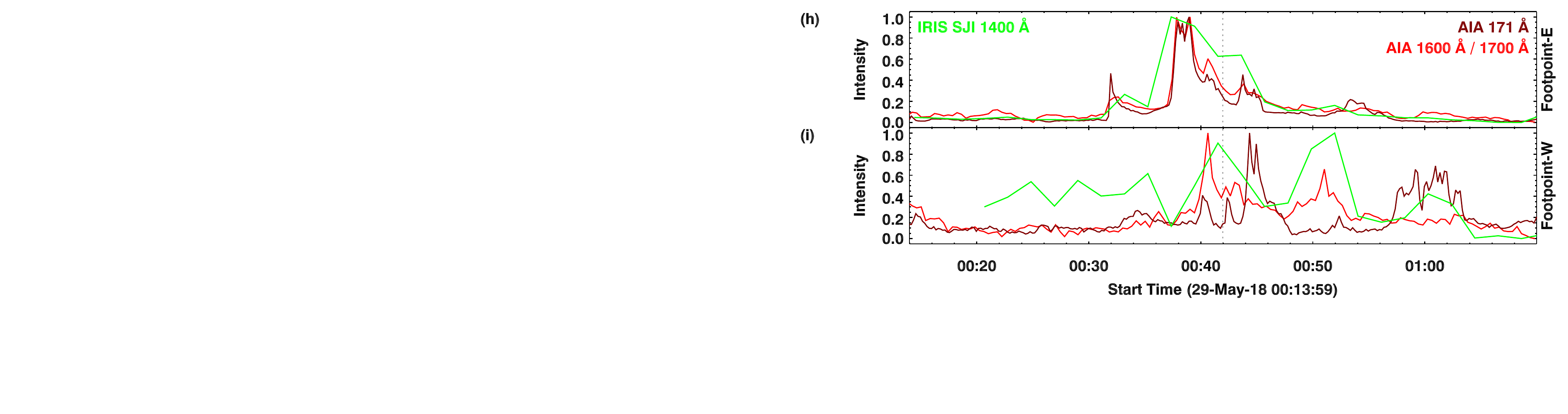}
{\vskip-4.75cm
\includegraphics[width=0.5025\textwidth]{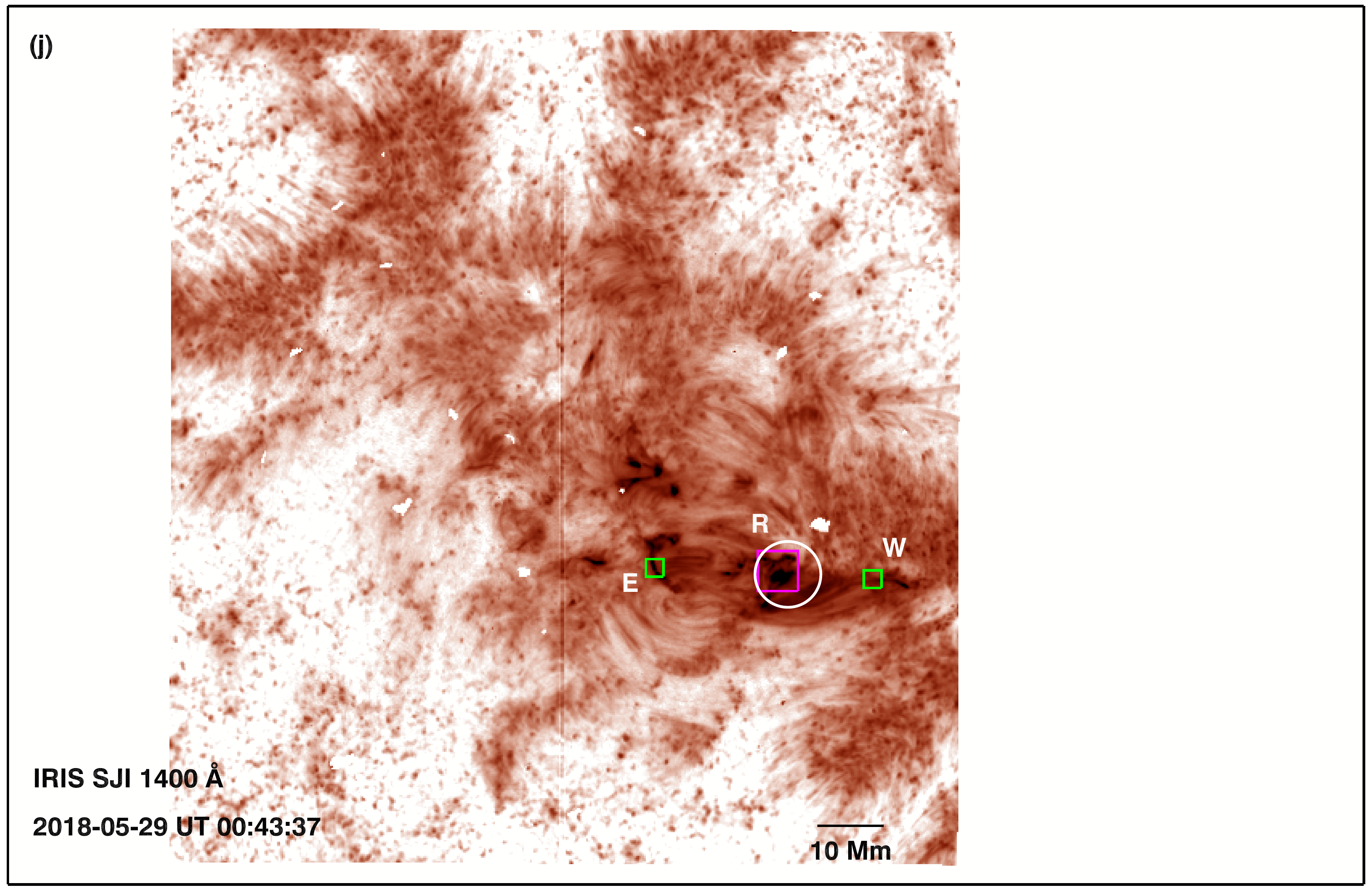}}
\caption{Impulsive heating in the core of AR 12712. Panels (a)-(i) and (f)-(g) are similar to Fig.\,\ref{fig:AR12665c1}. Panel (j) shows IRIS SJI 1400\,\AA\ map in inverted grey scale. Animation of panels (a) to (g) is available online. \label{fig:AR12712c1}}
\end{figure*}

\begin{figure*}
\begin{center}
\includegraphics[width=\textwidth]{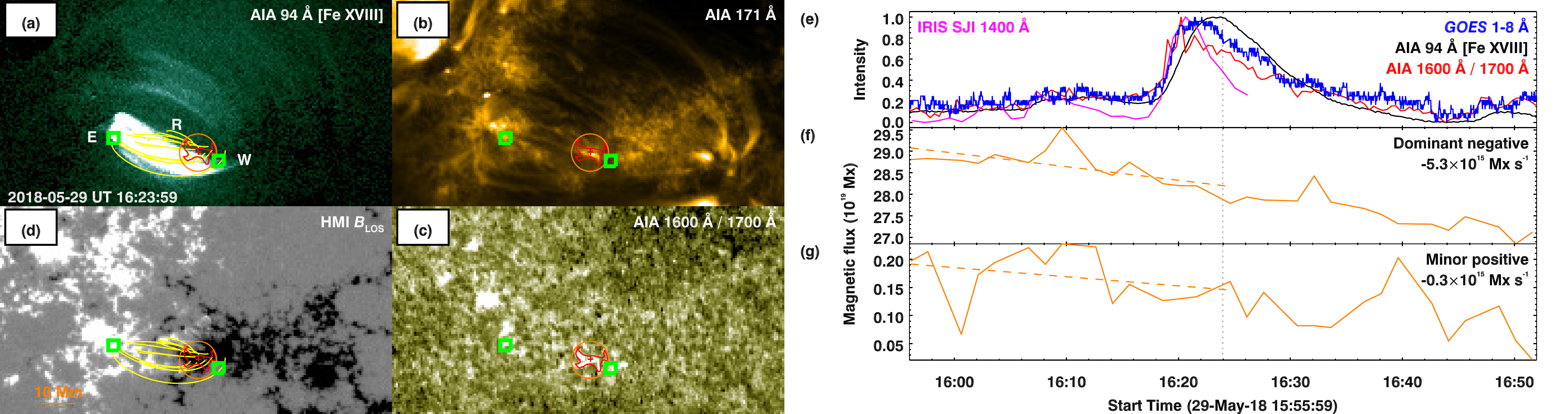}
\includegraphics[width=\textwidth]{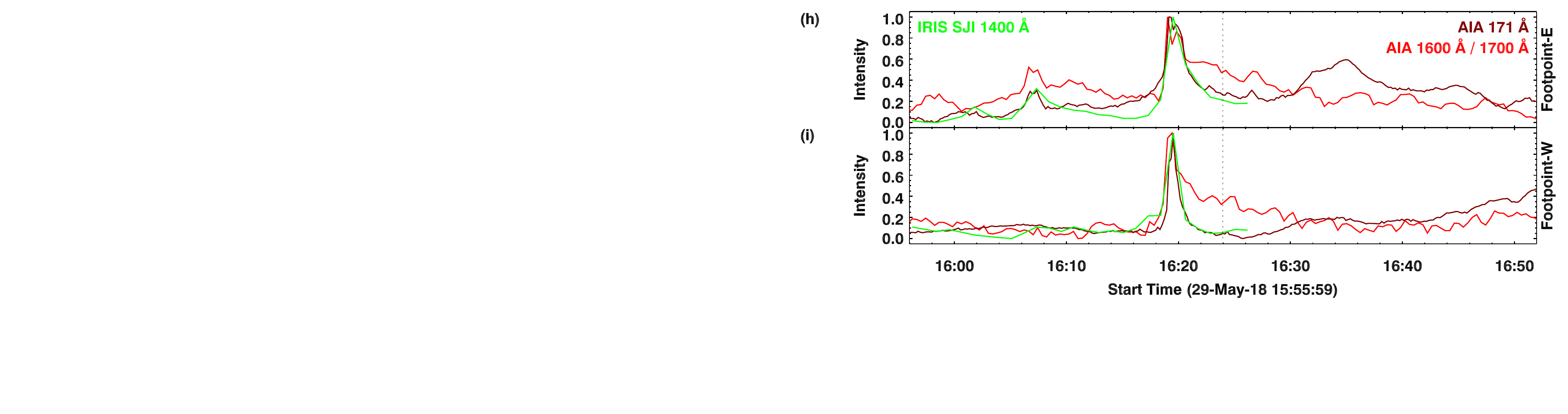}
{\vskip-4.75cm
\includegraphics[width=0.45\textwidth]{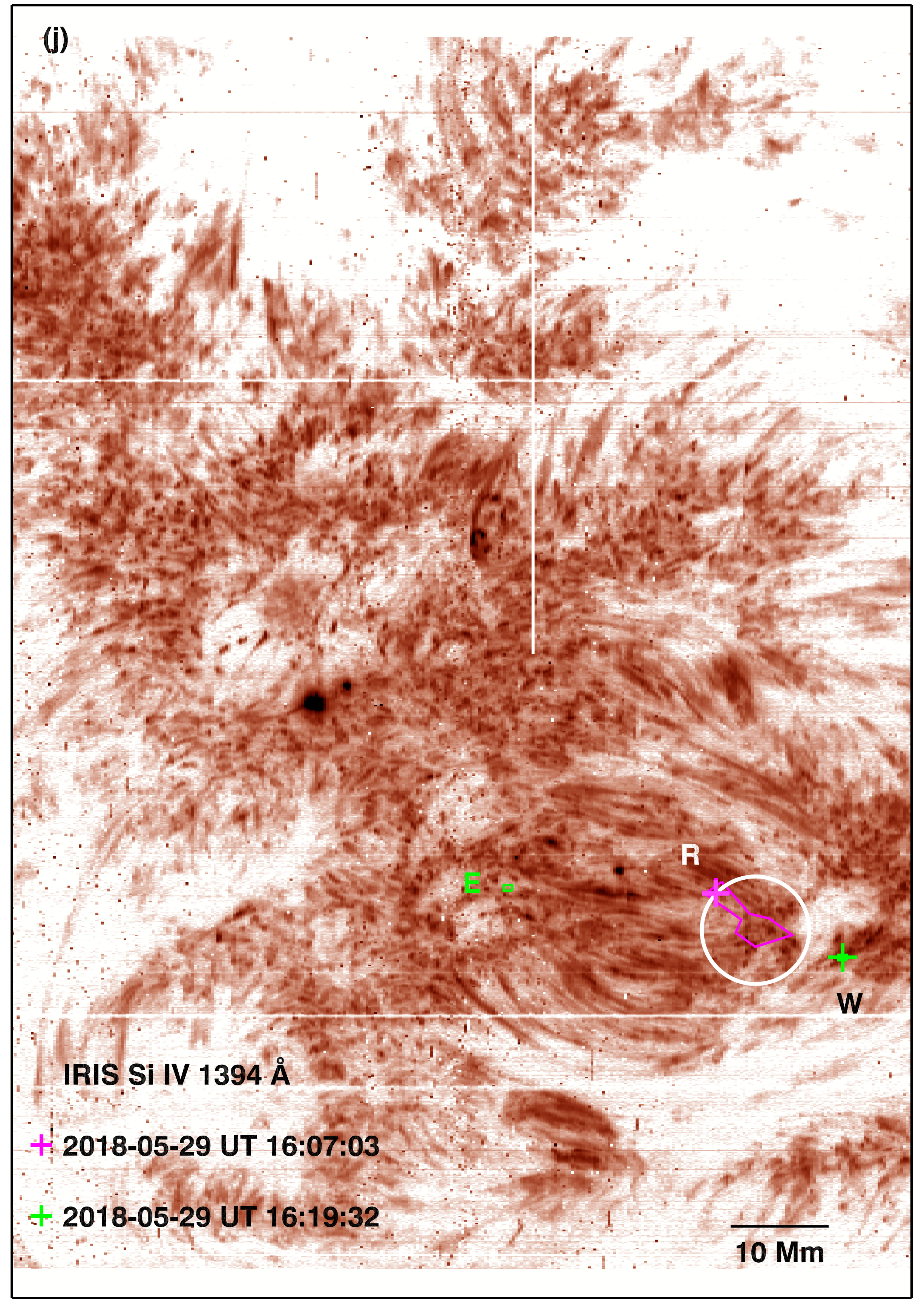}}
\includegraphics[width=0.45\textwidth]{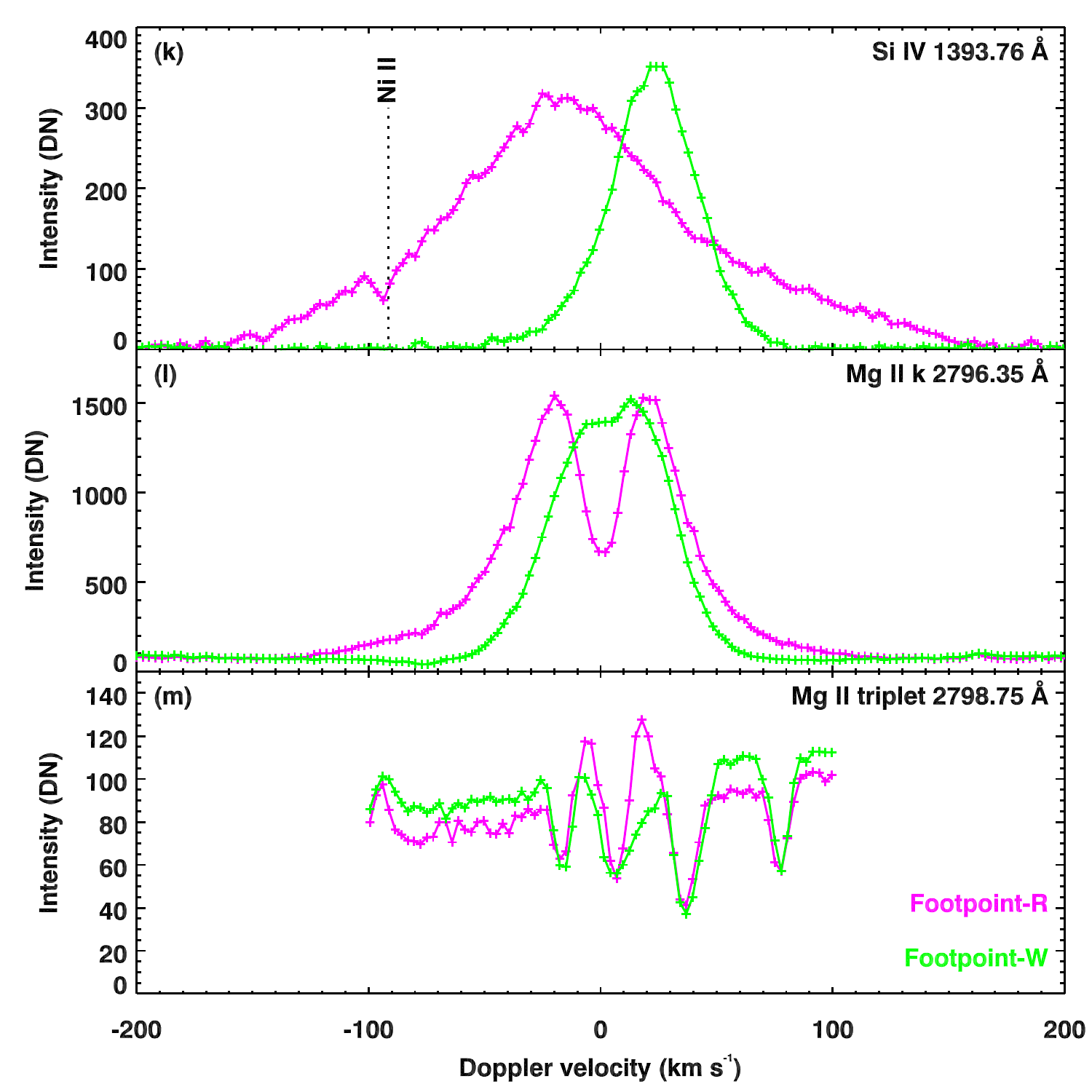}
\caption{Similar impulsive heating in AR 12712 as in Fig.\,\ref{fig:AR12712c1}, but after 16\,hours. The format is the same as in Fig.\,\ref{fig:AR12692c1}. Animation of panels (a) to (g) is available online. \label{fig:AR12712c2}}
\end{center}
\end{figure*}

\begin{table*}
\begin{center}
\caption{Overview of impulsive heating events in the core of active region AR12712.\label{tab:AR12712}}
\begin{tabular}{l c c  c c c}
\hline\hline
Event No. & Fe\,{\sc xviii} peak time (UT) & Solar (\textit{X}, \textit{Y}) & Event type & \textbf{D} (10$^{15}$ Mx s$^{-1}$) & \textbf{M} (10$^{15}$ Mx s$^{-1}$) \\
\hline
1   & 2018-05-27 UT 18:53:59   & (-526.0\arcsec, 268.5\arcsec)   & Unipolar   & \textendash   & \textendash   \\
2   & 2018-05-28 UT 11:57:59   & (-406.6\arcsec, 270.8\arcsec)   & Unipolar   & \textendash   & \textendash   \\
3   & 2018-05-28 UT 15:07:59   & (-402.4\arcsec, 271.4\arcsec)   & Mixed   &   -0.5   &    2.5   \\
4   & 2018-05-28 UT 17:15:59   & (-381.9\arcsec, 269.2\arcsec)   & Mixed   &    2.6   &   -5.6   \\
5   & 2018-05-28 UT 19:23:59   & (-358.8\arcsec, 266.9\arcsec)   & Mixed   &   -8.7   &    0.4   \\
6   & 2018-05-28 UT 21:43:59   & (-344.6\arcsec, 262.8\arcsec)   & Mixed   &   -5.4   &    1.1   \\
7   & 2018-05-29 UT 00:41:59   & (-265.5\arcsec, 248.1\arcsec)   & Mixed   &    4.6   &  -11.3   \\
8   & 2018-05-29 UT 03:13:59   & (-246.8\arcsec, 251.0\arcsec)   & Mixed   &   -6.7   &   -3.9   \\
9   & 2018-05-29 UT 16:23:59   & (-127.7\arcsec, 244.3\arcsec)   & Mixed   &   -5.3   &   -0.3   \\
10   & 2018-05-30 UT 20:09:59   & (88.8\arcsec, 264.6\arcsec)   & Mixed   &    2.5   &   -3.0   \\
11   & 2018-05-31 UT 01:11:59   & (138.7\arcsec, 282.9\arcsec)   & Unipolar   & \textendash   & \textendash   \\
12   & 2018-05-31 UT 04:35:59   & (230.6\arcsec, 261.1\arcsec)   & Unipolar   & \textendash   & \textendash   \\
13   & 2018-05-31 UT 07:01:59   & (178.4\arcsec, 278.1\arcsec)   & Unipolar   & \textendash   & \textendash   \\
14   & 2018-05-31 UT 08:45:59   & (199.3\arcsec, 270.8\arcsec)   & Mixed   &    9.3   &   -4.8   \\
15   & 2018-05-31 UT 11:01:59   & (207.6\arcsec, 267.5\arcsec)   & Mixed   &   -2.7   &   -0.9   \\
16   & 2018-05-31 UT 12:57:59   & (297.4\arcsec, 241.2\arcsec)   & Unipolar   & \textendash   & \textendash   \\
17   & 2018-05-31 UT 16:13:59   & (260.1\arcsec, 264.8\arcsec)   & Mixed   &   12.5   &    2.3   \\
18   & 2018-05-31 UT 19:15:59   & (293.0\arcsec, 271.7\arcsec)   & Mixed   &   -1.5   &    3.9   \\
19   & 2018-05-31 UT 21:09:59   & (294.5\arcsec, 265.9\arcsec)   & Mixed   &   -1.2   &    3.7   \\
20   & 2018-05-31 UT 23:03:59   & (312.6\arcsec, 267.2\arcsec)   & Mixed   &    7.9   &   -3.1   \\
21   & 2018-06-01 UT 02:41:59   & (355.8\arcsec, 269.7\arcsec)   & Mixed   &    0.4   &   -1.6   \\
22   & 2018-06-01 UT 06:13:59   & (379.3\arcsec, 276.3\arcsec)   & Mixed   &    3.2   &   -3.5   \\
23   & 2018-06-01 UT 07:57:59   & (372.6\arcsec, 270.5\arcsec)   & Mixed   &   -1.1   &   -3.4   \\
24   & 2018-06-01 UT 12:23:59   & (508.8\arcsec, 227.9\arcsec)   & Unipolar   & \textendash   & \textendash   \\
25   & 2018-06-01 UT 17:35:59   & (530.1\arcsec, 243.2\arcsec)   & Unipolar   & \textendash   & \textendash   \\
26   & 2018-06-01 UT 20:39:59   & (552.8\arcsec, 238.0\arcsec)   & Unipolar   & \textendash   & \textendash   \\
27   & 2018-06-02 UT 02:29:59   & (577.7\arcsec, 238.3\arcsec)   & Unipolar   & \textendash   & \textendash   \\
28   & 2018-06-02 UT 04:33:59   & (601.0\arcsec, 234.5\arcsec)   & Unipolar   & \textendash   & \textendash   \\
29   & 2018-06-02 UT 06:29:59   & (590.0\arcsec, 241.9\arcsec)   & Mixed   &    4.3   &   -0.0   \\
30   & 2018-06-02 UT 08:13:59   & (571.6\arcsec, 254.0\arcsec)   & Mixed   &   -0.1   &    0.0   \\
31   & 2018-06-02 UT 11:27:59   & (620.2\arcsec, 263.0\arcsec)   & Mixed   &    0.1   &   -0.1   \\
32   & 2018-06-02 UT 14:55:59   & (643.4\arcsec, 259.5\arcsec)   & Mixed   &   -0.4   &   -0.2   \\
\hline
\end{tabular}
\end{center}
\end{table*}

\clearpage

\newpage
\subsection{AR 12713}

\begin{figure*}
\begin{center}
\includegraphics[width=\textwidth]{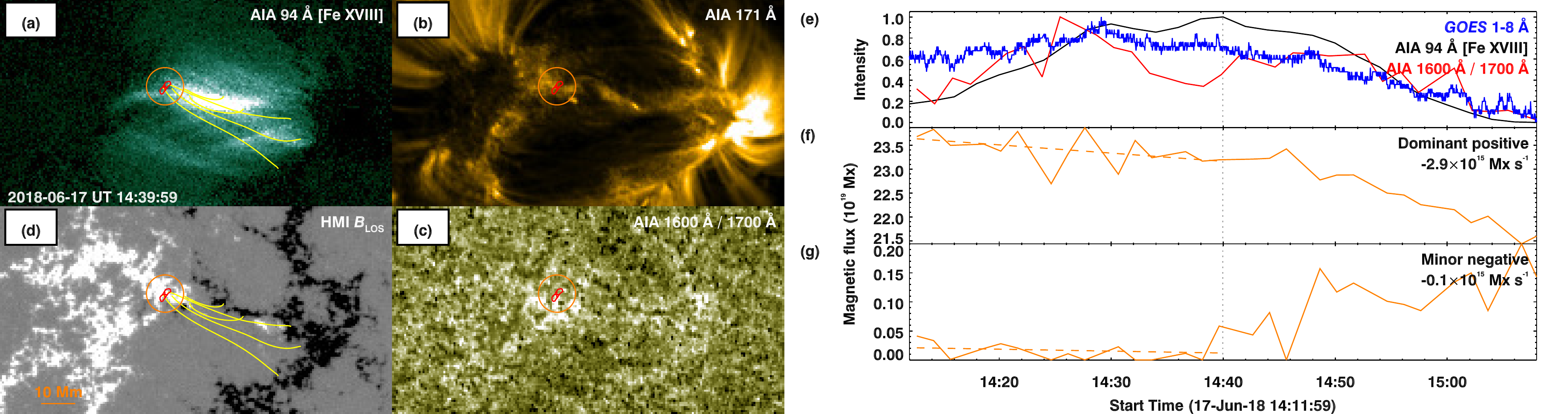}
\includegraphics[width=0.45\textwidth]{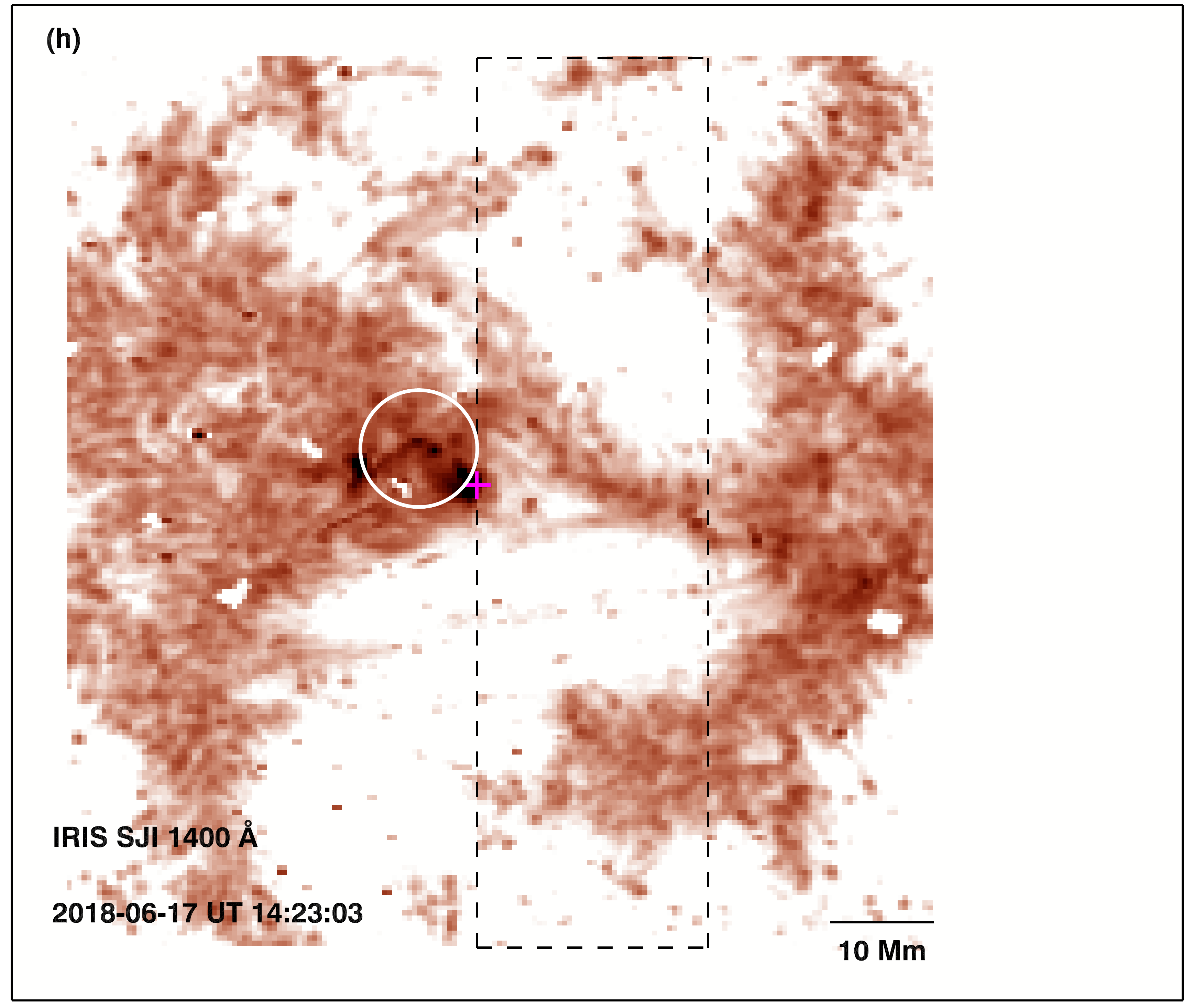}
\includegraphics[width=0.45\textwidth]{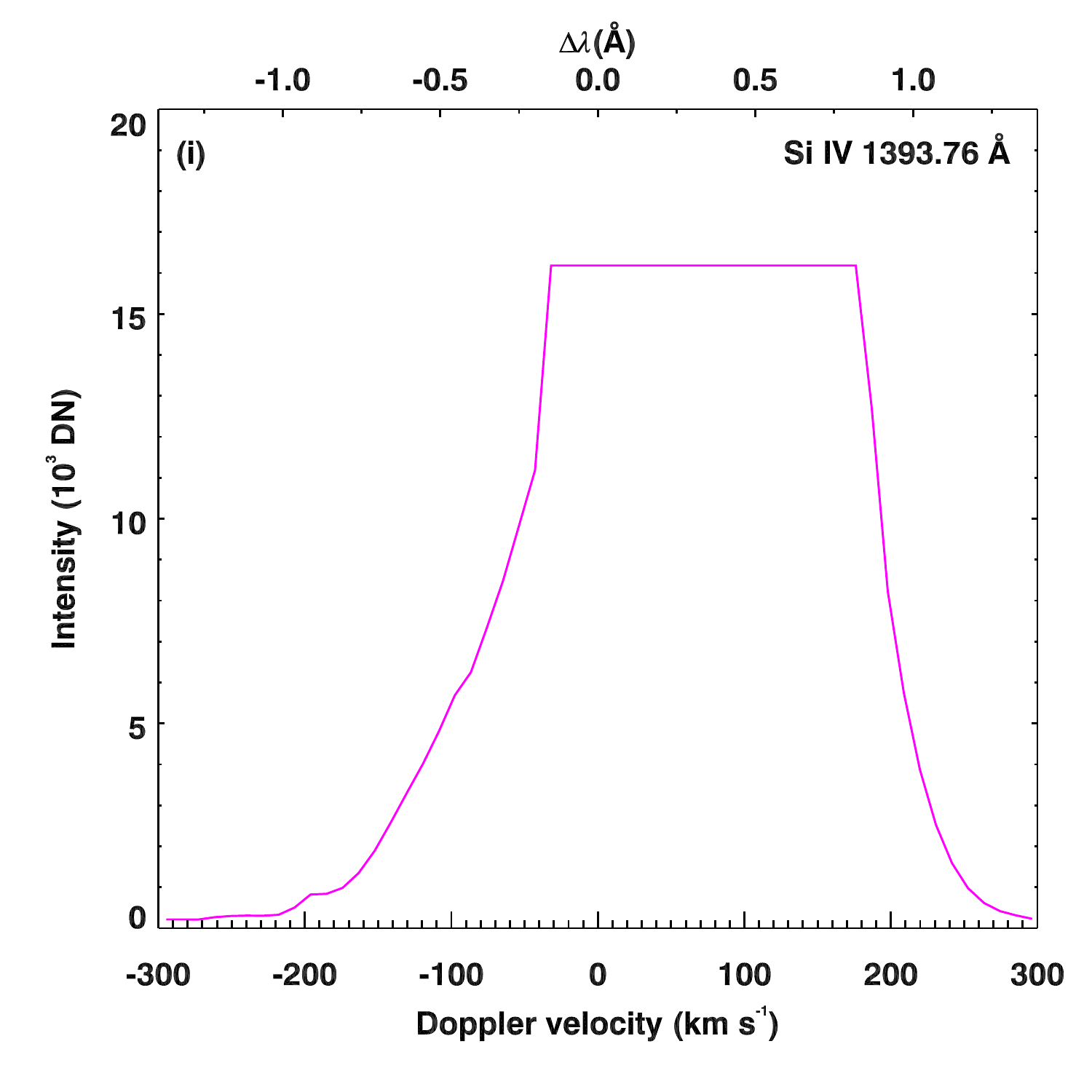}
\caption{Impulsive heating in the core of AR 12713. The format is the same as in Fig.\,\ref{fig:AR12712c1}. The bottom left panel displays IRIS SJI 1400\,\AA\ map in an inverted grey scale. The dashed region outlines the spatial extent of raster observations in this case. The circular region from the top panels is overlaid. The magenta plus symbol marks a feature at the edge of the circle. In the lower right panel we show the Si\,{\sc iv} 1394\,\AA\ spectral profile at the location of plus symbol. The profile is saturated due mainly to the 60\,s long exposure time used in this observation. AIA light curves are shown at lower cadence. Animation of panels (a) to (g) is available online. \label{fig:AR12713c1}}
\end{center}
\end{figure*}

\begin{table*}
\begin{center}
\caption{Overview of impulsive heating events in the core of active region AR12713.\label{tab:AR12713}}
\begin{tabular}{l c c  c c c}
\hline\hline
Event No. & Fe\,{\sc xviii} peak time (UT) & Solar (\textit{X}, \textit{Y}) & Event type & \textbf{D} (10$^{15}$ Mx s$^{-1}$) & \textbf{M} (10$^{15}$ Mx s$^{-1}$) \\
\hline
1   & 2018-06-17 UT 04:25:59   & (-\arcsec, -\arcsec)   & Unclear   & \textendash   & \textendash   \\
2   & 2018-06-17 UT 05:33:59   & (-231.4\arcsec, 63.5\arcsec)   & Mixed   &    5.0   &    0.5   \\
3   & 2018-06-17 UT 07:01:59   & (-178.2\arcsec, 59.1\arcsec)   & Mixed   &   -0.6   &   -0.0   \\
4   & 2018-06-17 UT 08:29:59   & (-195.9\arcsec, 68.1\arcsec)   & Mixed   &   -1.7   &    0.0   \\
5   & 2018-06-17 UT 10:59:59   & (-163.7\arcsec, 29.9\arcsec)   & Unipolar   & \textendash   & \textendash   \\
6   & 2018-06-17 UT 14:39:59   & (-154.9\arcsec, 66.1\arcsec)   & Mixed   &   -2.9   &   -0.1   \\
7   & 2018-06-17 UT 16:47:59   & (-117.2\arcsec, 94.7\arcsec)   & Mixed   &    1.6   &    0.0   \\
8   & 2018-06-17 UT 18:19:59   & (-126.7\arcsec, 58.3\arcsec)   & Mixed   &    3.1   &   -0.4   \\
9   & 2018-06-17 UT 20:17:59   & (-120.5\arcsec, 33.2\arcsec)   & Unipolar   & \textendash   & \textendash   \\
10   & 2018-06-17 UT 22:57:59   & (-18.5\arcsec, 56.8\arcsec)   & Unipolar   & \textendash   & \textendash   \\
11   & 2018-06-18 UT 04:05:59   & (29.8\arcsec, 54.5\arcsec)   & Unipolar   & \textendash   & \textendash   \\
12   & 2018-06-18 UT 05:11:59   & (-\arcsec, -\arcsec)   & Unclear   & \textendash   & \textendash   \\
13   & 2018-06-18 UT 06:23:59   & (-15.6\arcsec, 65.8\arcsec)   & Mixed   &    2.9   &    1.6   \\
14   & 2018-06-18 UT 08:45:59   & (33.0\arcsec, 62.4\arcsec)   & Unipolar   & \textendash   & \textendash   \\
\hline
\end{tabular}
\end{center}
\end{table*}

\clearpage

\newpage

\subsection{AR 12733}

\begin{figure*}
\begin{center}
\includegraphics[width=0.5\textwidth]{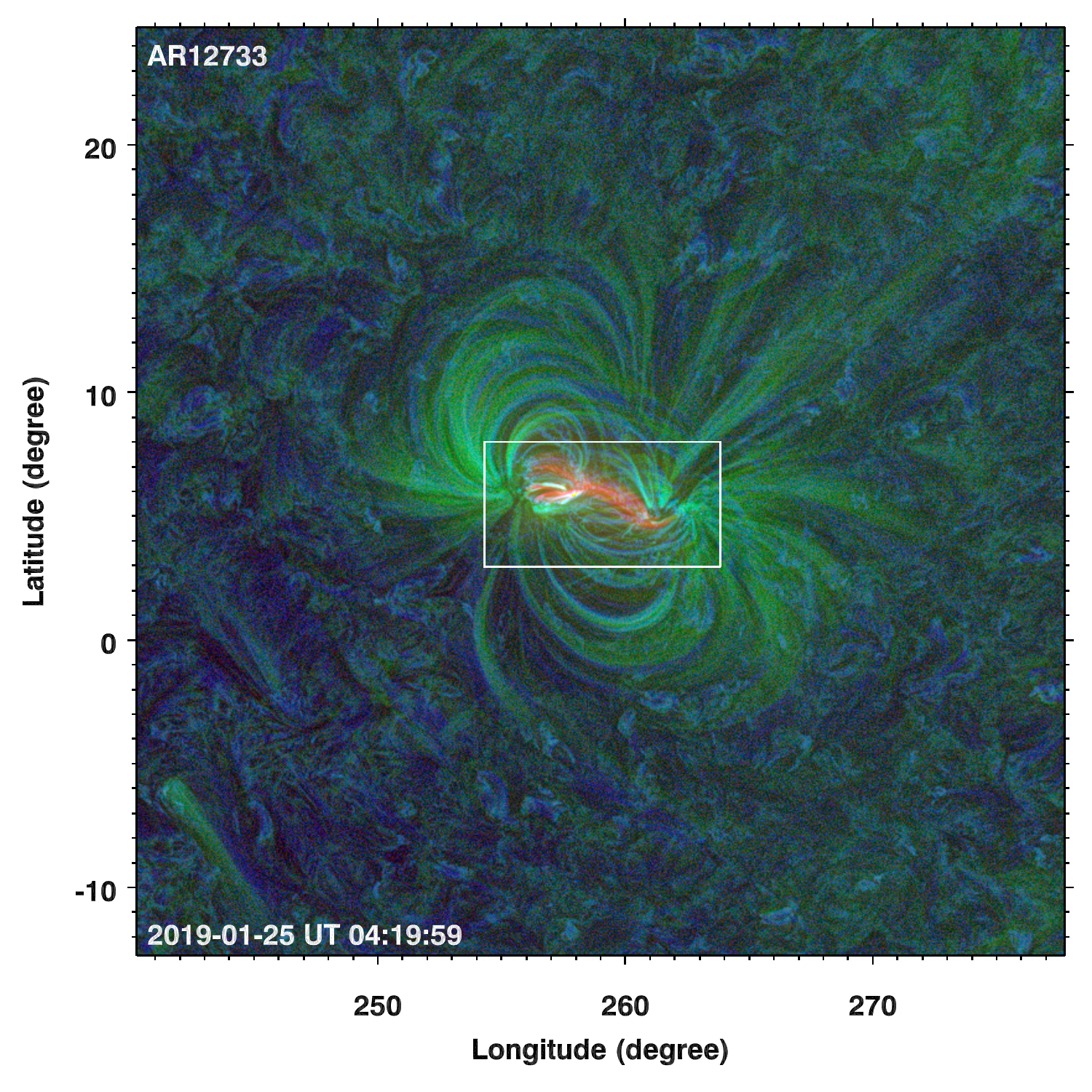}
\includegraphics[width=\textwidth]{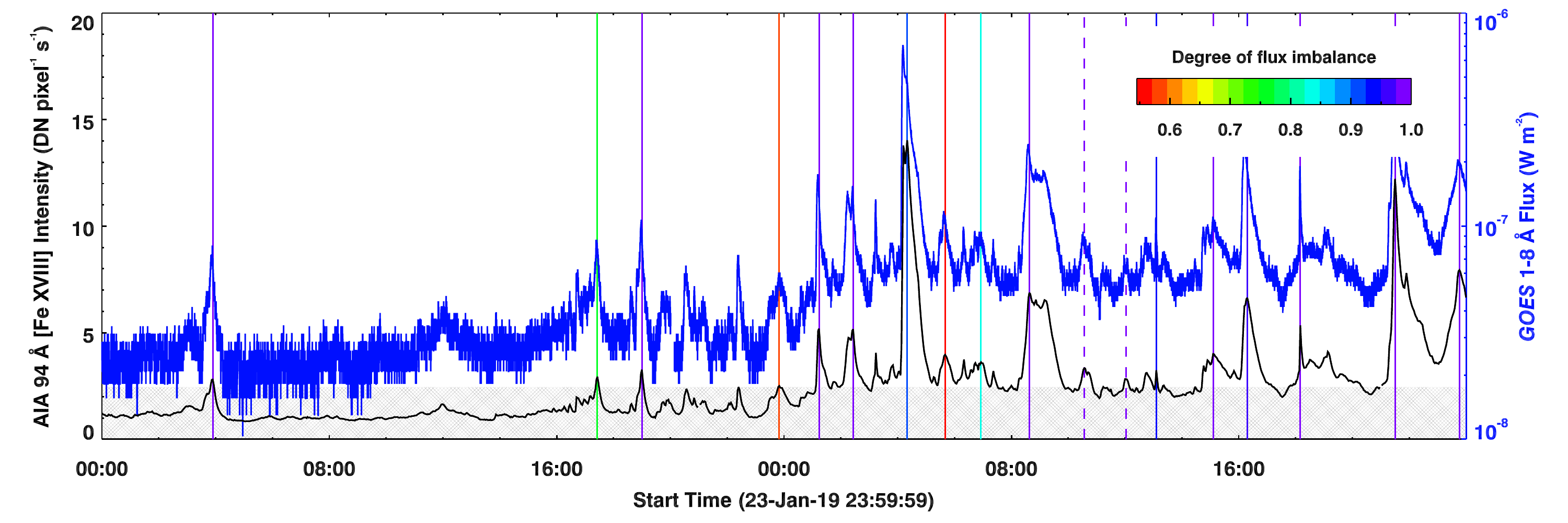}
\caption{Overview of impulsive heating events observed in AR 12733. The format is the same as in Fig.\,\ref{fig:over1}. \label{fig:over6}}
\end{center}
\end{figure*}

\begin{figure*}
\begin{center}
\includegraphics[width=\textwidth]{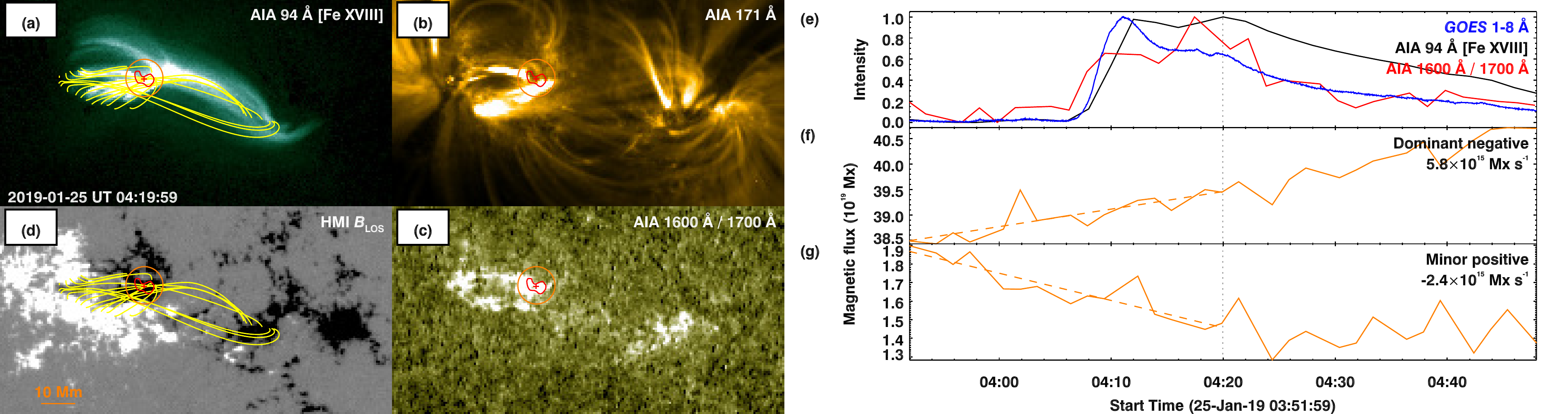}
\caption{Impulsive heating in the core of AR 12733. The format is the same as in Fig.\,\ref{fig:AR12665c1}. AIA light curves are shown at lower cadence. Animation of panels (a) to (g) is available online. \label{fig:AR12733c1}}
\end{center}
\end{figure*}

\begin{table*}
\begin{center}
\caption{Overview of impulsive heating events in the core of active region AR12733.\label{tab:AR12733}}
\begin{tabular}{l c c  c c c}
\hline\hline
Event No. & Fe\,{\sc xviii} peak time (UT) & Solar (\textit{X}, \textit{Y}) & Event type & \textbf{D} (10$^{15}$ Mx s$^{-1}$) & \textbf{M} (10$^{15}$ Mx s$^{-1}$) \\
\hline
1   & 2019-01-24 UT 03:53:59   & (-73.9\arcsec, 189.1\arcsec)   & Mixed   &   -4.8   &    0.1   \\
2   & 2019-01-24 UT 17:25:59   & (29.8\arcsec, 194.8\arcsec)   & Mixed   &    8.2   &    2.8   \\
3   & 2019-01-24 UT 18:59:59   & (24.0\arcsec, 189.1\arcsec)   & Mixed   &   -3.9   &    0.4   \\
4   & 2019-01-24 UT 23:49:59   & (98.2\arcsec, 187.5\arcsec)   & Mixed   &    4.1   &    2.1   \\
5   & 2019-01-25 UT 01:13:59   & (79.7\arcsec, 186.9\arcsec)   & Mixed   &   -2.4   &    0.1   \\
6   & 2019-01-25 UT 02:25:59   & (157.9\arcsec, 172.6\arcsec)   & Mixed   &   27.4   &   -0.2   \\
7   & 2019-01-25 UT 04:19:59   & (131.1\arcsec, 194.9\arcsec)   & Mixed   &    5.8   &   -2.4   \\
8   & 2019-01-25 UT 05:39:59   & (144.3\arcsec, 184.7\arcsec)   & Mixed   &   -2.6   &    2.8   \\
9   & 2019-01-25 UT 06:55:59   & (153.2\arcsec, 193.8\arcsec)   & Mixed   &   -4.8   &   -1.7   \\
10   & 2019-01-25 UT 08:37:59   & (224.3\arcsec, 173.2\arcsec)   & Mixed   &   -8.8   &    1.4   \\
11   & 2019-01-25 UT 10:33:59   & (195.6\arcsec, 194.8\arcsec)   & Unipolar   & \textendash   & \textendash   \\
12   & 2019-01-25 UT 12:01:59   & (173.3\arcsec, 212.7\arcsec)   & Unipolar   & \textendash   & \textendash   \\
13   & 2019-01-25 UT 13:05:59   & (261.5\arcsec, 162.3\arcsec)   & Mixed   &    1.1   &   -0.5   \\
14   & 2019-01-25 UT 15:05:59   & (281.0\arcsec, 167.7\arcsec)   & Mixed   &    2.9   &    0.3   \\
15   & 2019-01-25 UT 16:17:59   & (255.6\arcsec, 177.0\arcsec)   & Mixed   &   -1.4   &   -0.3   \\
16   & 2019-01-25 UT 18:09:59   & (250.0\arcsec, 185.6\arcsec)   & Mixed   &   -1.6   &    0.2   \\
17   & 2019-01-25 UT 21:29:59   & (360.2\arcsec, 168.9\arcsec)   & Mixed   &  -25.4   &    0.1   \\
18   & 2019-01-25 UT 23:45:59   & (383.8\arcsec, 163.9\arcsec)   & Mixed   &  -10.9   &    0.3   \\
\hline
\end{tabular}
\end{center}
\end{table*}

\clearpage
\newpage

\subsection{AR 12738}

\begin{figure*}
\begin{center}
\includegraphics[width=0.5\textwidth]{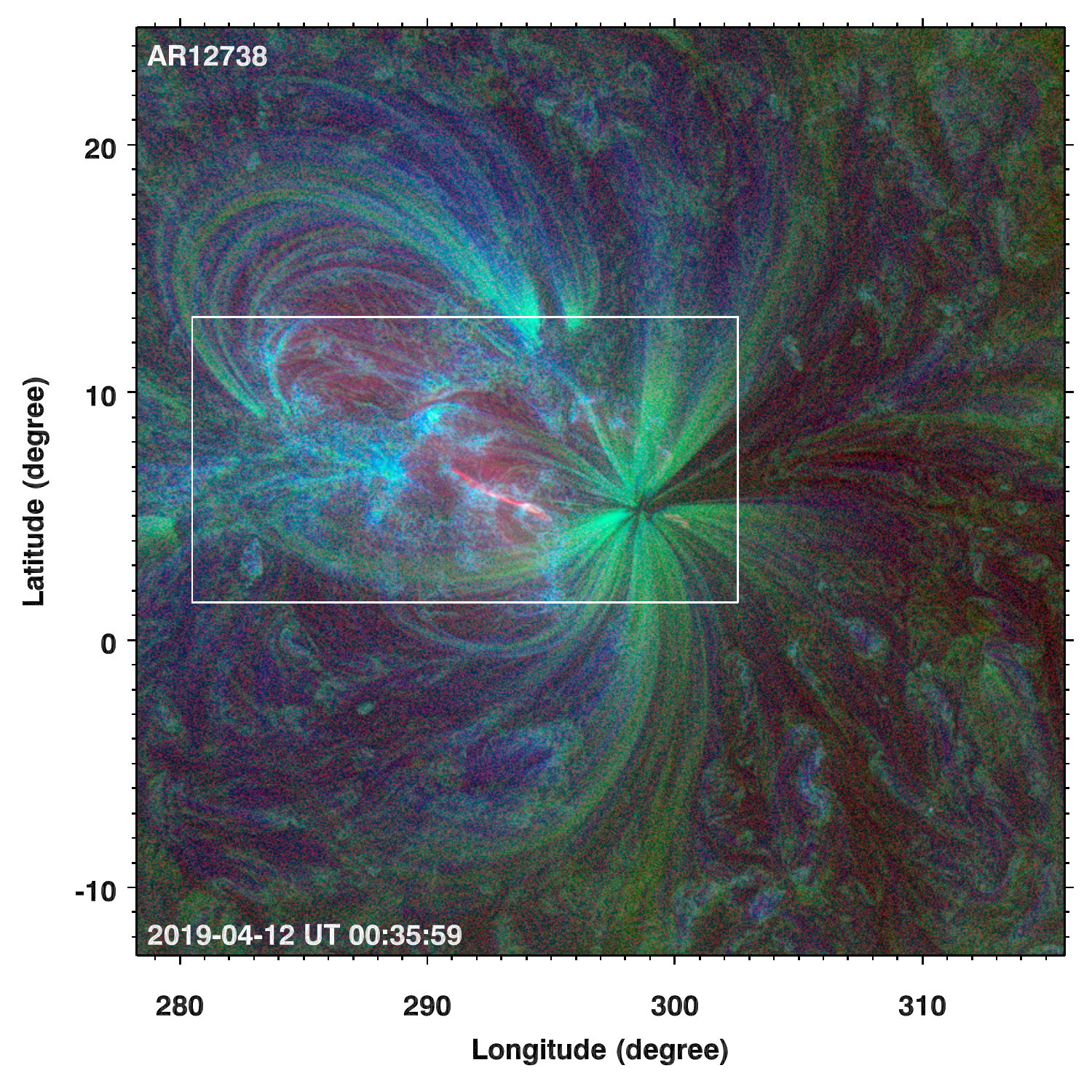}
\includegraphics[width=\textwidth]{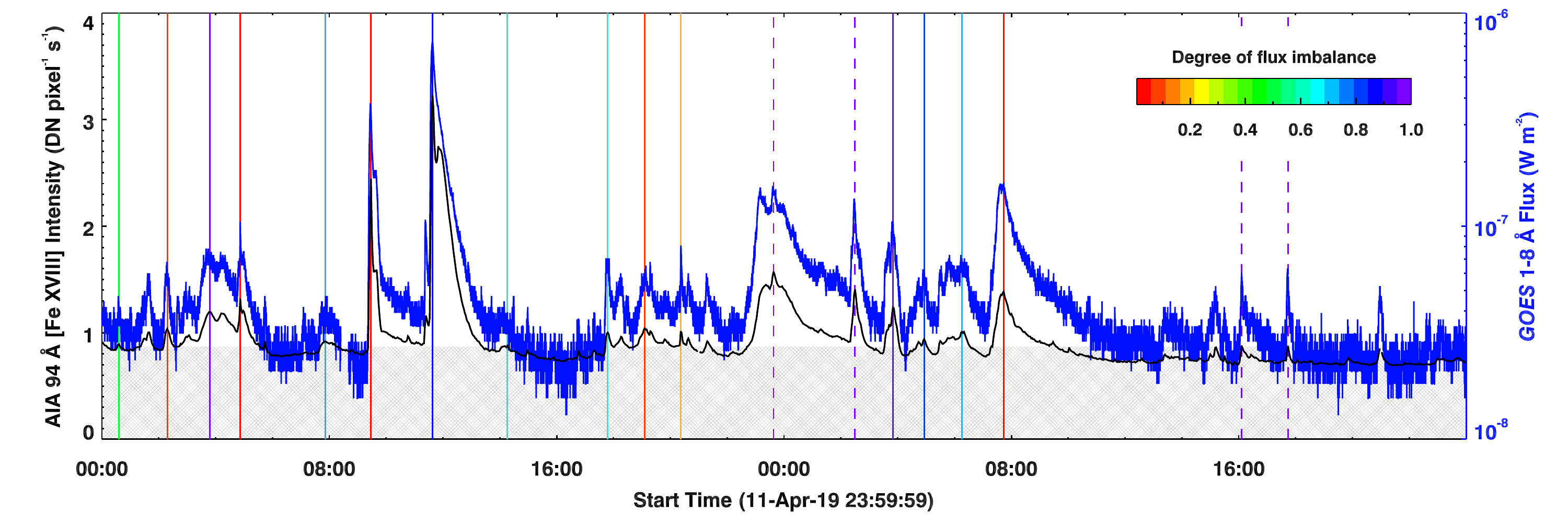}
\caption{Overview of impulsive heating events observed in AR 12738. The format is the same as in Fig.\,\ref{fig:over1}. \label{fig:over7}}
\end{center}
\end{figure*}

\begin{table*}
\begin{center}
\caption{Overview of impulsive heating events in the core of active region AR12738.\label{tab:AR12738}}
\begin{tabular}{l c c  c c c}
\hline\hline
Event No. & Fe\,{\sc xviii} peak time (UT) & Solar (\textit{X}, \textit{Y}) & Event type & \textbf{D} (10$^{15}$ Mx s$^{-1}$) & \textbf{M} (10$^{15}$ Mx s$^{-1}$) \\
\hline
1   & 2019-04-12 UT 00:35:59   & (-372.4\arcsec, 177.3\arcsec)   & Mixed   &    2.2   &   -0.9   \\
2   & 2019-04-12 UT 02:17:59   & (-294.6\arcsec, 209.7\arcsec)   & Mixed   &    2.2   &    2.2   \\
3   & 2019-04-12 UT 03:47:59   & (-345.9\arcsec, 200.2\arcsec)   & Mixed   &    2.3   &   -0.8   \\
4   & 2019-04-12 UT 04:51:59   & (-332.3\arcsec, 168.6\arcsec)   & Mixed   &    3.5   &   -3.3   \\
5   & 2019-04-12 UT 07:51:59   & (-290.7\arcsec, 159.2\arcsec)   & Mixed   &   -2.2   &   -1.3   \\
6   & 2019-04-12 UT 09:27:59   & (-230.1\arcsec, 214.2\arcsec)   & Mixed   &    0.6   &    3.5   \\
7   & 2019-04-12 UT 11:37:59   & (-242.1\arcsec, 206.5\arcsec)   & Mixed   &    0.9   &    1.6   \\
8   & 2019-04-12 UT 14:15:59   & (-255.1\arcsec, 181.4\arcsec)   & Mixed   &   -2.5   &    3.4   \\
9   & 2019-04-12 UT 17:47:59   & (-305.8\arcsec, 237.8\arcsec)   & Mixed   &    1.0   &    0.6   \\
10   & 2019-04-12 UT 19:05:59   & (-191.3\arcsec, 188.0\arcsec)   & Mixed   &   -3.3   &    5.0   \\
11   & 2019-04-12 UT 20:21:59   & (-263.7\arcsec, 252.3\arcsec)   & Mixed   &   -2.8   &   -0.7   \\
12   & 2019-04-12 UT 23:37:59   & (-133.4\arcsec, 243.5\arcsec)   & Unipolar   & \textendash   & \textendash   \\
13   & 2019-04-13 UT 02:29:59   & (-179.4\arcsec, 194.0\arcsec)   & Unipolar   & \textendash   & \textendash   \\
14   & 2019-04-13 UT 03:49:59   & (-89.4\arcsec, 183.1\arcsec)   & Mixed   &   -2.6   &    0.1   \\
15   & 2019-04-13 UT 04:55:59   & (-88.6\arcsec, 249.1\arcsec)   & Mixed   &   -3.1   &    1.2   \\
16   & 2019-04-13 UT 06:15:59   & (-96.5\arcsec, 178.6\arcsec)   & Mixed   &   -8.3   &   -2.3   \\
17   & 2019-04-13 UT 07:43:59   & (-21.8\arcsec, 222.4\arcsec)   & Mixed   &    2.1   &    0.8   \\
18   & 2019-04-13 UT 16:05:59   & (10.8\arcsec, 253.2\arcsec)   & Unipolar   & \textendash   & \textendash   \\
19   & 2019-04-13 UT 17:43:59   & (86.3\arcsec, 177.5\arcsec)   & Unipolar   & \textendash   & \textendash   \\
\hline
\end{tabular}
\end{center}
\end{table*}

\clearpage
\newpage

\section{Unipolar case \label{sec:uni}}

\begin{figure*}
\begin{center}
\includegraphics[width=\textwidth]{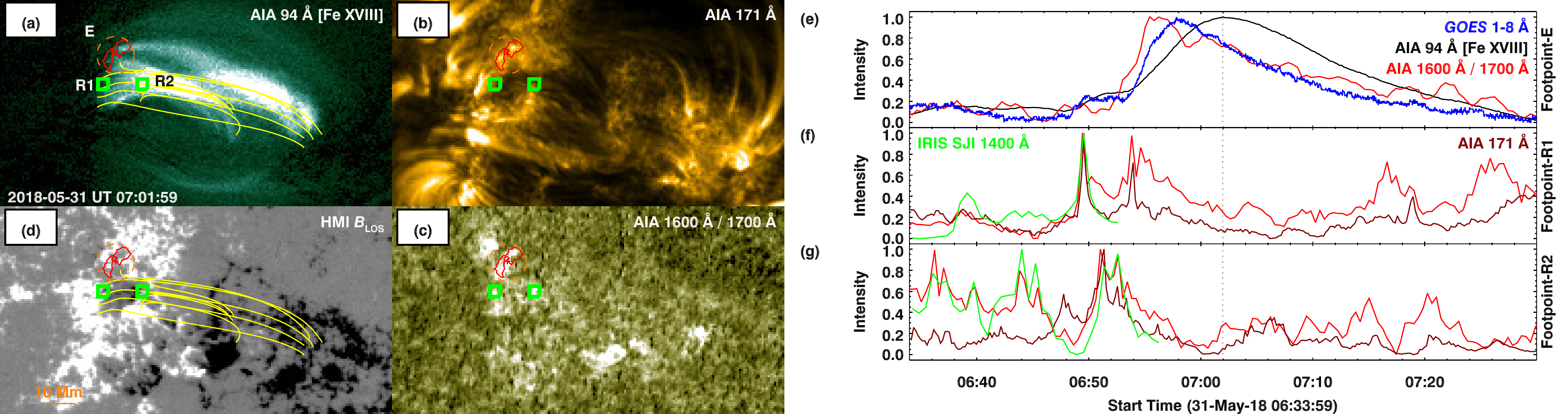}
\includegraphics[width=0.45\textwidth]{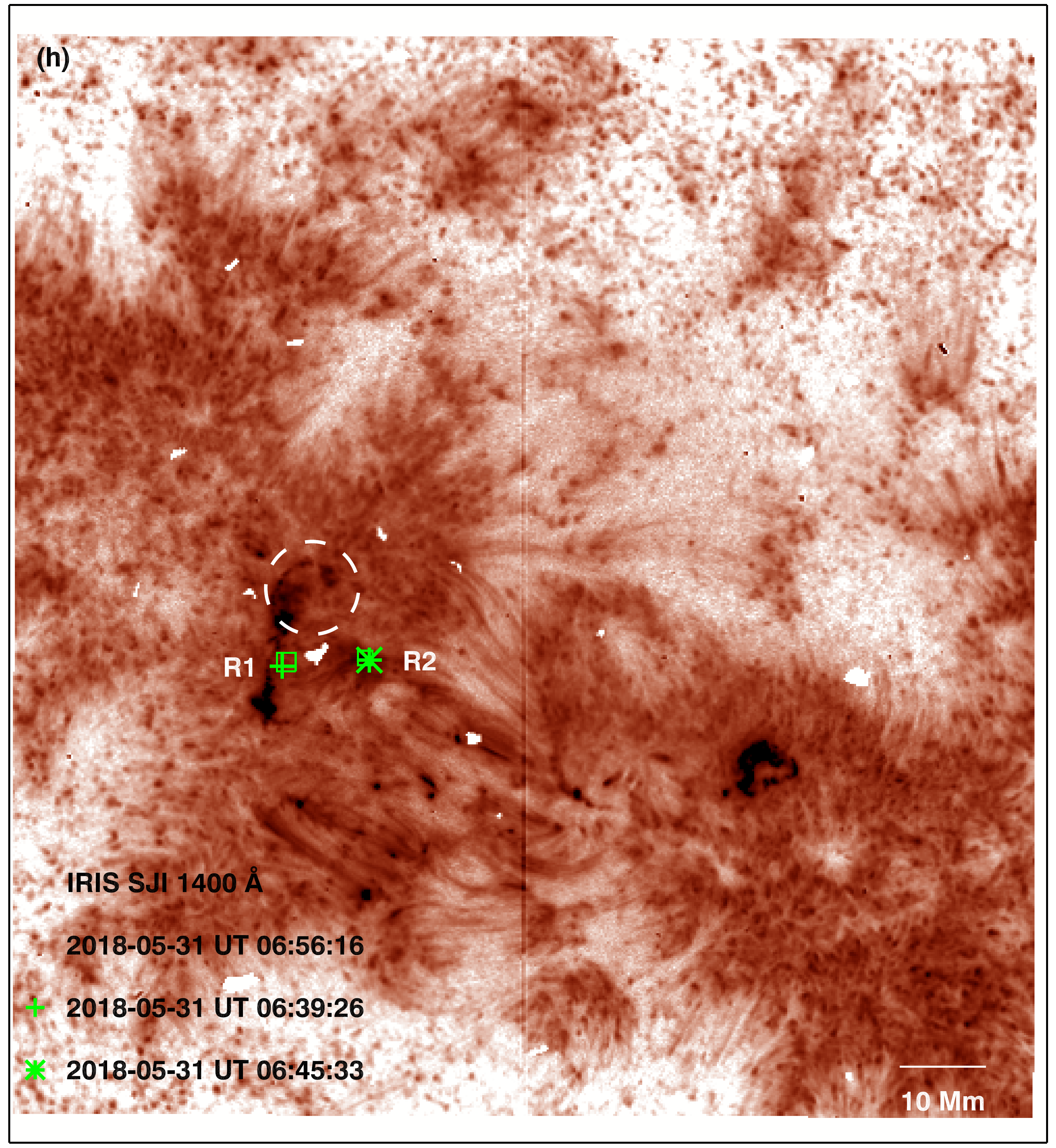}
\includegraphics[width=0.45\textwidth]{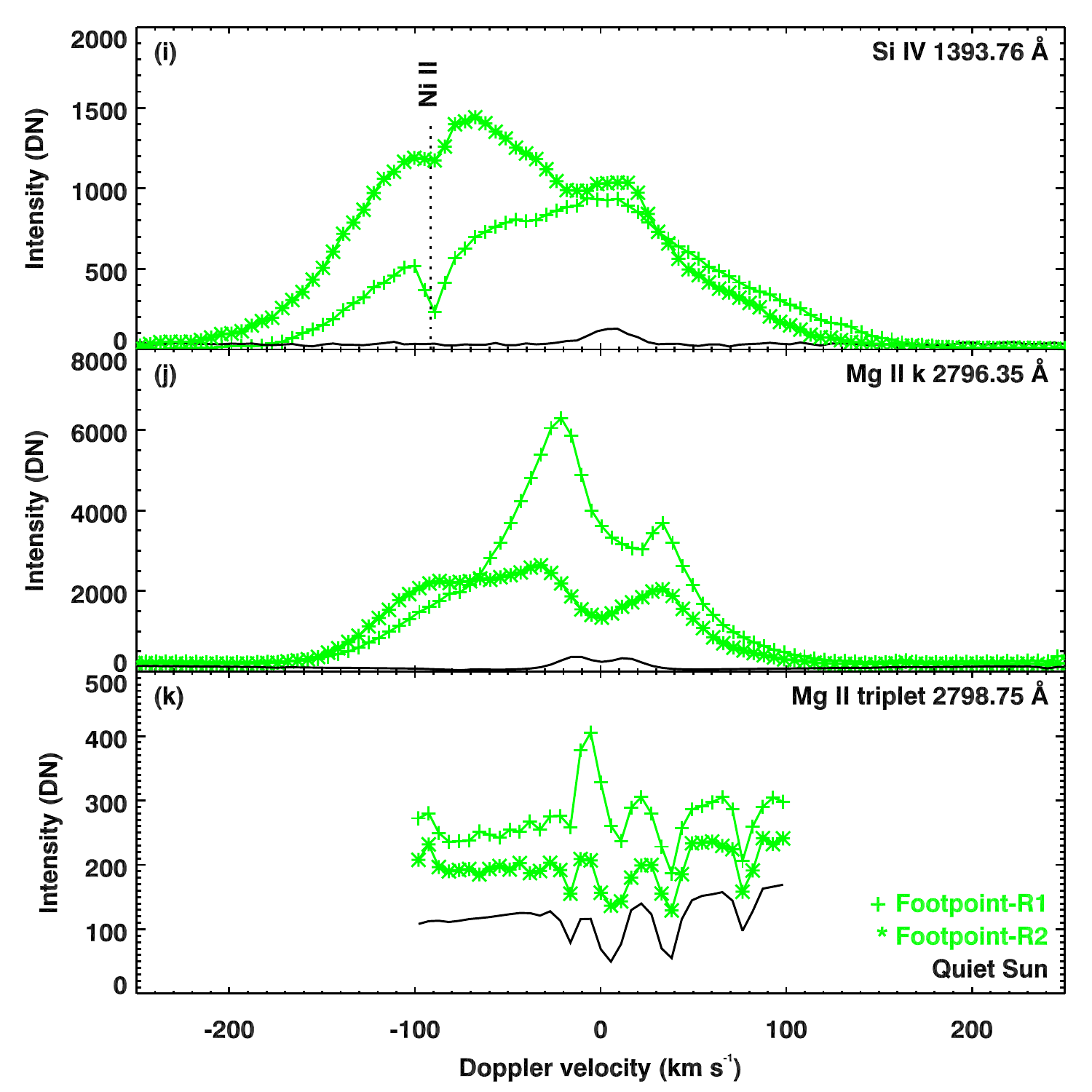}
\caption{Impulsive heating in the core of AR 12712. This is an example of a hot loop system with the detected footpoint rooted in an apparent magnetic unipolar region (footpoint E, lying in the dashed circle), whereas the undetected Footpoints R1 and R2 lie in mixed polarity locations. The format is the same as in Fig.\,\ref{fig:AR12692c1}. Animation of panels (a) to (g) is available online. See Appendix\,\ref{sec:uni} for details. \label{fig:AR12712c3}}
\end{center}
\end{figure*}

Based on an automated identification of the footpoints of hot loops (see Sect.\,\ref{sec:obs} and Appendix\,\ref{sec:met}), we found that in 34 of the 137 analysed cases, the detected footpoint is rooted in unipolar regions (i.e. the footpoint is devoid of a significant opposite-polarity magnetic patch within a circular zone of radius 5.4\,Mm). Post-analysis, through visual examination of these so-called unipolar cases, we found examples in which the undetected footpoints were rooted in magnetic mixed-polarity regions at the solar surface. In Fig.\,\ref{fig:AR12712c3} we discuss one such clear example. The hot loop system in Fig.\,\ref{fig:AR12712c3}(a) has a complex morphology with multiple, extended footpoint regions. The footpoint detected by our method is labelled as footpoint-E, which lies over a (positive) unipolar surface magnetic field (Fig.\,\ref{fig:AR12712c3}d)). Indeed, the UV ratio signal from footpoint-E goes along with the core-integrated Fe\,{\sc xviii} emission and the GOES X-ray flux (panel e). Nevertheless, it is clear that the detected footpoint-E covers only a small area along a rather diffused footpoint region. For instance, magnetic field lines traced from the lfff extrapolations show that region R1 is also connected to the loop and so is the region between E and R1. Similarly, extrapolations reveal that a part of the loop system ends in region R2. Thus the loop system ends in closely spaced footpoints E, R1, and R2 on one side and at distinct locations on the other side.

HMI magnetograms reveal that patches of minor negative polarity magnetic field are embedded in the region between the footpoints R1-R2 (see online animation accompanying Fig.\,\ref{fig:AR12712c3}). IRIS diagnostics at R1 show broad line profiles (panels i-k) that are consistent with signatures of reconnection at the loop footpoint. In addition, magnetic interactions at R2 also resulted in similar broad spectral profiles. Intriguingly, in this case, the UV ratio signals from R1 and R2 are markedly different from that of footpoint-E and do not correlate well with the core-integrated Fe\,{\sc xviii} emission (panels e-g). It is apparent that due to this poor correlation, our method did not detect either R1 or R2 to be footpoints of the hot loop. This example reveals that different parts of the same (extended) footpoint of a hot loop might exhibit different emission characteristics. This example suggests that the percentage of hot loops with mixed magnetic polarities at their footpoints that we identified using a cross-correlation technique is an underestimate.

\end{appendix}

\end{document}